\documentclass[proof]{pasj00}

\begin{document}
\SetRunningHead{Author(s) in page-head}{Running Head}

\title{Near-Infrared Spectroscopy of Quasars at $z\sim3$ and Estimates of Their Supermassive Black Hole Masses}

\author{Yuriko \textsc{Saito},\altaffilmark{1,2}
        Masatoshi \textsc{Imanishi},\altaffilmark{1,2,3}
        Yosuke \textsc{Minowa},\altaffilmark{1,2}
        Tomoki \textsc{Morokuma},\altaffilmark{4}
        Toshihiro \textsc{Kawaguchi},\altaffilmark{5}
        Hiroaki \textsc{Sameshima},\altaffilmark{6}
        Takeo \textsc{Minezaki},\altaffilmark{4}
        Nagisa \textsc{Oi},\altaffilmark{7}
        Tohru \textsc{Nagao},\altaffilmark{8}
        Nozomu \textsc{Kawatatu},\altaffilmark{9}
        and
        Kenta \textsc{Matsuoka}\altaffilmark{10}
        }

\altaffiltext{1}{Department of Astronomy, School of Science, Graduate
University for Advanced Studies (SOKENDAI), Mitaka, Tokyo 181-8588, Japan}
\email{yuriko.saitou@nao.ac.jp}

\altaffiltext{2}{Subaru Telescope, 650 North A'ohoku Place, Hilo,
Hawaii, 96720, U.S.A.}

\altaffiltext{3}{National Astronomical Observatory, 2-21-1, Osawa,
Mitaka, Tokyo 181-8588, Japan}

\altaffiltext{4}{Institute of Astronomy, Graduate School of Science, The University of Tokyo, 2-21-1, 
Osawa, Mitaka, Tokyo 181-0015, Japan}

\altaffiltext{5}{Department of Physics, Sapporo Medical University, S1 W17,
Chuo-ku, Sapporo 060-556, Japan}

\altaffiltext{6}{Laboratory of Infrared High-resolution spectroscopy (LIH), Koyama Astronomical Observatory, Kyoto Sangyo University,
Motoyama, Kamigamo, Kita-ku, Kyoto 603-8555, Japan}

\altaffiltext{7}{Japan Aerospace Exploration Agency, Sagamihara, Kanagawa, 252-5210, Japan}

\altaffiltext{8}{Research Center for Space and Cosmic Evolution (RCSCE), Ehime University, 
Bunkyo-cho 2-5, Matsuyama, Ehime 790-8577, Japan}

\altaffiltext{9}{Kure National College of Technology, Kure, Hiroshima, 737-8506, Japan}


\altaffiltext{10}{Department of Astronomy, Kyoto University, Kitashirakawa-Oiwake-cho, Sakyo-ku, Kyoto 606-8502, Japan}


\maketitle

\begin{abstract}
We present the results of new infrared spectroscopic observations of 37 quasars at $z\sim3$, selected based on the optical $r'$-band magnitude and the availability of nearby bright stars for future imaging follow-up with Adaptive Optics system. The supermassive black hole (SMBH) masses ($M_\mathrm{BH}$) were successfully estimated in 28 out of 37 observed objects from the combination of the H$\beta$ emission linewidth and continuum luminosity at rest-frame 5100 $\mathrm{\AA}$. Comparing these results with those from previous studies of quasars with similar redshift, our sample exhibited slightly lower ($\sim-0.11$ dex in median) Eddington ratios; and, the SMBH masses are slightly ($\sim0.38$ dex in median) higher. The SMBH growth time, $t_\mathrm{grow}$, was calculated by dividing the estimated SMBH mass by the mass accretion rate measured using optical luminosity. We found, given reasonable assumptions, that $t_\mathrm{grow}$ was smaller than the age of the universe at the redshift of individual quasars for a large fraction of observed sources, suggesting that the SMBHs in many of our observed quasars are in growing phase with high accretion rates.
A comparison of the SMBH masses derived from our H$\beta$ data and archived C\emissiontype{IV} data indicated considerable scattering, as indicated in previous studies.
All quasars with measured SMBH masses have at least one nearby bright star, such that they are suitable targets for adaptive optics observations to study the mass relationship between SMBHs and host galaxies' stellar component at high redshift.
\end{abstract}

\section{Introduction}
The tight correlations between the properties of galaxy spheroidal components (e.g., mass $M_\mathrm{spheroid}$, velocity dispersion, and luminosity $L_\mathrm{spheroid}$) and the supermassive black hole (SMBH) mass ($M_\mathrm{BH}$) in the local universe (\cite{Kormendy1995}; \cite{Marconi2003}; \cite{McConnell2013}; see also \cite{Kormendy2013} for a review) indicate that the formation/growth of galaxies and central SMBHs are closely related (so-called coevolution).
Recent multi-wavelength deep surveys have revealed similar downsizing evolution among galaxies, active galactic nuclei (AGNs), and SMBHs in AGNs
(\cite{Cowie1996}; \cite{Kodama2004}; \cite{Ueda2003}; \cite{Vestergaard2008}), where more luminous, massive galaxies, more luminous AGNs, and more massive active SMBHs in AGNs show their number density peaks at higher redshift. This implies that the histories of cosmic star formation and the mass accretion of SMBHs may be synchronized (\cite{Franceschini1999}; \cite{Silverman2008}; \cite{Zheng2009}). 
It is important to observationally better constrain how SMBHs and galaxies have coevolved from the early to the current universe.

Depending on various galaxy- and SMBH-growth mechanisms, 
different redshift (z) evolutions of the $M_\mathrm{BH}$/$M_\mathrm{spheroid}$ ratio ($\equiv R_\mathrm{BH/spheroid}$) can be predicted.
(1) If the outflow from the central SMBH plays a major role in coevolution (\cite{Silk1998}), then
the $R_\mathrm{BH/spheroid}$ ratio is predicted to increase at higher redshift, due to the strong suppression of
SMBH accretion at a later age (\cite{Wyithe2003}).
(2) When the SMBH and spheroid growths are governed by major merging
of gas-rich galaxies, no significant redshift evolution in $R_\mathrm{BH/spheroid}$ is expected (e.g., \cite{Robertson2006}; see also \cite{Kawakatu2003}).
(3) If the transformation of disk stars to bulge stars during major mergers is taken into account,
in addition to the star formation triggered by mergers, the growth of the
$M_\mathrm{spheroid}$ is enhanced in the late stage, relative to SMBH growth. Hence, the $R_\mathrm{BH/spheroid}$
ratio is larger at higher redshift, but smaller than (1) (\cite{Croton2006}).
The differences among models are larger at higher redshift. Therefore, observational investigation of the $R_\mathrm{BH/spheroid}$ ratio at high redshift is essential to
apply constraints to the coevolution models and/or the allowable parameter ranges of the key processes. 

In the local universe, the $M_\mathrm{BH}$ estimate is obtained via spatially resolved spectroscopy of normal galaxies (e.g., \cite{Kuo2011}; \cite{Bosch2010}; \cite{Genzel2010}; \cite{Walsh2010}; \cite{Bender2005}; \cite{Cappellari2002}; \cite{Miyoshi1995}, see also \cite{Kormendy2013}; \cite{Kormendy1995}; \cite{Genzel1994} for reviews). However, it is impossible to apply the same method to distant normal galaxies, due to the lack of spatial resolution. 
Quasars (QSOs; highly luminous AGNs) are useful objects for studying SMBH masses at high redshift, because various strong emission lines from gas clouds, whose dynamics are dominated by the gravitational force of the central SMBHs, can be detected and used to estimate SMBH masses from single epoch spectroscopy, based on the combination of emission linewidths and nearby continuum luminosity (\cite{McLure2002}; \cite{Vestergaard2002}; \cite{Shemmer2004}; \cite{Wu2004}; \cite{Vestergaard2006}; \cite{Netzer2007}; \cite{Wang2009}; \cite{Trakhtenbrot2012}).

Previous studies of the $R_\mathrm{BH/spheroid}$ ratio at high redshift using QSOs were performed mainly at $z<2$ (e.g., \cite{Haring2004}; \cite{Jahnke2004}; \cite{Jahnke2009}; \cite{Decarli2010}; \cite{Merloni2010}; \cite{Bennert2011}; \cite{Cisternas2011}; \cite{Schramm2013}; \cite{Matsuoka2014}). However, to better distinguish among various coevolution scenarios, data at higher redshift may be preferable. On the other hand, 
detection of the QSO host galaxy to measure the $M_\mathrm{spheroid}$ becomes more difficult at higher redshift, because the surface brightness of the host galaxy stellar emission becomes faint in proportion to $(1+z)^{4}$. Thus, the optimal redshift range must be determined by taking into account the practical observational limitations.

To estimate $M_\mathrm{BH}$ of distant QSOs at $z>2$, the C\emissiontype{IV} $\lambda$1549 emission line is commonly used, because it is redshifted into the optical wavelength range, facilitating observation. However, it often shows a large blueshift and an asymmetric profile with respect to low ionization lines (e.g., \cite{Gaskell1982}; \cite{Tytler1992}; \cite{Richards2002}; \cite{Shen2008}, \yearcite{Shen2011}). 
Therefore, it is not obvious whether or not the C\emissiontype{IV} linewidths precisely reflect the gravitational potential well of the SMBH. In fact, $M_\mathrm{BH}$ estimations based on C\emissiontype{IV} may have large uncertainty, due to the scatter observed in the C\emissiontype{IV}-derived SMBH mass distribution, compared to the better-calibrated Mg\emissiontype{II} $\lambda$2800- and H$\beta$ $\lambda$4861-based SMBH mass estimate (e.g., \cite{Baskin2005}; \cite{Netzer2007}; \cite{Sulentic2007}; \cite{Shen2008}; \cite{Marziani2012}). 
On the other hand, H$\beta$ emission shows a smooth line profile, dominated by the gravitational potential of the SMBH, and is best calibrated to estimate $M_\mathrm{BH}$ in the local universe, based on the reverberation mapping method (e.g., \cite{Peterson1999}; \cite{Kaspi2000}; \cite{Peterson2004}; \cite{Bentz2007}, \yearcite{Bentz2009}).
In this study, the H$\beta$ emission linewidth and nearby continuum luminosity at 5100 $\mathrm{\AA}\:(L_{5100})$ were used to estimate $M_\mathrm{BH}$ of distant QSOs in the most reliable manner.
To cover the redshifted H$\beta$ emission line within the $K$-band (2.2 $\mu$m), the longest high-sensitivity atmospheric window for ground-based observations, the redshift range of target QSOs was limited to $z<3.5$. Detection of the QSOs' host galaxy stellar emission at $z\sim3.5$ was technically feasible using the latest powerful 8--10 m ground-based observation facilities (e.g., \cite{Falomo2005}; \cite{Peng2006}; \cite{Schramm2008}; \cite{McLeod2009}; \cite{Targett2011}). For these reasons, QSOs at $z<3.5$ were targeted to constrain their SMBH and galaxy stellar mass ratio, and to observationally better constrain the coevolution of SMBHs and galaxies in the early universe. 

In this paper, we report the results of our spectroscopic observations of $z<3.5$ QSOs for their $M_\mathrm{BH}$ estimate. The imaging results related to the host galaxies' stellar component will be presented in separate papers (Kawaguchi et al. 2015; in preparation, Saito et al. 2015; in preparation). We describe the sample selection in Section 2 and the observations and data analysis in Section 3. The method used for the $M_\mathrm{BH}$ estimate is described in detail in Section 4, along with the main results. Section 5 provides a discussion of our main findings, which are summarized in Section 6. Throughout this paper, we adopt the Vega magnitude system for all infrared data and the standard $\Lambda$CDM cosmology, with $\Omega_{\Lambda}=0.7$, $\Omega_{M}=0.3$, and $H_{0}=70\mathrm{\:km\:s^{-1}\:Mpc^{-1}}$.

\section{Observations and Data Analysis}
\subsection{Sample Selection}
All of our samples are radio-quiet QSOs drawn from the seventh data release of the Sloan Digital Sky Survey (SDSS DR7; \cite{Abazajian2009}). We first selected QSOs with redshift shown in SDSS DR7 $\sim$ 3.11--3.50, so that the H$\beta$ emission line and 5100 $\mathrm{\AA}$ continuum could be observed within the $K$-band. Next, we set limitations on the optical $r'$-band (0.62 $\mu$m) magnitude, 18.5 $< r'(SDSS) <$ 19. The faint limit was set to obtain sufficient quality spectra for our discussion within a reasonable amount of exposure time. Assuming that observed luminosity cannot exceed the Eddington luminosity limit of $L_\mathrm{Edd}=3.2\times10^{4}(M/$\MO)\LO,
the minimum required $M_\mathrm{BH}$ is higher for brighter QSOs. The bright limit was set, because observing only the brightest end of the QSOs could strongly bias the data to intrinsically large $M_\mathrm{BH}$ systems, resulting in a biased view of the $R_\mathrm{BH/spheroid}$ ratio at high redshift. Radio-loud QSOs were almost excluded because they often show jet-induced extended narrow emission line regions, which could induce considerable uncertainty to the $M_\mathrm{BH}$ estimate using the emission linewidth. 
We roughly estimate radio-loudness of our targets from the ratio of their rest-frame luminosities at $5\:\mathrm{GHz}$ to those at $2500\:\mathrm{\AA}$ ($L(5\:\mathrm{GHz})/L(2500\:\mathrm{\AA})$; \cite{Stocke1992}). The $L(2500\:\mathrm{\AA})$ and $L(5\:\mathrm{GHz})$ are derived from the $z'$-band magnitude from the SDSS and the 20cm flux from Faint Images of the Radio Sky at Twenty-Centimeters (FIRST) survey (\cite{Becker1994}), respectively. Since the wavelength of $z'$-band and 20 cm correspond to $2500\:\mathrm{\AA}$ and $5\:\mathrm{GHz}$, respectively, in the rest frame of $z\sim3$ QSOs, we do not take into account the K correction. Most of our target QSOs (35 out of 37) used in this paper have $L(5\:\mathrm{GHz})/L(2500\:\mathrm{\AA}) < 10$, except for 2 target QSOs (J0847+3831 and J1337+3152) with larger radio loudness.

To achieve our final goal for studying the coevolution of SMBH and galaxies at high redshift, adaptive optics (AO) imaging data are going to be used to derive the host galaxy's spheroidal stellar mass, using the Subaru 8.2 m telescope atop Mauna Kea, Hawaii (latitude $\sim$ $+20^{\circ}$). For AO observation, an AO guide star with $R\:(0.64\:\mu m)<$18 mag within 60" from the target is required. In addition, to observe each object for longer than 4 hours at a higher elevation than $50^{\circ}$ from Mauna Kea, with good AO performance and high spatial resolution, we selected QSOs with declination $-5^{\circ}$ $<$ Dec $<$ +45$^{\circ}$. Finally, to subtract the central bright AGN glare with high accuracy, we chose targets that had at least one nearby PSF reference star with a magnitude similar to that of the target QSOs. 
Approximately 120 $z\sim3.5$ QSOs met all of the requirements, among which 37 were chosen for near-infrared spectroscopic observation in this study (Table~\ref{table:1}).

\subsection{Observations and Data Reduction}
The near-infrared $K$-band (2.2 $\mu$m) spectra were obtained using the NASA Infrared Telescope Facility (IRTF 3 m), the United Kingdom Infrared Telescope (UKIRT 3.6 m), the William Herschel Telescope (WHT 4.2 m), and the Subaru Telescope (8.2 m). Bright targets were observed primarily with the SpeX instrument (\cite{Rayner2003}) on the IRTF, the UIST (\cite{Ramsay2004}) on the UKIRT, and the LIRIS (\cite{Manchado1998}) on the WHT. Fainter QSOs were observed with the IRCS instrument (\cite{Kobayashi2000}) assisted with the Adaptive Optics system AO188 (\cite{Hayano2010}) on the Subaru. With the exception of the Subaru IRCS, these instruments enabled us to obtain $H$- (1.6 $\mu$m) and $K$-band (2.2 $\mu$m) spectra simultaneously. For all targets, H$\beta$ $\lambda$4861 and [O\emissiontype{III}] $\lambda\lambda$4959, 5007 emission lines were observed within the $K$-band; however, $H$-band spectra were used when available to better determine the continuum flux level at the shorter part of these emission lines. We also obtained spectroscopic data of standard star (spectral type is A, G, or F) for each target (Table 1) for telluric correction and flux calibration. To estimate the telluric correction, we divided the quasar spectra with standard star spectra, and then multiplied simple blackbody spectra with the temperature corresponding to the spectral type of the standard star derived from Allen's astrophysical quantities (\cite{Tokunaga2000}).
 
\subsubsection{IRTF/SpeX}
IRTF/SpeX spectra of 20 targets (Table~\ref{table:1}) were obtained in the 0.8--2.5 $\mu$m cross-dispersed mode with a 1$\farcs$6 $\times$ 15$\farcs$0 slit. This mode provided spectral resolution with $R\sim375$. Although the $R$ was relatively low comparing to other observations, we choose this mode for safer tracking/guiding and for obtaining better S/N ratios. Spectra were taken at two different positions (A and B) along the slit. Each exposure time ranged from 60--300 s, depending on the magnitude of the targets and the weather conditions. Also, 1--2 coadds were used at each slit position. 

The reduction was carried out using the Spectral Extraction Package for SpeX (Spextool; \cite{Cushing2004}) that works on IDL. The Spextool goes through almost all data reduction processes, including spectral flat fielding, sky emission subtraction, bad pixel correction, 
extraction of one-dimensional (1D) spectra, and wavelength calibration (using argon lines). 
We first created A$-$B data and then median-combined multiple A$-$B data sets to increase the S/N ratios. After extraction of a 1D spectrum, the data at the A and B slit positions were summed using the IDL task {\it xcombspec}. Then the {\it xtellcor-basic} task in IDL was used for telluric correction and flux calibration. Finally, different order spectra were merged into a single spectrum corresponding to the $H$- and $K$-bands (1.41--2.42 $\mu$m). 

\subsubsection{UKIRT/UIST}
UKIRT/UIST spectra of five targets (Table~\ref{table:1}) were obtained through the 0$\farcs$6 $\times$ 120$\farcs$0 slit with an $H+K$ grism. The achieved spectral resolution is $R \sim1000$. The spectra were taken at A and B positions along the slit. Each exposure time was 240 s, and 1 coadd was adopted at each position for all targets.

Data were reduced using standard IRAF tasks. Initially, frames taken with an A (or B) beam were subtracted from frames subsequently taken with a B (or A) beam. The resulting subtracted frames were added and divided by a spectroscopic flat image. Then bad pixels and pixels impacted by cosmic rays were replaced with the interpolated values from the surrounding pixels. Finally, the spectra of the QSOs and the standard stars were extracted using the IRAF task {\it apall}. After wavelength calibration (using argon lines), telluric correction, and flux calibrations, we obtained the final spectra.

\subsubsection{WHT/LIRIS}
WHT/LIRIS spectra of five targets (Table~\ref{table:1}) were taken with the 0$\farcs$75 $\times$ 252$\farcs$0 slit and the $H+K$ grism. The spectral resolution was $R\sim945$. The exposure time and coadd at each slit position (A or B) were 300 s and 1 coadd, respectively. Data reduction was carried out in the same manner as for UKIRT/UIST.

\subsubsection{Subaru/IRCS}
Subaru/IRCS spectra of seven targets (Table~\ref{table:1}) were obtained using the 0$\farcs$45 $\times$ 18$\farcs$0 slit in the 52-mas mode. We used the $K$-grism that covers the 1.93--2.48 $\mu$m range. The spectral resolution was $R\sim400$. The exposure time and coadd at each slit position (A and B) were 600 s and 1, respectively. 
We used IRAF for data reduction similar to that used for UKIRT/UIST and WHT/LIRIS data.

The details of the observations are summarized in Table~\ref{table:1}. Redshift and $r'$-band magnitude are from SDSS catalog. Final spectra are shown in Figures~\ref{fig:1} and \ref{fig:2}.
We also obtained imaging data for all targets and standard stars just before or after spectroscopic observations by using the same instruments used for the spectroscopic observations, and measured photometric magnitudes of the targets. The magnitudes derived from our slit spectra that were calibrated by spectroscopic standard star (possibly affected by slit loss) and those from imaging data generally agreed within 0.4 mag (Table~\ref{table:2} and Figure~\ref{fig:3}). We adopted the photometric magnitude from the imaging data for flux calibration.

\section{Spectral Analysis}
To estimate $M_\mathrm{BH}$, we fitted the observed spectra with a model combining the linear continuum, the underlying very broad Fe\emissiontype{II} emission line complex (4500--5600 $\mathrm{\AA}$), the broad and narrow H$\beta$ $\lambda$4861 emission lines, and two narrow [O\emissiontype{III}] $\lambda\lambda$4959, 5007 emission lines (as the forbidden [O\emissiontype{III}] emission line originates primarily from the narrow line regions). The fitting of the spectra of the $z\sim3$ QSOs was performed individually using QDP (Tennant 1991)\footnote{The detailed information is available at http://heasarc.gsfc.nasa.gov/ftools/others/qdp/node3.html.} and IDL. First, a tentative linear continuum was determined using several data points that were not strongly affected by H$\beta$, [O\emissiontype{III}], and Fe\emissiontype{II} emission lines. For the IRTF, UKIRT, WHT targets, data points at rest-frame 4000--4050 $\mathrm{\AA}$ or 4150--4200 $\mathrm{\AA}$, and 5080--5120 $\mathrm{\AA}$ were used. For the Subaru targets, because the $H$-band spectral data were not available, data from rest-frame 4700--4750 $\mathrm{\AA}$, 5080--5120 $\mathrm{\AA}$, and 5470--5500 $\mathrm{\AA}$ were adopted. Several data points in the wavelength range considered for continuum determination were significantly affected by the Earth's atmospheric absorption, depending on the redshifts of individual QSOs, and were considerably noisy. Thus, these noisy data points were excluded from the continuum determination. 

Next, we fitted the [O\emissiontype{III}] $\lambda\lambda$4959, 5007 doublet emission lines with two narrow Gaussian components. The linewidth and the redshift of the two components were set to be the same. The relative 5007 $\mathrm{\AA}$-to-4959 $\mathrm{\AA}$ strength was fixed at 3.0 (\cite{Dimitrijevic2007}). Then the H$\beta$ emission line was fit with two Gaussian (broad and narrow) components. Although some previous studies adopt more complex model (\cite{Shemmer2004}; \cite{NetzerTrak2007}; \cite{Schulze2010}), we use only one broad Gaussian and one narrow Gaussian models for H$\beta$ line fitting, because of limited spectral resolution and S/N of our data. For the narrow H$\beta$ emission component, the same linewidth as the [O\emissiontype{III}] line was adopted. Another Gaussian component with a larger linewidth, corresponding to the broad components of the H$\beta$ emission line, was added. All parameters of this component was set as free parameters. We allowed the velocity shift of the peak wavelength between the narrow H$\beta$ and [O\emissiontype{III}] lines to be up to 200 km $\mathrm{s^{-1}}$, following \citet{Netzer2007}. Then Fe\emissiontype{II} emission line fitting was carried out using the template derived from the nearby well-studied QSO, I Zw 1, by \citet{Tsuzuki2006}. We note that our aim is not to try to accurately fit the Fe\emissiontype{II} emission features, but to obtain a reliable estimate of the H$\beta$ line width and nearby continuum flux, unaffected by the contamination from the Fe\emissiontype{II} emission lines. The template was convolved with a Gaussian that had the same linewidth as the broad H$\beta$ emission line, as determined above, because the Fe\emissiontype{II} emission-line complex originates predominantly from the broad line regions. 
The fitting wavelength ranges for the Fe\emissiontype{II} features were rest-frame 4500--4650 $\mathrm{\AA}$ and 5100--5600 $\mathrm{\AA}$. Because all QSOs do not have exactly the same Fe\emissiontype{II} emission line profile as the template used (Fe\emissiontype{II} spectrum extracted from I Zw 1), we divided the Fe\emissiontype{II} template into $\lambda<5008\:\mathrm{\AA}$ and $\lambda>5008\:\mathrm{\AA}$ (the Fe\emissiontype{II} feature shows a minimum value at 5008 $\mathrm{\AA}$ in the template), and varied their relative strengths to better fit the spectral features actually observed near the Fe\emissiontype{II} emission in our QSOs spectra. The scaled template was subtracted from the observed spectra. Finally, the Fe\emissiontype{II} subtracted spectrum was refit using the approach described above. This time, the peak wavelength shift between the narrow H$\beta$ and [O\emissiontype{III}] was allowed to be up to 450 km $\mathrm{s^{-1}}$. We determined the final fitting parameters for a linear continuum, H$\beta$, and the [O\emissiontype{III}] emission lines. The contribution of the instrumental resolution was removed from the fitted linewidth to obtain the actual H$\beta$ emission linewidth. 

Some QSOs were fit with only a broad Gaussian component for the H$\beta$ emission line, because the ${\chi}^{2}$ values became larger when both broad and narrow Gaussian components were added. For J2134+0011, the H$\beta$ line and [O\emissiontype{III}] lines were not clearly deblended (Figure~\ref{fig:1}). Therefore, only wavelengths shorter than 4861 $\mathrm{\AA}$ were used to fit the broad H$\beta$ emission line. For some sources, continuum fit at the longest wavelength is problematic, because our continuum fitting ranges do not cover longer wavelength part. Also, longer wavelength part of  Fe\emissiontype{II} emission is not well fitted for some sources. However these problematic fit do not affect $L_\mathrm{5100}$ value significantly.

We succeeded in fitting the H$\beta$ emission lines for 28 out of 37 observed objects.
Tables~\ref{table:3} and \ref{table:4} show the flux and luminosity of H$\beta$ and [O\emissiontype{III}] emission lines, respectively, based on the above best Gaussian fit. In Table~\ref{table:5}, the redshift values estimated from H$\beta$ and [O\emissiontype{III}] lines are compared to that from the SDSS; they show good agreement.

\section{Results}
\subsection{$M_\mathrm{BH}$ and $L_\mathrm{bol}/L_\mathrm{Edd}$ Estimation}
The SMBH mass ($M_\mathrm{BH}$) of the 28 QSOs (with successful H$\beta$ fit) were estimated from the following formula (\cite{Vestergaard2006}): 

\begin{equation}
\mathrm{log_{10}} (M_\mathrm{BH}/\MO)= \mathrm{log_{10}} \left\{ \left[ \mathrm{\frac{FWHM(H\beta)}{1000\:km\:s^{-1}}} \right]^{2} \left[\frac{\lambda L_{\lambda}(5100\:\mathrm{\AA})}{10^{44}\:\mathrm{erg\:s^{-1}}} \right]^{0.5} \right\}+(6.91 \pm 0.02),
\end{equation}
and are summarized in Table~\ref{table:6}. 

We employed a resampling approach to obtain realistic uncertainties of $L_\mathrm{5100}$, the full-width at half maximum (FWHM) of the H$\beta$ broad emission line, and $M_\mathrm{BH}$ (e.g., \cite{Schulze2010}, \cite{Assef2011}, \cite{Shen2012}). Namely, we artificially added Gaussian random noise that is scaled by the observed S/N and refit them. We attempted this procedure for 100 simulated spectra for each target. The $L_\mathrm{5100}$, FWHM, and $M_\mathrm{BH}$ error was estimated from the resulting scatter of the derived  $L_\mathrm{5100}$, FWHM, and $M_\mathrm{BH}$ values from 100 spectra.

For comparison, we also calculated $M_\mathrm{BH}$ using the different formula proposed by \citet{McLure2002}. 
\begin{equation}
\mathrm{log_{10}} (M_\mathrm{BH}/\MO)= \mathrm{log_{10}} \left\{4.74 \left[ \mathrm{\frac{FWHM(H\beta)}{km\:s^{-1}}} \right]^{2} \left[\frac{\lambda L_{\lambda}(5100\:\mathrm{\AA})}{10^{44}\:\mathrm{erg\:s^{-1}}} \right]^{0.61} \right\}.
\end{equation}
The difference of $M_\mathrm{BH}$ related to choice of estimator is typically (median) 0.03 dex (range of $-$0.36$\sim+$0.03 dex). Table~\ref{table:6} and Figure~\ref{fig:4} show a comparison of the $M_\mathrm{BH}$ values derived from both methods, which mutually agree within statistical error.

The AGN bolometric luminosity $L_\mathrm{bol}$ was calculated from the observed $L_{5100}$ luminosity, using a bolometric correction, $f_{L} (L_\mathrm{bol} = f_{L}\times\lambda\times L_\mathrm{5100})$. For Type 1 unobscured luminous AGNs, $f_{L}$ was estimated to be 5--13 (e.g., \cite{Elvis1994}; \cite{Kaspi2000}; \cite{Netzer2003}; \cite{Marconi2004}; \cite{Richards2006}).
We adopt a constant ratio of $f_{L}=7$, following \citet{Netzer2007}. The AGN bolometric luminosity, relative to the Eddington luminosity for a given $M_\mathrm{BH}$ ($L_\mathrm{Edd}=3.2\times10^{4}(M_\mathrm{BH}/\MO)\LO$), the so-called Eddington ratio (${L_\mathrm{bol}/L_\mathrm{Edd}}$), is often used to estimate the activity of SMBHs (i.e., the SMBH-mass normalized accretion rate). In the case of $f_{L}=7$, the Eddington ratio is given by
\begin{equation}
L_\mathrm{bol}/L_\mathrm{Edd}=\frac{7\times\lambda L_\mathrm{5100}}{1.5\times10^{38}(M_\mathrm{BH}/\MO)}.
\end{equation}
The estimated $L_\mathrm{bol}/L_\mathrm{Edd}$ of 28 QSOs is summarized in Table~\ref{table:6}.

\subsection{$M_\mathrm{BH}$ and $L_\mathrm{bol}/L_\mathrm{Edd}$ distributions}
The upper panels of Figure~\ref{fig:5}(a) and \ref{fig:5}(b) show the distribution of $M_\mathrm{BH}$ and the Eddington ratio ($L_\mathrm{bol}/L_\mathrm{Edd}$) for our sample, respectively. \citet{Netzer2007} and \citet{Shemmer2004} performed near-infrared spectroscopy of QSOs at $z=2$--$4$, and measured the SMBH masses in 15 and 29 sources, respectively, based on the H$\beta$ method, of which 14 QSOs in total (8 sources in Netzer's sample, and 6 sources in Shemmer's sample) were at $z>3$. 
Our near-infrared spectroscopy tripled the number of $z>3$ QSOs, with H$\beta$-based reliably estimated $M_\mathrm{BH}$ information.

The $M_\mathrm{BH}$ and $L_\mathrm{bol}/L_\mathrm{Edd}$ distributions of these 14 QSOs at $z>3$ studied by \citet{Netzer2007} and \citet{Shemmer2004} are shown in the lower panels of Figure~\ref{fig:5}(a) and \ref{fig:5}(b), respectively, for comparison. The $M_\mathrm{BH}$ and $L_\mathrm{bol}/L_\mathrm{Edd}$ of the comparison sample are re-calculated using Eq.(1) and (3) with FWHM(H$\beta$) and $L_\mathrm{5100}$ drawn from the literature (Table 2 in \citet{Netzer2007}, and Table 2 in \citet{Shemmer2004}). The difference of $M_\mathrm{BH}$ due to choice of estimator (Eq.(1) in this paper and Eq.(1) in \citet{Netzer2007}, or Eq.(1) in \citet{Shemmer2004}) is typically 0.25 dex (range of 0.19--0.30 dex) for Netzer's sample and 0.15 dex (range of 0.06--0.23 dex) for Shemmer's sample. The comparison sample has slightly fainter luminosity than our sample (median value of $L_\mathrm{bol} =7.88\times10^{46}$ erg $\mathrm{s^{-1}}$ for Netzer's sample and $L_\mathrm{bol} = 1.61\times10^{47}$ erg $\mathrm{s^{-1}}$ for our sample), and $M_\mathrm{BH}$ of the comparison sample is smaller (range of $10^{8.59}$--$10^{9.59}\:M_\mathrm{\odot}$) than our sample (range of $10^{8.81}$--$10^{10.13}\:M_\mathrm{\odot}$) as shown in Figure~\ref{fig:5}(a). On the other hand, the Eddington ratios of our sample are systematically smaller than those of the comparison sample, as shown in Figure~\ref{fig:5}(b). We performed Kolmogorov-Smirnov test to check if those distributions are statistically the same or not. P-values were calculated to be $P(M_\mathrm{BH})=0.005$ for the $M_\mathrm{BH}$ distributions, and $P(L_\mathrm{bol}/L_\mathrm{Edd})=0.045$ for the $L_\mathrm{bol}/L_\mathrm{Edd}$ distributions, respectively. The result of K-S test shows that two samples (our sample and the reference sample) are drawn from different parent distributions for both $M_\mathrm{BH}$ mass distributions and $L_\mathrm{bol}/L_\mathrm{Edd}$ distributions (K-S probability of being drawn from the same population $<0.05$).

There are two possible reasons to produce the high AGN luminosity, as observed for our QSO sample: (1) the QSO has a modest SMBH mass and a large Eddington ratio, or (2) the QSO has a large SMBH mass and a normal Eddington ratio. If we pick up only the second sample (i.e., QSOs at the higher end of the $M_\mathrm{BH}$ distribution), $M_\mathrm{BH}/M_\mathrm{spheroid}$ ratios could be systematically larger than the typical values (e.g., \cite{Lauer2007}; \cite{Schulze2011}, \yearcite{Schulze2014}), possibly providing systematically biased results regarding the redshift evolution of the $M_\mathrm{BH}/M_\mathrm{spheroid}$ ratios. $M_\mathrm{BH}$ of our sample and the comparison sample agree within a factor of a few.
As shown in Table~\ref{table:6} and Figure~\ref{fig:5}, it is unlikely that majority of our sample have the $M_\mathrm{BH}$ much larger than the break (cut-off) of the SMBH mass function at $z=3.2$ ($\sim$\,$10^{9.7}M_\mathrm{\odot}$; \citet{Kelly2013}). The cut-off mass of $\log M_\mathrm{BH}=9.7$ was estimated by eye using the mass function
at $z=3.2$ in Figure 4 by \citet{Kelly2013}, although it might be risky to believe the mass function estimated by C\emissiontype{IV}-based $M_\mathrm{BH}$ at face values.
We compared the H$\beta$-based and C\emissiontype{IV}-based $M_\mathrm{BH}$ (Figure~\ref{fig:6}, left) and found that C\emissiontype{IV}-based $M_\mathrm{BH}$ is not always biased toward large $M_\mathrm{BH}$.
Therefore, we consider that our QSO sample corresponds to the case (1) and that the Lauer-bias is not affecting our sample so severely for a study of redshift evolution of $M_\mathrm{BH}/M_\mathrm{spheroid}$.
This suggests that we can use our QSO sample to discuss the redshift evolution of the $M_\mathrm{BH}/M_\mathrm{spheroid}$ ratios, without obvious strong bias, using the combination of the near-infrared spectroscopy of the $M_\mathrm{BH}$ estimate (this paper) and ongoing near-infrared, multi-band, high-spatial-resolution AO imaging observations to estimate $M_\mathrm{spheroid}$. If the local scaling relations hold all the way to $z>3$, the expected $M_\mathrm{spheroid}$ for our observed QSOs with $M_\mathrm{BH} = 6.5\times10^8$ -- $1.4\times10^{10}\:\MO$, are $4.3\times10^{11}$ -- $9.0\times10^{12}\:\MO$, which are detectable with 8 -- 10 m telescopes, and have actually been detected in our high-spatial-resolution infrared $J$- and $K'$-band imaging observations using Subaru 8.2 m telescope and AO (Kawaguchi et al. 2015; in preparation, Saito et al. 2015; in preparation).

\section{Discussion}
\subsection{Comparison of $M_\mathrm{BH}$ estimated from $\mathrm{H\beta}$ $\lambda$4861 and $\mathrm{C\emissiontype{IV}}\:\lambda1549$}
The combination of the C\emissiontype{IV} $\lambda$1549 emission line and continuum luminosity at 1450 $\mathrm{\AA}$ has often been used to estimate SMBH mass (hereafter, the C\emissiontype{IV} method) for distant QSOs, because the C\emissiontype{IV} emission line is redshifted into the optical wavelength range where spectroscopic observations are easier to attain. However, as mentioned in Section 1, the C\emissiontype{IV} emission line often shows an asymmetric profile. This suggests that, in addition to the motion dominated by the gravitational potential of SMBH, some other non-gravitational component may be contaminated, such as outflow 
(\cite{Vestergaard2006}; \cite{Marziani2012}). In fact, several previous studies indicated significant scatter about the comparison of SMBH masses estimated using the H$\beta$ method and the C\emissiontype{IV} method (e.g., \cite{Netzer2007}; \cite{Shen2008}; \cite{Ho2012}; \cite{Shen2012}; \cite{Trakhtenbrot2012}). 

\citet{Shen2011} reported C\emissiontype{IV}-based BH masses for our QSO sample, using the following formula (\cite{Vestergaard2006}):
\begin{equation}
\mathrm{log_{10}} (M_\mathrm{BH}/\MO)= \mathrm{log_{10}} \left\{ \left[ \mathrm{\frac{FWHM(C\emissiontype{IV})}{1000\:km\:s^{-1}}} \right]^{2} \left[\frac{\lambda L_{\lambda}(1350\:\mathrm{\AA})}{10^{44}\:\mathrm{erg\:s^{-1}}} \right]^{0.53} \right\}+(6.66 \pm 0.01).
\end{equation}
We adopt their estimates to compare with our $M_\mathrm{BH}$ using the H$\beta$ method. The FWHM of the C\emissiontype{IV} emission line and C\emissiontype{IV}-based $M_\mathrm{BH}$ of our targets are shown in Table~\ref{table:7}. For two QSOs, C\emissiontype{IV} data are not available in \citet{Shen2011}. The left panel of Figure~\ref{fig:6} shows a comparison between SMBH masses obtained by the two methods for our sample. The scatter is large (0.41 dex) and no significant correlation between H$\beta$-based and C\emissiontype{IV}-based $M_\mathrm{BH}$ is observed.
\citet{Shen2008} compared BH masses estimated using different methods for $\sim$60,000 QSOs at $0.1 \leq z \leq 4.5$, and found that while $M_\mathrm{BH}$ obtained by the H$\beta$- and Mg\emissiontype{II}-methods are tightly correlated, a comparison between C\emissiontype{IV}-based and Mg\emissiontype{II}-based $M_\mathrm{BH}$ shows large scatter with $\sim0.34$ dex (see also \cite{Shen2011}). Given these results, the large scatter shown in the left panel of Figure~\ref{fig:6} for our QSO sample is most likely due to the large uncertainty associated with C\emissiontype{IV}-based $M_\mathrm{BH}$. For a reliable $M_\mathrm{BH}$ estimate for $z\sim3$ QSOs, the H$\beta$-method based on near-infrared spectroscopy is preferred over the C\emissiontype{IV}-method based on optical spectroscopy.

The right panel of Figure~\ref{fig:6} is the same plot as that shown in the left panel of Figure~\ref{fig:6}; however, the marks are distinguished, depending on the Eddington ratio. The samples with high Eddington ratios appear to be distributed slightly to the upper side than those with low Eddington ratios, indicating that the C\emissiontype{IV}-based $M_\mathrm{BH}$ was larger than H$\beta$-based $M_\mathrm{BH}$ for sources with high Eddington ratios. Figure~\ref{fig:7} shows the relationship between the Eddington ratio and the linewidth ratio (FWHM(C\emissiontype{IV})/FWHM(H$\beta$)). A weak correlation is seen log(FWHM(C\emissiontype{IV})/FWHM(H$\beta$))=(0.32$\pm$0.15)$\times$log($L_\mathrm{bol}$/$L_\mathrm{Edd}$)$+$(0.03$\pm$0.08) for our sample only or log(FWHM(C\emissiontype{IV})/FWHM(H$\beta$))=(0.35$\pm$0.11)$\times$log($L_\mathrm{bol}$/$L_\mathrm{Edd}$)$+$(0.05$\pm$0.06) for all plotted samples. This correlation indicates that objects with a larger Eddington ratio have a wider C\emissiontype{IV} linewidth, compared to H$\beta$. QSOs with higher Eddington ratios can have stronger radiation pressure and thereby stronger outflow motion of gas than those with lower Eddington ratios. The C\emissiontype{IV} line-emitting region is more inside than the H$\beta$ line-emitting region (\cite{Peterson1999}). It may be possible that this outflow-origin motion broadens the C\emissiontype{IV} line profile, compared to the SMBH's gravitational motion alone, resulting in a larger $M_\mathrm{BH}$ estimate than the H$\beta$-based method.

\subsection{Growth Time of SMBHs}
From the AGN bolometric luminosity, we can obtain information on the mass accretion rate. When the measured SMBH mass is divided by the accretion rate, the time scale of SMBH growth can be derived ($t_\mathrm{grow}$).
Following \citet{Netzer2007}, we adopt the following formula:
\begin{equation}
t_\mathrm{grow}=t_\mathrm{Edd}\frac{\eta/(1-\eta)}{f_{L}\lambda L_\mathrm{5100}/L_\mathrm{Edd}}\ln\left(\frac{M_\mathrm{BH}}{M_\mathrm{seed}}\right)\frac{1}{f_\mathrm{active}},
\end{equation}
where $t_{Edd}$ is the Eddington time scale ($= 3.8\times10^{8}$ yr), $\eta$ is the accretion efficiency, $M_\mathrm{seed}$ is the seed SMBH mass, $f_\mathrm{active}$ is the duty cycle (the fractional activity time) of the SMBH, and $f_{L}$ is the bolometric correction (Section 4.2). 
We assume fiducial values for $\eta = 0.1$, $M_\mathrm{seed} = 10^{4} \MO$, $f_\mathrm{active} = 1$, and $f_{L} = 7$, similar to \citet{Netzer2007}. We now define $t_\mathrm{universe}$, which is the age of the universe at the redshift of each QSO (calculated from our adopted cosmology).
The $t_\mathrm{grow}/t_\mathrm{universe}$ value is summarized in Table~\ref{table:6}. Since we have identical choices for $f_\mathrm{L}$, $M_\mathrm{seed}$ and $f_\mathrm{active}$ for all samples, the $t_\mathrm{growth}$ difference is due to $M_\mathrm{BH}$ and $L_\mathrm{bol}/L_\mathrm{Edd}$. The comparison of these values in our and other samples is shown in Figure 5. For Netzer's and Shemmer's sample, we calculated $t_\mathrm{grow}/t_\mathrm{universe}$ value by using Eq.(5) by adopting recalculated $M_\mathrm{BH}$ and Eddington ratio in Section 4.2. (1) If $t_\mathrm{grow}/t_\mathrm{universe} < 1$, then the measured $M_\mathrm{BH}$ can be reproduced with the estimated mass accretion rate. On the other hand, (2) if $t_\mathrm{grow}/t_\mathrm{universe} > 1$, the measured $M_\mathrm{BH}$ cannot be reproduced with the estimated mass accretion rate, requiring a more active phase in the past at higher redshift than the QSO's redshift. This means that QSOs are not sufficiently active at the QSO redshift. 
Most of our sample have less than 1 of $t_\mathrm{grow}/t_\mathrm{universe}$ values. The range of $t_\mathrm{grow}/t_\mathrm{universe}$ values for our sample is 0.22--3.52, and most of our sample have similar value as \citet{Netzer2007} (0.45--2.27). While, QSOs in \citet{Shemmer2004} have systematically smaller $t_\mathrm{grow}/t_\mathrm{universe}$ values (0.05--0.16) than ours, which could be due to their sample being selected only from luminous sources ($L\gtrsim 10^{46}\:\mathrm{erg\:s^{-1}}$). Some of  objects, with $t_\mathrm{grow}/t_\mathrm{universe}>1$, in our and Netzer's samples should have experienced a rapidly growing phase in the past, while, Shemmer's sample and most of our sample are in rapidly growing phase at $z\sim3$.

\section{Summary}
We present new near-infrared spectroscopic observations of 37 QSOs at $z\sim3$. We successfully estimated the SMBH masses of 28 out of 37 observed QSOs, using the well-calibrated H$\beta$-method, based on a broad H$\beta$ emission linewidth and nearby continuum luminosity. A summary of the main results is given below.

1. A comparison of our work to similar studies of $z=2$--$4$ QSOs by \citet{Netzer2007} and \citet{Shemmer2004} indicated that our sample had slightly larger $M_\mathrm{BH}$ and smaller Eddington ratios than the comparison sample. Also, it is unlikely that most of our sample have the $M_\mathrm{BH}$ much larger than the break of the BH mass function at that redshift.
Given that all our QSOs have at least one nearby bright star, high-spatial-resolution AO observations to investigate the detailed properties of the host galaxies are possible and on-going. Our sample is suited to an investigation of $M_\mathrm{BH}$/$M_\mathrm{spheroid}$ evolution at $z\sim3$, without obvious selection bias.

2. A comparison of the H$\beta$-based SMBH mass estimate through near-infrared spectroscopy and previous C\emissiontype{IV}-based SMBH mass estimates using optical spectroscopy showed large scatter and no significant correlation. QSOs with higher Eddington ratios tended to display higher C\emissiontype{IV}-derived $M_\mathrm{BH}$ than H$\beta$-derived $M_\mathrm{BH}$, due possibly to effects other than the motion dominated by the gravitational potential of SMBHs. As argued in previous studies, the use of C\emissiontype{IV} for the $M_\mathrm{BH}$ estimate could introduce large uncertainty. H$\beta$-based $M_\mathrm{BH}$ estimation using near-infrared spectroscopy is desirable for reliable $M_\mathrm{BH}$ estimates of $z\sim3$ QSOs.

\bigskip
We appreciate the anonymous referee for his/her useful comment. We are grateful to S. Hayashi for the observation time with the Subaru and K. Imase for supporting observations in the early phase of the study. We wish to thank K. Aoki for general discussions about QSOs, and A. Schulze for helpful discussions about possible bias of our sample. This work was partly supported by the Grants-in-Aid of the Ministry of Education, Science, Culture, and Sport [23540273(MI), 19740105(TK), 09J07156(TM)], and the Grants-in-Aid for Young Scientists [25800099(NK)]. This work was also supported in part by a Japanese Society for the Promotion of Science (JSPS) Core-to-Core Program gInternational Research Network for Dark Energy.h Use of the UKIRT for observations was supported by the National Astronomy Organization of Japan (NAOJ).

\clearpage
\begin{table}[h]
\begin{center}
\caption{Observation log.}
\label{table:1}
\small
   \begin{tabular}{cccccccc}
   \hline
   (1) & (2) & (3) & (4) & (5) & (6) & (7) & (8)\\
   Object ID (SDSS J) & $z$ & $r'$ mag & Instrument & Date & Exposure time[sec] & Standard star & Spectral type\\
   \hline \hline
  
   014619.97$-$004628.7 & 3.17 & 18.86 & IRTF/SpeX & 2009 Jan 19, 21 & 1800+1600 & HD13936 & A0\\ 
   031845.17$-$001845.3 & 3.22 & 18.80 & WHT/LIRIS & 2009 Jan 10 & 7200 & HD15004 & A0\\ 
   072554.52$+$392243.4 & 3.25 & 19.10 & Subaru/IRCS & 2013 Mar 11 & 2400 & SAO42621 & G2\\ 
   074628.71$+$301419.0 & 3.11 & 18.43 & IRTF/SpeX & 2009 Jan 19 & 2120 & HD56386 & A0\\ 
   074939.01$+$433217.6 & 3.14 & 18.56 & UKIRT/UIST & 2009 Jan 20 & 1920 & HD63160 & A0\\ 
   075515.93$+$154216.6 & 3.30 & 19.63 & WHT/LIRIS & 2009 Jan 10 & 10500 & HD87737 & A0\\ 
   075841.66$+$174558.0 & 3.17 & 18.98 & WHT/LIRIS & 2010 Mar 3 & 7200 & HR2692 & G9\\ 
   082645.88$+$071647.0 & 3.13 & 18.28 & IRTF/SpeX & 2009 Jan 20, 21 & 920+3120 & HD65241 & A0\\ 
   083700.82$+$350550.2 & 3.31 & 18.38 & UKIRT/UIST & 2009 Jan 20 & 2880 & HD71906 & A0\\ 
   084715.16$+$383110.0 & 3.18 & 18.42 & UKIRT/UIST & 2009 Jan 20 & 2880 & HD63160 & A0\\ 
   094202.04$+$042244.5 & 3.28 & 17.18 & IRTF/SpeX & 2009 Jan 19 & 2120 & HR3906 & A0\\ 
   095406.40$+$290208.0 & 3.24 & 18.57 & IRTF/SpeX & 2013 Mar 21 & 5200 & HD91163 & G2\\ 
   095735.37$+$353520.6 & 3.27 & 18.13 & IRTF/SpeX & 2013 Mar 20 & 4800 & HD77930 & F6\\ 
   100610.55$+$370513.8 & 3.20 & 17.83 & IRTF/SpeX & 2009 Jan 19 & 4920 & HD88960 & A0\\ 
   103456.31$+$035859.4 & 3.36 & 17.86 & IRTF/SpeX & 2009 Jan 20 & 3400 & HR3906 & A0\\ 
   111137.72$+$073305.9 & 3.46 & 18.68 & WHT/LIRIS & 2010 Mar 3 & 7200 & HR4079 & F6\\ 
   111656.89$+$080829.4 & 3.23 & 18.22 & IRTF/SpeX & 2009 Jan 21 & 2800 & HD97585 & A0\\ 
   113002.35$+$115438.3 & 3.39 & 18.41 & UKIRT/UIST & 2009 Jan 20 & 3840 & HD101060 & A0\\ 
   114412.77$+$315800.8 & 3.23 & 18.56 & UKIRT/UIST & 2009 Jan 20 & 7440 & HD98989 & A0\\ 
   123815.03$+$443026.2 & 3.25 & 18.17 & IRTF/SpeX & 2013 Mar 22 & 6000 & HD111859 & F2\\
   133724.69$+$315254.5 & 3.18 & 18.53 & IRTF/SpeX & 2013 Mar 21 & 4080 & HD121149 & G0\\ 
   133757.87$+$021820.9 & 3.33 & 18.13 & IRTF/SpeX & 2009 Jan 21, Apr 2 & 3000+6480 & HD124224 & A0\\ 
   140745.50$+$403702.2 & 3.20 & 18.47 & IRTF/SpeX & 2013 Mar 20 & 4800 & HD134169 & G1\\ 
   142755.85$-$002951.1 & 3.36 & 18.23 & IRTF/SpeX & 2011 Feb 19 & 2640 & HD116960 & A0\\ 
   150238.38$+$030228.2 & 3.35 & 18.53 & IRTF/SpeX & 2013 Mar 22 & 6000 & HD34495 & F6\\ 
   150726.32$+$440649.2 & 3.12 & 17.85 & IRTF/SpeX & 2009 Jan 20 & 800 & HD127304 & A0\\ 
   151044.66$+$321712.9 & 3.48 & 19.83 & Subaru/IRCS & 2013 Apr 18 & 2400 & HIP81272 & F5\\ 
   155036.80$+$053749.9 & 3.15 & 17.99 & IRTF/SpeX & 2013 Mar 20 & 5760 & HD145436 & F6\\ 
   155137.22$+$321307.5 & 3.18 & 20.15 & Subaru/IRCS & 2013 Apr 17 & 2400 & HIP87556 & F2\\ 
   155823.22$+$353252.2 & 3.20 & 19.90 & Subaru/IRCS & 2013 Apr 19 & 2400 & HIP81272 & F5\\ 
   162508.09$+$265052.2 & 3.44 & 18.92 & WHT/LIRIS & 2010 Mar 3 & 6000 & HR5728 & G2\\ 
   165523.09$+$184708.4 & 3.32 & 17.81 & IRTF/SpeX & 2013 Mar 21 & 4560 & HD145228 & F0\\ 
   211936.77$+$104623.9 & 3.24 & 19.19 & Subaru/IRCS & 2012 Oct 14 & 3600 & HR8041 & G1\\ 
   213023.61$+$122252.2 & 3.26 & 18.04 & IRTF/SpeX & 2008 Aug 28 & 4800 & HD208108 & A0\\ 
   213455.08$+$001056.8 & 3.24 & 18.77 & Subaru/IRCS & 2012 Oct 14 & 3600 & HR8041 & G1\\ 
   231858.56$-$005049.6 & 3.20 & 19.56 & Subaru/IRCS & 2012 Oct 14 & 3600 & HR8041 & G1\\ 
   234150.01$+$144906.0 & 3.18 & 18.44 & IRTF/SpeX & 2008 Aug 29 & 8000 & BD+14 4774 & A0\\ 
   \hline
  \end{tabular}
\end{center}
{\bf Notes.} Column (1): Object name. Column (2): Sloan Digital Sky Survey (SDSS) redshift. Column (3): SDSS $r'$-band magnitude (PSF magnitude). Column (4): Used instrument. Column (5): Observation date in UT. Column (6): Exposure time. Column (7): Standard star used for flux calibration and telluric correction. Column (8): Spectral type of the standard star. 
\end{table}

\clearpage
\begin{table}[htbp]
\caption{$K$-band magnitude.}
 \label{table:2}
 \begin{center}
 \small
  \begin{tabular}{ccccc}
   \hline
   (1) & (2) & (3) & (4) & (5)\\
   Object ID (SDSS J) & Spectroscopic $K$mag & Imaging $K$mag & UKIDSS $K$mag & 2MASS $K$mag\\
   \hline \hline 
   014619.97$-$004628.7 & 16.21 & 16.69 & 16.59 & --\\
   031845.17$-$001845.3 & 16.57 & 16.38 & 16.47 & --\\
   072554.52$+$392243.4 & 17.00 & 16.95 & -- & --\\
   074939.01$+$433217.6 & 16.24 & 16.07 & -- & 15.66\\
   075515.93$+$154216.6 & 16.53 & 16.68 & -- & --\\
   075841.66$+$174558.0 &15.56 & 16.08 & -- & --\\
   083700.82$+$350550.2 & 16.03 & 16.06 & -- & --\\
   084715.16$+$383110.0 & 16.07 & 15.99 & -- & --\\
   094202.04$+$042244.5 & 13.71 & 14.58 & 14.58 & 14.62\\
   095735.37$+$353520.6 & 16.31 & 16.14 & -- & --\\
   100610.55$+$370513.8 & 14.45 & 15.15 & -- & 15.27\\
   111656.89$+$080829.4 & 15.62 & 15.63 & 15.56 & 15.39\\
   113002.35$+$115438.3 & 16.19 & 16.29 & 16.18 & 15.98\\
   133724.69$+$315254.5 & 16.08 & 16.15 & 16.04 & --\\
   133757.87$+$021820.9 & 15.15 & 15.74 & 15.72 & $>$15.17\\
   140745.50$+$403702.2 & 14.59 & 14.48 & -- & 14.63\\
   142755.85$-$002951.1 & 14.65 & 15.49 & 15.39 & 15.27\\
   150238.38$+$030228.2 & 16.21 & 16.54 & 16.15 & --\\
   151044.66$+$321712.9 & 17.38 & 17.34 & -- & --\\
   155036.80$+$053749.9 & 15.51 & 15.76 & 15.84 & $>$15.59\\
   155137.22$+$321307.5 & 18.29 & --\footnotemark[$*$] & -- & --\\
   155823.22$+$353252.2 & 17.43 & --\footnotemark[$*$] & -- & --\\
   165523.09$+$184708.4 & 15.38 & 15.43 & -- & --\\
   211936.77$+$104623.9 & 16.88 & 16.84 & -- & --\\
   213023.61$+$122252.2 & 15.06 & --\footnotemark[$*$] & -- & 15.29\\
   213455.08$+$001056.8 & 16.72 & 16.97 & 16.69 & --\\
   231858.56$-$005049.6 & 16.85 & 17.09 & 16.95 & --\\
   234150.01$+$144906.0 & 16.15 & 15.87 & 15.92 & --\\
   \hline
  \end{tabular}
 \end{center}
 {\bf Notes.} Column (1): Object name. Column (2): Spectroscopic magnitude based on our data. Flux calibration is carried out by using spectroscopic standard star. Column (3): Imaging magnitude based on our data (PSF magnitude). Column (4): UKIDSS magnitude (aperture magnitude). Column (5): 2MASS magnitude (PSF magnitude). Only sources whose SMBH masses ($M_\mathrm{BH}$) were estimated are listed. Imaging magnitudes were used for flux calibration.
 
 \footnotemark[$*$] Imaging data quality is not good enough to obtain reliable photometric magnitude.
\end{table}

\clearpage
\begin{table}[h]
 \caption{$5100\mathrm{\AA}$ continuum flux densities and H$\beta$ flux and luminosity.}
  \label{table:3}
\begin{center}
\begin{tabular}{cccc|ccc}
\hline
(1) & (2) &
 \multicolumn{2}{c|}{Flux [$10^{-15}\mathrm{erg/s/cm^{2}}$]} & 
 \multicolumn{2}{c}{Luminosity [$10^{44}$erg/s]} \\ \cline{3-6}
 Object ID (SDSS J) & $5100\mathrm{\AA}$ continuum flux & (3) & (4) & (5) & (6)\\
   & [$10^{-17}\mathrm{erg/s/cm^{2}/\AA}$] & H$\beta$ broad & H$\beta$ narrow & H$\beta$ broad & H$\beta$ narrow \\ \hline \hline
   014619.97$-$004628.7 & 0.860 $\pm$ 0.047 & 0.661 $\pm$ 0.097 & 0.065 $\pm$ 0.025 & 0.584 $\pm$ 0.086 & 0.057 $\pm$ 0.022 \\
   031845.17$-$001845.3 & 0.727 $\pm$ 0.121 & 1.270 $\pm$ 0.144 & -- & 1.164 $\pm$ 0.132 & -- \\
   072554.52$+$392243.4 & 0.631 $\pm$ 0.014 & 0.415 $\pm$ 0.016 & -- & 0.386 $\pm$ 0.015 & --\\ 
   074939.01$+$433217.6 & 1.278 $\pm$ 0.064 &1.033 $\pm$ 0.116 & 0.075 $\pm$ 0.045 & 0.891 $\pm$ 0.100 & 0.065 $\pm$ 0.039 \\
   075515.93$+$154216.6 & 0.690 $\pm$ 0.151 & 0.216 $\pm$ 0.085 & -- & 0.210 $\pm$ 0.083 & -- \\
   075841.66$+$174558.0 & 0.932 $\pm$ 0.087 & 0.857 $\pm$ 0.012 & -- & 0.757 $\pm$ 0.107 & -- \\
   083700.82$+$350550.2 & 1.268 $\pm$ 0.093 & 0.922 $\pm$ 0.148 & 0.078 $\pm$ 0.070 & 0.903 $\pm$ 0.145 & 0.077 $\pm$ 0.069 \\
   084715.16$+$383110.0 & 1.405 $\pm$ 0.061 & 1.017 $\pm$ 0.059 & -- & 0.905 $\pm$ 0.053 & -- \\
   094202.04$+$042244.5 & 5.473 $\pm$ 0.110 & 2.900 $\pm$ 0.179 & 0.435 $\pm$ 0.064 & 2.779 $\pm$ 0.171 & 0.417 $\pm$ 0.061 \\
   095735.37$+$353520.6 & 1.354 $\pm$ 0.034 & 0.497 $\pm$ 0.029 & -- & 0.473 $\pm$ 0.028 & --  \\
   100610.55$+$370513.8 & 3.223 $\pm$ 0.043 & 1.383 $\pm$ 0.144 & 0.148 $\pm$ 0.048 & 1.250 $\pm$ 0.130 & 0.134 $\pm$ 0.043 \\
   111656.89$+$080829.4 & 2.692 $\pm$ 0.119 & 2.235 $\pm$ 0.174 & 0.109 $\pm$ 0.053 & 2.064 $\pm$ 0.160 & 0.101 $\pm$ 0.049 \\
   113002.35$+$115438.3 & 0.981 $\pm$ 0.065 & 0.559 $\pm$ 0.059 & 0.062 $\pm$ 0.028 & 0.580 $\pm$ 0.061 & 0.064 $\pm$ 0.029 \\
   133724.69$+$315254.5 & 1.163 $\pm$ 0.025 & 0.887 $\pm$ 0.056 & 0.448 $\pm$ 0.030 & 0.789 $\pm$ 0.050 & 0.399 $\pm$ 0.027 \\
   133757.87$+$021820.9 & 1.601 $\pm$ 0.079 & 0.790 $\pm$ 0.114 & 0.194 $\pm$ 0.066 & 0.785 $\pm$ 0.113 & 0.193 $\pm$ 0.066 \\
   140745.50$+$403702.2 & 7.128 $\pm$ 0.031 & 2.378 $\pm$ 0.050 & -- & 2.148 $\pm$ 0.045 & -- \\
   142755.85$-$002951.1 &  2.274 $\pm$ 0.082 & 1.240 $\pm$ 0.018 & 0.370 $\pm$ 0.057 & 1.259 $\pm$ 0.181 & 0.376 $\pm$ 0.058 \\  
   150238.38$+$030228.2 & 0.665 $\pm$ 0.030 & 0.644 $\pm$ 0.031 & -- & 0.649 $\pm$ 0.031 & -- \\
   151044.66$+$321712.9 & 0.265 $\pm$ 0.016 & 0.295 $\pm$ 0.028 & -- & 0.325 $\pm$ 0.031 & --\\ 
   155036.80$+$053749.9 & 1.793 $\pm$ 0.032 & 1.203 $\pm$ 0.046 & 0.242 $\pm$ 0.015 & 1.046 $\pm$ 0.040 & 0.210 $\pm$ 0.013\\ 
   155137.22$+$321307.5 & 0.506 $\pm$ 0.035 & 0.487 $\pm$ 0.105 & 0.102 $\pm$ 0.024 & 0.434 $\pm$ 0.093 & 0.090 $\pm$ 0.021\\
   155823.22$+$353252.2 & 0.413 $\pm$ 0.019 & 0.243 $\pm$ 0.116 & 0.047 $\pm$ 0.014 & 0.219 $\pm$ 0.105 & 0.042 $\pm$ 0.013\\
   165523.09$+$184708.4 & 2.388 $\pm$ 0.039 & 0.691 $\pm$ 0.047 & 0.331 $\pm$ 0.031 & 0.682 $\pm$ 0.046 & 0.326 $\pm$ 0.031\\
   211936.77$+$104623.9 & 0.684 $\pm$ 0.015 & 0.229 $\pm$ 0.059 & 0.115 $\pm$ 0.050 & 0.213 $\pm$ 0.055 & 0.107 $\pm$ 0.046\\
   213023.61$+$122252.2 & 3.620 $\pm$ 0.117 & 1.692 $\pm$ 0.139 & 0.321 $\pm$ 0.040 & 1.597 $\pm$ 0.131 & 0.303 $\pm$ 0.038 \\
   213455.08$+$001056.8 & 0.664 $\pm$ 0.015 & 0.169 $\pm$ 0.017 & -- & 0.157 $\pm$ 0.016 & -- \\
   231858.56$-$005049.6 & 0.567 $\pm$ 0.015 & 0.189 $\pm$ 0.028 & 0.040 $\pm$ 0.018 & 0.171 $\pm$ 0.025 & 0.036 $\pm$ 0.016\\
   234150.01$+$144906.0 & 1.891 $\pm$ 0.046 & 0.604 $\pm$ 0.072 & -- & 0.537 $\pm$ 0.064 & -- \\ \hline
\end{tabular}
\end{center}
{\bf Notes.} Column (1): Object name. Column (2): Continuum flux at $5100\mathrm{\AA}$ (3): Flux of broad H$\beta$ emission line. Column (4): Flux of narrow H$\beta$ emission line. Column (5): Luminosity of broad H$\beta$ emission line. Column (6): Luminosity of narrow H$\beta$ emission line. The values were calculated from fitting results.
\end{table}

\clearpage
\begin{table}[h]
 \caption{[O\emissiontype{III}] flux and luminosity.}
   \label{table:4}
\begin{center}
\begin{tabular}{ccc|cc}
\hline
&
 \multicolumn{2}{c|}{Flux [$10^{-15}\mathrm{erg/s/cm^{2}}$]} &
  \multicolumn{2}{c}{Luminosity [$10^{44}$erg/s]} \\ \cline{2-5} 
  (1) & (2) & (3) & (4) & (5)\\
  Object ID (SDSS J) & [O\emissiontype{III}] $\lambda$5007 & [O\emissiontype{III}] $\lambda$4959 & [O\emissiontype{III}] $\lambda$5007 & [O\emissiontype{III}] $\lambda$4959 \\ \hline\hline
   014619.97$-$004628.7 & 0.361 $\pm$ 0.016 & 0.120 $\pm$ 0.005 & 0.318 $\pm$ 0.014 & 0.106 $\pm$ 0.005\\
   031845.17$-$001845.3 & 0.351 $\pm$ 0.084 & 0.117 $\pm$ 0.028 & 0.322 $\pm$ 0.078 & 0.107 $\pm$ 0.026\\
   072554.52$+$392243.4 & 0.170 $\pm$ 0.016 & 0.057 $\pm$ 0.005 & 0.158 $\pm$ 0.015 & 0.053 $\pm$ 0.005\\ 
   074939.01$+$433217.6 & 1.127 $\pm$ 0.040 & 0.376 $\pm$ 0.013 & 0.973 $\pm$ 0.035 & 0.324 $\pm$ 0.012\\
   075515.93$+$154216.6 & 0.185 $\pm$ 0.161 & 0.062 $\pm$ 0.054 & 0.180 $\pm$ 0.156 & 0.060 $\pm$ 0.052\\
   075841.66$+$174558.0 & 0.922 $\pm$ 0.098 & 0.307 $\pm$ 0.033 & 0.814 $\pm$ 0.087 & 0.271 $\pm$ 0.029\\
   083700.82$+$350550.2 & 0.422 $\pm$ 0.094 & 0.141 $\pm$ 0.031 & 0.413 $\pm$ 0.092 & 0.138 $\pm$ 0.031\\
   084715.16$+$383110.0 & 0.755 $\pm$ 0.031 & 0.252 $\pm$ 0.010 & 0.672 $\pm$ 0.027 & 0.224 $\pm$ 0.009\\
   094202.04$+$042244.5 & 1.137 $\pm$ 0.063 & 0.379 $\pm$ 0.021 & 1.090 $\pm$ 0.060 & 0.363 $\pm$ 0.020\\
   095735.37$+$353520.6 & 0.173 $\pm$ 0.030 & 0.058 $\pm$ 0.010 & 0.165 $\pm$ 0.028 & 0.055 $\pm$ 0.009\\
   100610.55$+$370513.8 & 1.171 $\pm$ 0.046 & 0.390 $\pm$ 0.015 & 1.058 $\pm$ 0.041 & 0.353 $\pm$ 0.014\\
   111656.89$+$080829.4 & 0.765 $\pm$ 0.058 & 0.255 $\pm$ 0.019 & 0.707 $\pm$ 0.053 & 0.236 $\pm$ 0.018\\
   113002.35$+$115438.3 & 0.212 $\pm$ 0.038 & 0.071 $\pm$ 0.013 & 0.219 $\pm$ 0.039 & 0.073 $\pm$ 0.013\\
   133724.69$+$315254.5 & 0.462 $\pm$ 0.022 & 0.154 $\pm$ 0.007 & 0.411 $\pm$ 0.020 & 0.137 $\pm$ 0.007\\
   133757.87$+$021820.9 & 0.436 $\pm$ 0.061 & 0.145 $\pm$ 0.020 & 0.433 $\pm$ 0.061 & 0.014 $\pm$ 0.020\\
   140745.50$+$403702.2 & -- & -- & -- & -- \\ 
   142755.85$-$002951.1 & 0.507 $\pm$ 0.050 & 0.169 $\pm$ 0.017 & 0.514 $\pm$ 0.050 & 0.171 $\pm$ 0.017\\
   150238.38$+$030228.2 & -- & -- & -- & -- \\
   151044.66$+$321712.9 & -- & -- & -- & -- \\
   155036.80$+$053749.9 & 0.993 $\pm$ 0.017 & 0.331 $\pm$ 0.006 & 0.864 $\pm$ 0.015 & 0.288 $\pm$ 0.005\\
   155137.22$+$321307.5 & 0.416 $\pm$ 0.027 & 0.139 $\pm$ 0.009 & 0.370 $\pm$ 0.024 & 0.123 $\pm$ 0.008\\
   155823.22$+$353252.2 & 0.110 $\pm$ 0.017 & 0.037 $\pm$ 0.006 & 0.099 $\pm$ 0.015 & 0.033 $\pm$ 0.005\\
   165523.09$+$184708.4 & 0.534 $\pm$ 0.028 & 0.178 $\pm$ 0.009 & 0.527 $\pm$ 0.027 & 0.176 $\pm$ 0.009\\
   211936.77$+$104623.9 & 0.159 $\pm$ 0.024 & 0.053 $\pm$ 0.008 & 0.148 $\pm$ 0.022 & 0.049 $\pm$ 0.007\\
   213023.61$+$122252.2 & 1.535 $\pm$ 0.051 & 0.512 $\pm$ 0.017 & 1.449 $\pm$ 0.048 & 0.483 $\pm$ 0.016 \\
   213455.08$+$001056.8 & -- & -- & -- & --\\
   231858.56$-$005049.6 & 0.049 $\pm$ 0.018 & 0.016 $\pm$ 0.006 & 0.044 $\pm$ 0.016 & 0.015 $\pm$ 0.005\\ 
   234150.01$+$144906.0 & 0.013 $\pm$ 0.005 & 0.004 $\pm$ 0.002 & 0.111 $\pm$ 0.046 & 0.037 $\pm$ 0.015\\ \hline
\end{tabular}
\end{center}
{\bf Notes.} Column (1): Object name. Column (2): Flux of [O\emissiontype{III}]$\lambda5007$ emission line. Column (3): Flux of [O\emissiontype{III}]$\lambda4959$ emission line. Column (4): Luminosity of [O\emissiontype{III}]$\lambda5007$ emission line. Column (5): Luminosity of [O\emissiontype{III}]$\lambda4959$ emission line. The values were calculated from fitting results.
\end{table}

\clearpage
\begin{table}[htbp]
  \caption{Redshift comparison.}
 \label{table:5}
 \begin{center}
 \small
  \begin{tabular}{cccc}
   \hline
   (1) & (2) & (3) & (4)\\
   Object ID (SDSS J) & H$\beta$ & [O\emissiontype{III}] $\lambda$5007 & SDSS \\
   \hline \hline
   014619.97$-$004628.7 & 3.173 &  3.170 & 3.173 \\
   031845.17$-$001845.3 & 3.225 & 3.222 & 3.224 \\
   072554.52$+$392243.4 & 3.258 & 3.251 & 3.249 \\
   074939.01$+$433217.6 & 3.144 & 3.134 & 3.141 \\
   075515.93$+$154216.6 & 3.288 & 3.294 & 3.298 \\
   075841.66$+$174558.0 & 3.182 & 3.166 & 3.170 \\
   083700.82$+$350550.2 & 3.322 & 3.305 & 3.311 \\
   084715.16$+$383110.0 &  3.189 & 3.184 & 3.180 \\
   094202.04$+$042244.5 &  3.287 & 3.277 & 3.276\\
   095735.37$+$353520.6 & 3.287 & 3.269 & 3.276 \\
   100610.55$+$370513.8 & 3.203 & 3.201 & 3.201 \\
   111656.89$+$080829.4 & 3.240 & 3.239 & 3.234\\
   113002.35$+$115438.3 & 3.434 & 3.392 & 3.394\\
   133724.69$+$315254.5 & 3.192 & 3.175 & 3.208\\
   133757.87$+$021820.9 & 3.358 & 3.334 & 3.333\\
   140745.50$+$403702.2 & 3.168 & -- & 3.200 \\
   142755.85$-$002951.1 & 3.373 & 3.362 & 3.365 \\
   150238.38$+$030228.2 & 3.370 & -- & 3.358\\
   151044.66$+$321712.9 & 3.478 & -- & 3.474\\
   155036.80$+$053749.9 & 3.159 & 3.147 & 3.153\\
   155137.22$+$321307.5 & 3.143 & 3.152 & 3.184\\
   155823.22$+$353252.2 & 3.191 & 3.186 & 3.198\\
   165523.09$+$184708.4 & 3.375 & 3.349 & 3.323 \\
   211936.77$+$104623.9 & 3.274 & 3.257 & 3.248\\
   213023.61$+$122252.2 & 3.267 & 3.274 & 3.272 \\
   213455.08$+$001056.8 & 3.289 & -- & 3.289\\
   231858.56$-$005049.6 & 3.211 & 3.208 & 3.209\\
   234150.01$+$144906.0 & 3.170 & 3.181 & 3.184\\ 
   \hline
  \end{tabular}
 \end{center}
  {\bf Notes.} Column (1): Object name. Column (2): Redshift measured by broad H$\beta$ emission line. Column (3): Redshift measured by [O\emissiontype{III}]$\lambda$5007 emission line. Column (4): SDSS redshift.
\end{table}

\clearpage
\begin{table}[htbp]
 \caption{Observed and derived properties related to SMBHs.}
 \label{table:6}
 \begin{center}
 \tiny
  \begin{tabular}{cccccccc}
   \hline
   (1) & (2) & (3) & (4) & (5) & (6) & (7)\\
   Object ID (SDSS J) & FWHM(H$\beta$)[$10^{3}$km/s] & $\lambda L_{\lambda}(5100\mathrm{\AA})$[$10^{46}$erg/s] & log $M_\mathrm{BH}$[\MO] & log $L_\mathrm{bol}/L_\mathrm{Edd}$ &$ t_\mathrm{grow}/t_\mathrm{universe} $ & log $M_\mathrm{BH}$[\MO] McL\\
   \hline \hline 
   014619.97$-$004628.7 & 6.92 $\pm$ 1.30 & 1.61 $\pm$ 0.09 & 9.68 $\pm$ 0.17 & $-$0.804 & 1.76 & 9.69 $\pm$ 0.17\\
   031845.17$-$001845.3 & 4.79 $\pm$ 0.48 & 1.43 $\pm$ 0.24 & 9.34 $\pm$ 0.09 & $-$0.516 & 0.87 & 9.34 $\pm$ 0.09\\
   072554.52$+$392243.4 & 4.32 $\pm$ 0.41 & 1.27 $\pm$ 0.03 & 9.23 $\pm$ 0.08 & $-$0.457 & 0.76 & 9.23 $\pm$ 0.08\\ 
   074939.01$+$433217.6 & 7.14 $\pm$ 0.53 & 2.33 $\pm$ 0.12 & 9.80 $\pm$ 0.06 & $-$0.764 & 1.63 & 9.82 $\pm$ 0.06\\
   075515.93$+$154216.6 & 2.70 $\pm$ 0.80 & 1.47 $\pm$ 0.32 & 8.81 $\pm$ 0.28 & 0.026 & 0.23 & 8.81 $\pm$ 0.28\\
   075841.66$+$174558.0 & 4.64 $\pm$ 1.03 & 1.75 $\pm$ 0.16 & 9.34 $\pm$ 0.20 & $-$0.428 & 0.70 & 9.35 $\pm$ 0.20\\
   083700.82$+$350550.2 & 5.18 $\pm$ 0.53 & 2.73 $\pm$ 0.20 & 9.55 $\pm$ 0.09 & $-$0.445 & 0.80 & 9.59 $\pm$ 0.09\\
   084715.16$+$383110.0 & 4.29 $\pm$ 0.46 & 2.67 $\pm$ 0.12 & 9.38 $\pm$ 0.09 & $-$0.284 & 0.51 & 9.41 $\pm$ 0.09\\
   094202.04$+$042244.5 & 5.34 $\pm$ 0.22 & 11.45 $\pm$ 0.23 & 9.89 $\pm$ 0.03 & $-$0.162 & 0.44 & 10.00 $\pm$ 0.03\\
   095735.37$+$353520.6 & 6.07 $\pm$ 0.71 & 2.80 $\pm$ 0.07 & 9.69 $\pm$ 0.11 & $-$0.574 & 1.08 & 9.73 $\pm$ 0.11\\
   100610.55$+$370513.8 & 5.94 $\pm$ 0.40 & 6.23 $\pm$ 0.08 & 9.85 $\pm$ 0.06 & $-$0.387 & 0.71 & 9.93 $\pm$ 0.06\\
   111656.89$+$080829.4 & 3.34 $\pm$ 0.22 & 5.36 $\pm$ 0.24 & 9.32 $\pm$ 0.05 & 0.078 & 0.22 & 9.39 $\pm$ 0.05\\
   113002.35$+$115438.3 & 6.05 $\pm$ 0.42 & 2.28 $\pm$ 0.15 & 9.65 $\pm$ 0.06 & $-$0.623 & 1.26 & 9.67 $\pm$ 0.06\\
   133724.69$+$315254.5 & 5.73 $\pm$ 0.24 & 2.21 $\pm$ 0.05 & 9.60 $\pm$ 0.04 & $-$0.587 & 1.06 & 9.62 $\pm$ 0.04\\
   133757.87$+$021820.9 & 5.18 $\pm$ 1.09 & 3.51 $\pm$ 0.17 & 9.59 $\pm$ 0.22 & $-$0.376 & 0.69 & 9.63 $\pm$ 0.22\\
   140745.50$+$403702.2 & 6.72 $\pm$ 0.17 & 13.79 $\pm$ 0.06 & 10.13 $\pm$ 0.02 & $-$0.389 & 0.74 & 10.24 $\pm$ 0.02\\
   142755.85$-$002951.1 & 5.49 $\pm$ 0.87 & 5.13 $\pm$ 0.19 & 9.73 $\pm$ 0.13 & $-$0.351 & 0.73 & 9.80 $\pm$ 0.13\\
   150238.38$+$030228.2 & 5.28 $\pm$ 0.21 & 1.49 $\pm$ 0.07 & 9.44 $\pm$ 0.03 & $-$0.598 & 1.13 & 9.45 $\pm$ 0.03\\
   151044.66$+$321712.9 & 5.04 $\pm$ 0.60 & 0.67 $\pm$ 0.04 & 9.22 $\pm$ 0.10 & $-$0.725 & 1.51 & 9.19 $\pm$ 0.10\\ 
   155036.80$+$053749.9 & 6.11 $\pm$ 0.29 & 3.30 $\pm$ 0.06 & 9.74 $\pm$ 0.04 & $-$0.552 & 1.00 & 9.78 $\pm$ 0.04\\
   155137.22$+$321307.5 & 6.70 $\pm$ 1.12 & 0.96 $\pm$ 0.07 & 9.54 $\pm$ 0.18 & $-$0.889 & 2.11 & 9.52 $\pm$ 0.18\\
   155823.22$+$353252.2 & 8.49 $\pm$ 2.98 & 0.80 $\pm$ 0.04 & 9.67 $\pm$ 0.28 & $-$1.098 & 3.52 & 9.65 $\pm$ 0.28\\ 
   165523.09$+$184708.4 & 5.93 $\pm$ 0.21 & 5.19 $\pm$ 0.08 & 9.81 $\pm$ 0.03 & $-$0.426 & 0.80 & 9.88 $\pm$ 0.03\\
   211936.77$+$104623.9 & 4.72 $\pm$ 0.78 & 1.38 $\pm$ 0.03 & 9.31 $\pm$ 0.16 & $-$0.501 & 0.85 & 9.31 $\pm$ 0.16\\
   213023.61$+$122252.2 & 4.39 $\pm$ 0.35 & 7.43 $\pm$ 0.24 & 9.63 $\pm$ 0.07 & $-$0.090 & 0.35 & 9.71 $\pm$ 0.06\\
   213455.08$+$001056.8 & $>$3.09 $\pm$ 0.31\footnotemark[$*$] & 1.34 $\pm$ 0.03 & $>$8.95 $\pm$ 0.09 & $<-$0.157 & $>$0.36 & $>$8.95 $\pm$ 0.08\\
   231858.56$-$005049.6 & 5.21 $\pm$ 0.75 & 1.10 $\pm$ 0.03 & 9.35 $\pm$ 0.13 & $-$0.650 & 1.19 & 9.35 $\pm$ 0.13\\
   234150.01$+$144906.0 & 7.49 $\pm$ 4.36 & 3.59 $\pm$ 0.09 & 9.72 $\pm$ 0.69 & $-$0.496 & 0.88 & 9.77 $\pm$ 0.69\\
   \hline
  \end{tabular}
 \end{center}
 {\bf Notes.} Column (1): Object name. Column (2): Full-width at half maximum (FWHM) of broad H$\beta$ emission line. Column (3): Continuum luminosity at 5100 $\mathrm{\AA}$. Column (4): Decimal logarithm of SMBH mass. Statistical errors were estimated by resampling approach (Section 4.1; \cite{Schulze2010}, \cite{Assef2011}, \cite{Shen2012}). Column (5): Decimal logarithm of Eddington ratio. Column (6): Ratio of SMBH growth time ($t_\mathrm{grow}$) and the age of the universe ($t_\mathrm{universe}$). $t_\mathrm{grow}$ is the time scale of SMBH growth, calculated using the measured $M_\mathrm{BH}$ and the Eddington ratio. $t_\mathrm{universe}$ is dependent on the redshift of each target. See Section 5.2 for more details. Column (7): Decimal logarithm of SMBH masses derived from the formula in \citet{McLure2002}.
 
\footnotemark[$*$] For this object, we cannot measure the [O\emissiontype{III}] emission linewidth, therefore, the H$\beta$ narrow component is not subtracted. If the H$\beta$ narrow component exists, then the H$\beta$ broad emission linewidth broadens and the estimated $M_\mathrm{BH}$ becomes larger.
\end{table}

\clearpage
\begin{table}[htbp]
 \caption{C\emissiontype{IV}-based SMBH mass in our QSO sample.}
 \label{table:7}
 \begin{center}
 \small
  \begin{tabular}{ccc}
   \hline
   (1) & (2) & (3)\\
   Object ID (SDSS J) & FWHM(C\emissiontype{IV})[$10^{3}$km/s] & log $M_\mathrm{BH}$[\MO] \\
   \hline \hline 
   014619.97$-$004628.7 & 2.70 $\pm$ 0.15 &  8.91 $\pm$ 0.05\\
   031845.17$-$001845.3 & 3.96 $\pm$ 0.14 &  9.25 $\pm$ 0.03\\
   072554.52$+$392243.4 & 3.39 $\pm$ 0.18 &  9.11 $\pm$ 0.05\\ 
   074939.01$+$433217.6 & 2.52 $\pm$ 0.07 &  8.92 $\pm$ 0.03\\
   075515.93$+$154216.6 & 2.87 $\pm$ 0.24 &  8.78 $\pm$ 0.07\\
   075841.66$+$174558.0 & 3.37 $\pm$ 0.11 &  9.04 $\pm$ 0.03\\
   083700.82$+$350550.2 & 3.89 $\pm$ 0.30 &  9.36 $\pm$ 0.07\\
   084715.16$+$383110.0 & 2.25 $\pm$ 0.11 &  8.87 $\pm$ 0.04\\
   094202.04$+$042244.5 & 3.32 $\pm$ 0.10 &  9.49 $\pm$ 0.03\\
   095735.37$+$353520.6 & 7.56 $\pm$ 0.55 &  9.96 $\pm$ 0.06\\
   100610.55$+$370513.8 & 4.23 $\pm$ 0.11 &  9.42 $\pm$ 0.02\\
   111656.89$+$080829.4 & 2.81 $\pm$ 0.20 &  9.07 $\pm$ 0.06\\
   113002.35$+$115438.3 & 6.28 $\pm$ 1.38 &  9.76 $\pm$ 0.19\\
   133724.69$+$315254.5 & 4.03 $\pm$ 0.33 &  9.34 $\pm$ 0.07\\
   133757.87$+$021820.9 & 7.39 $\pm$ 0.41 &  9.96 $\pm$ 0.03\\
   140745.50$+$403702.2 & 4.54 $\pm$ 1.75 &  9.41 $\pm$ 0.34\\
   142755.85$-$002951.1 & 3.04 $\pm$ 0.06 &  9.20 $\pm$ 0.02\\
   150238.38$+$030228.2 & 6.45 $\pm$ 0.70 &  9.80 $\pm$ 0.09\\
   151044.66$+$321712.9 & 5.84 $\pm$ 0.61 &  9.44 $\pm$ 0.09\\ 
   155036.80$+$053749.9 & 2.76 $\pm$ 0.09 &  9.12 $\pm$ 0.03\\
   155137.22$+$321307.5 & --\footnotemark[$*$] &  -- \\
   155823.22$+$353252.2 & 2.05 $\pm$ 0.08 &  8.43 $\pm$ 0.04\\ 
   165523.09$+$184708.4 & 9.56 $\pm$ 0.59 &  10.32 $\pm$ 0.05\\
   211936.77$+$104623.9 & 5.63 $\pm$ 0.26 &  9.53 $\pm$ 0.04\\
   213023.61$+$122252.2 & 2.52 $\pm$ 0.10 &  9.11 $\pm$ 0.03\\
   213455.08$+$001056.8 & 6.10 $\pm$ 0.41 &  9.66 $\pm$ 0.06\\
   231858.56$-$005049.6 & 3.21 $\pm$ 0.37 &  8.87 $\pm$ 0.10\\
   234150.01$+$144906.0 & --\footnotemark[$*$] &  -- \\
   \hline
  \end{tabular}
 \end{center}
 {\bf Notes.} Column (1): Object name. Column (2): FWHM of C\emissiontype{IV} emission line from SDSS (\cite{Shen2011}). Column (3): C\emissiontype{IV}-based SMBH mass from SDSS (\cite{Shen2011}).
 
 \footnotemark[$*$]  For these objects, C\emissiontype{IV} data are not available in \citet{Shen2011}.
\end{table}

\newpage
   \begin{figure}[htbp]
  \begin{center}
   \begin{minipage}{.45\linewidth} 
   \FigureFile(75mm, 75mm){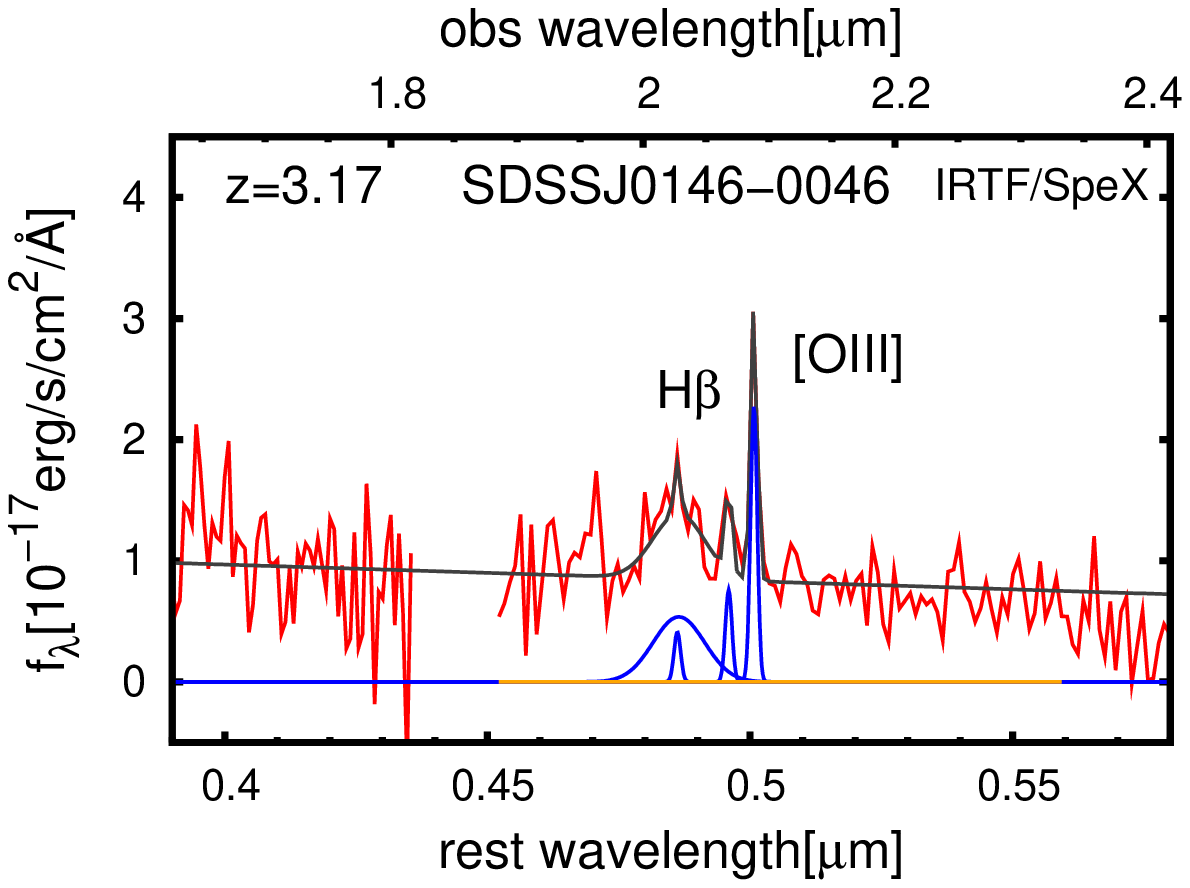} 
  \end{minipage} 
  \hspace{2.0pc} 
   \begin{minipage}{.45\linewidth} 
   \FigureFile(75mm, 75mm){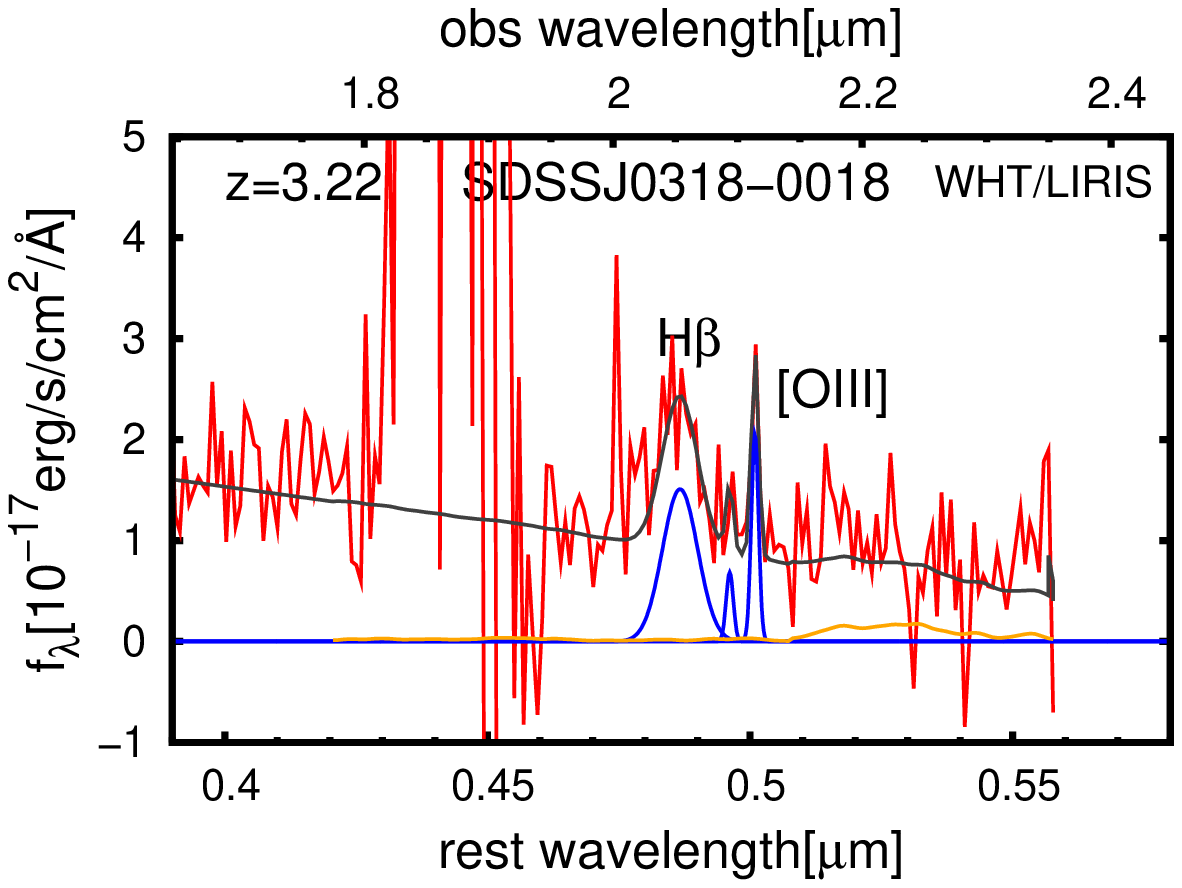} 
  \end{minipage}
  \end{center}
  
 \hspace{20mm} 
  
  \begin{center}
  \begin{minipage}{.45\linewidth} 
   \FigureFile(75mm, 75mm){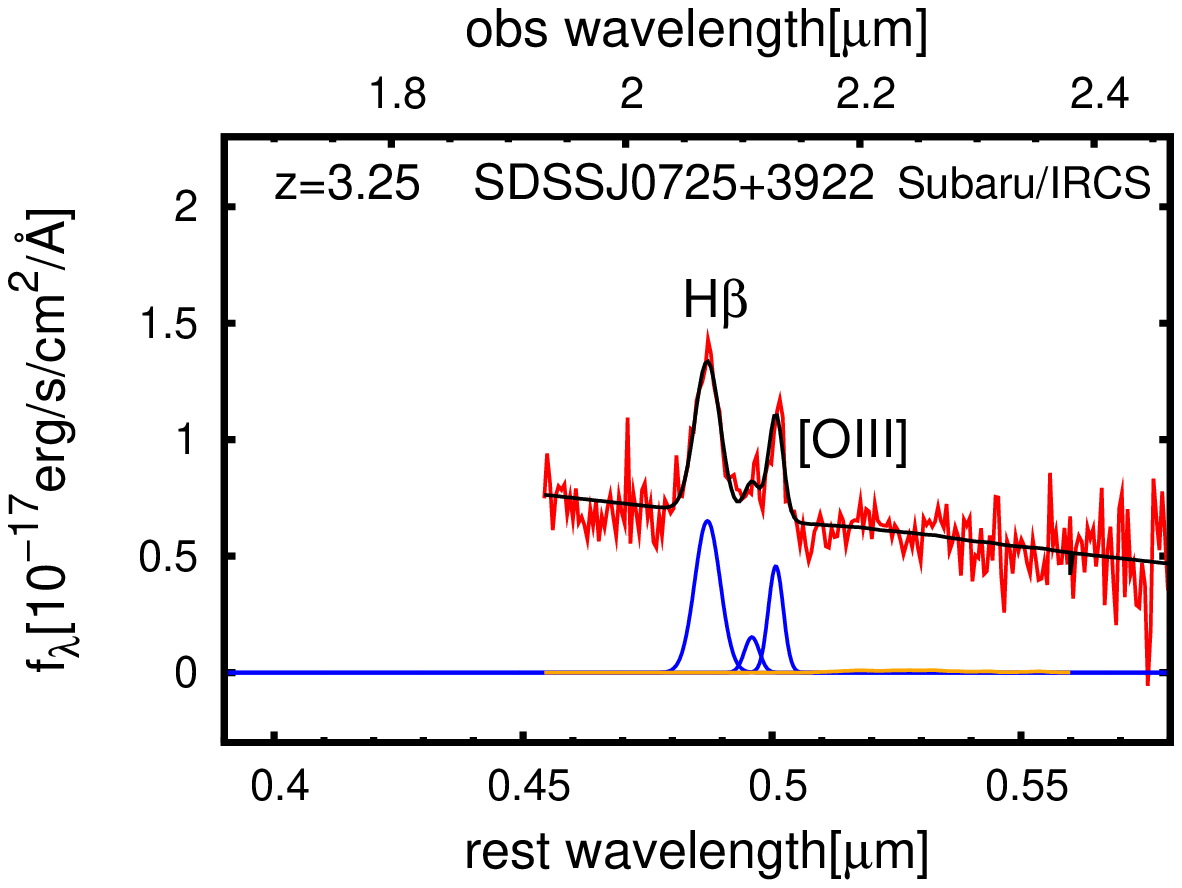} 
  \end{minipage}
   \hspace{2.0pc} 
  \begin{minipage}{.45\linewidth} 
   \FigureFile(75mm, 75mm){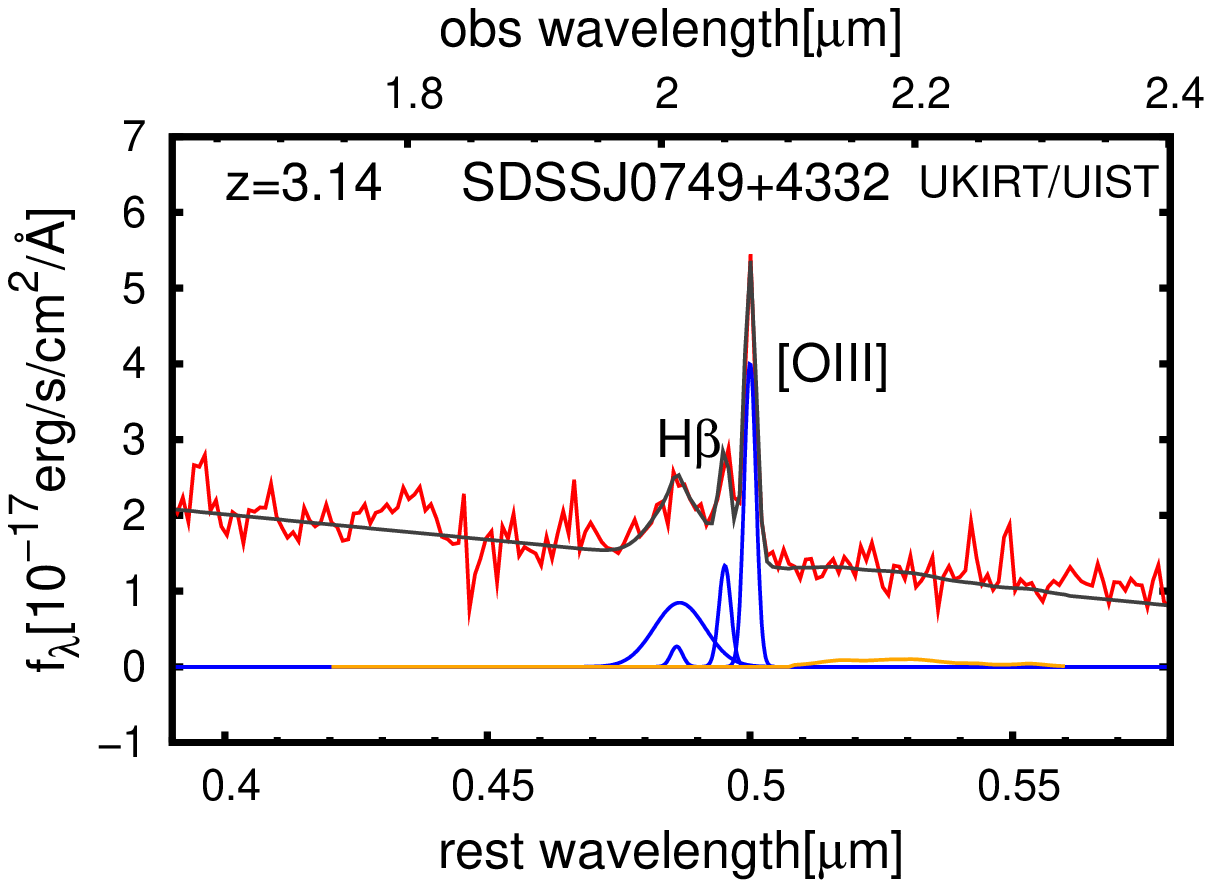} 
  \end{minipage}
  \end{center}

\hspace{20mm}

  \begin{center}
  \begin{minipage}{.45\linewidth} 
   \FigureFile(75mm, 75mm){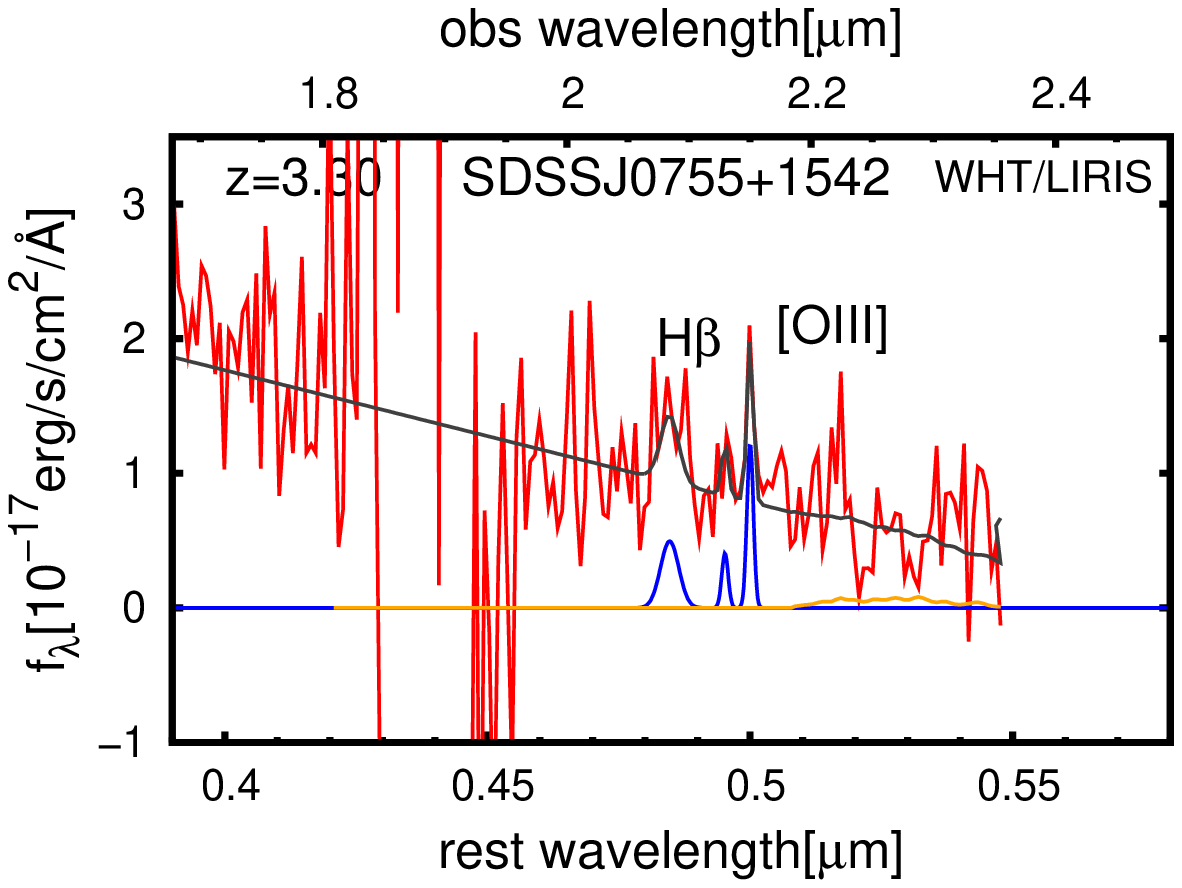} 
  \end{minipage} 
   \hspace{2.0pc} 
  \begin{minipage}{.45\linewidth} 
   \FigureFile(75mm, 75mm){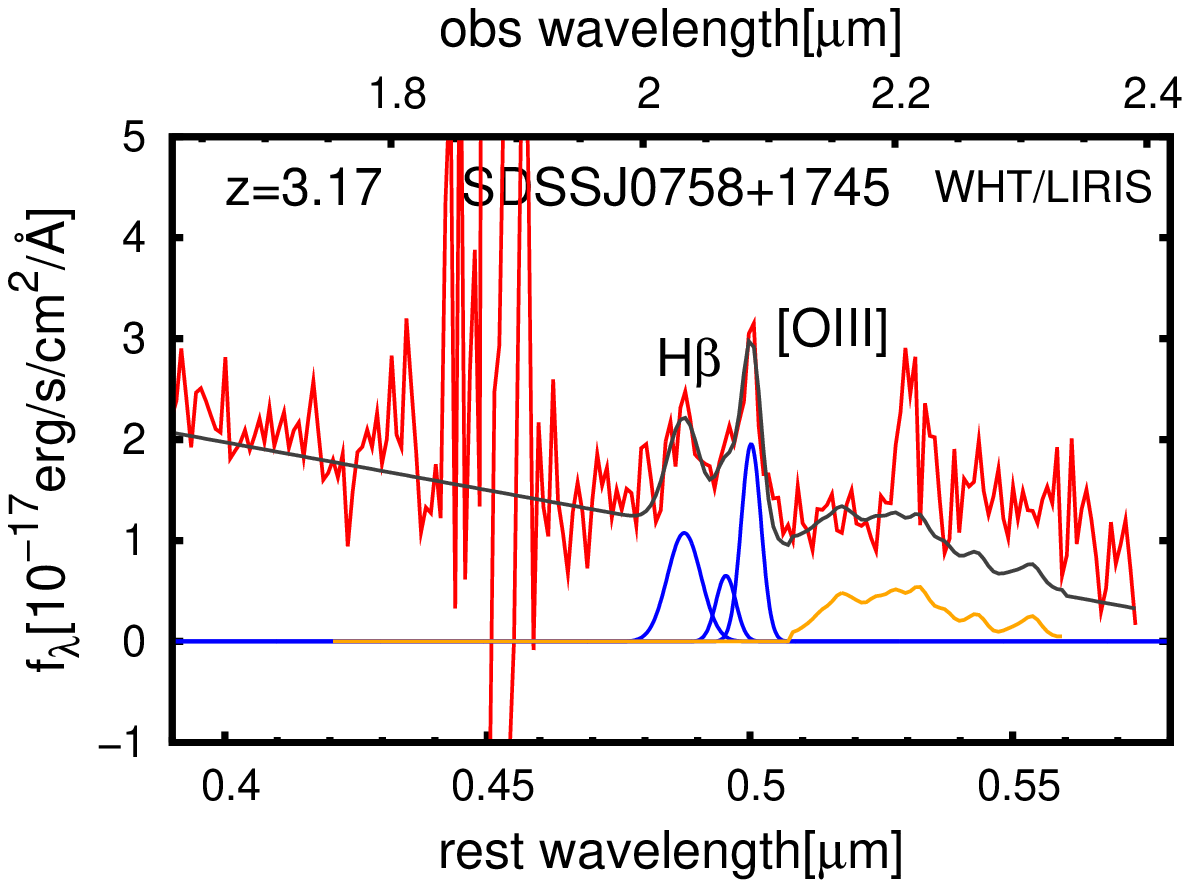} 
  \end{minipage}
  \end{center}
 \caption{All spectra of 28 $z\sim3$ QSOs with fitted H$\beta$ and the [O\emissiontype{III}] emission lines in our sample. The abscissa is the rest-frame (bottom) and the observed (top) wavelength in [$\mu$m]. The ordinate is the flux in [$\mathrm{10^{-17} erg/s/cm^{2}/ \mathrm{\AA}}$]. The best-fit model (black line) in each panel is composed of a continuum component, Fe\emissiontype{II} emission (orange line), and H$\beta$ and [O\emissiontype{III}] emission lines (blue lines).} \label{fig:1}
  \end{figure} 

\newpage  
\setcounter{figure}{0}
  \begin{figure}[htbp]    
  \begin{center}
  \begin{minipage}{.45\linewidth} 
   \FigureFile(75mm, 75mm){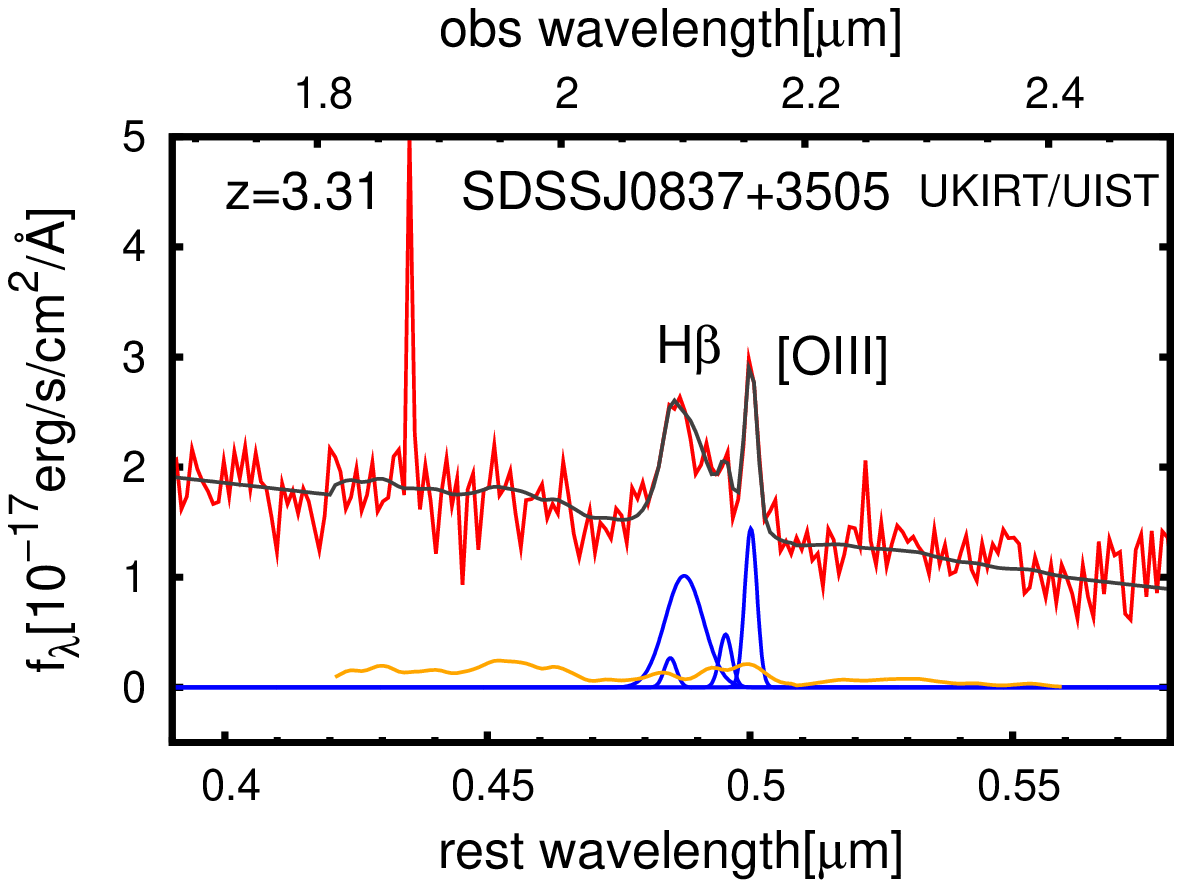} 
  \end{minipage}
  \hspace{2.0pc} 
  \begin{minipage}{.45\linewidth} 
   \FigureFile(75mm, 75mm){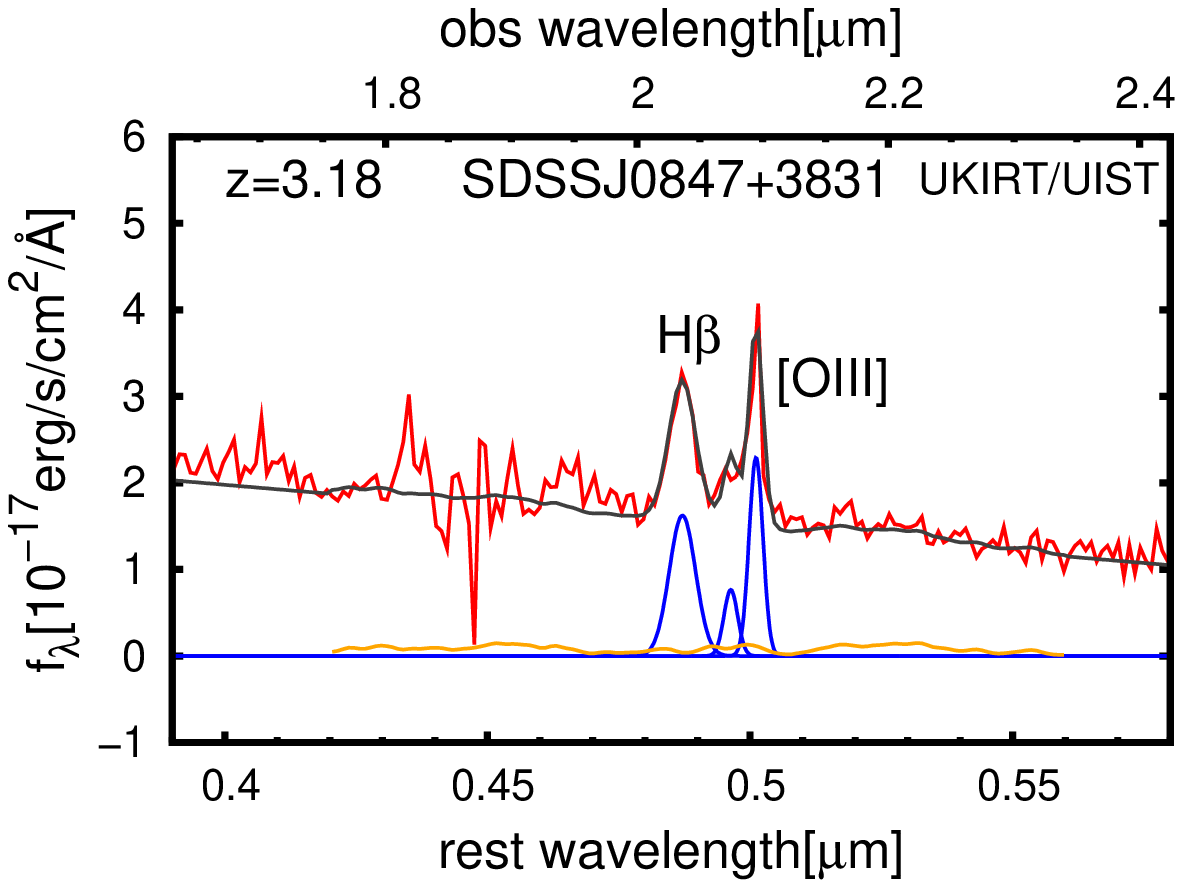} 
  \end{minipage}
  \end{center}

\hspace{20mm}    

\begin{center}
 \begin{minipage}{.45\linewidth} 
   \FigureFile(75mm, 75mm){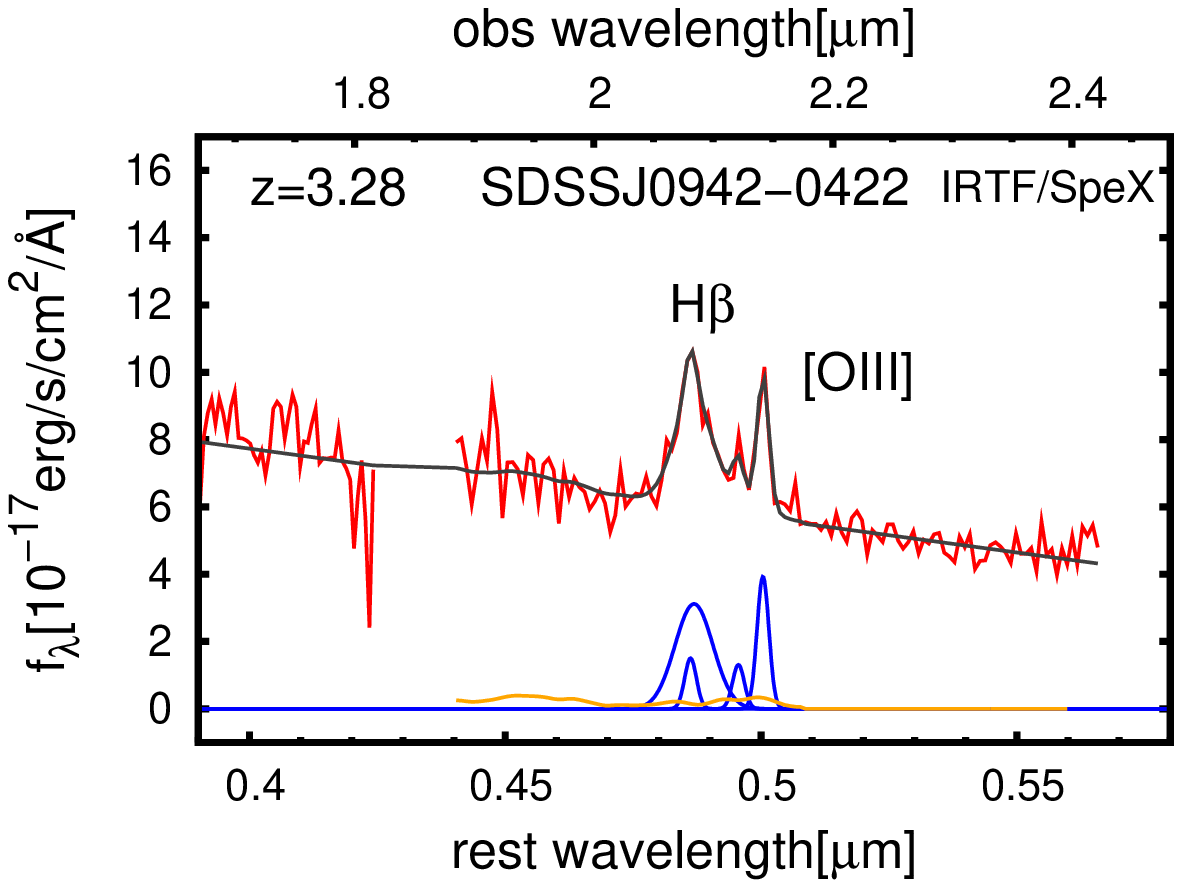} 
  \end{minipage}
\hspace{2.0pc} 
   \begin{minipage}{.45\linewidth} 
   \FigureFile(75mm, 75mm){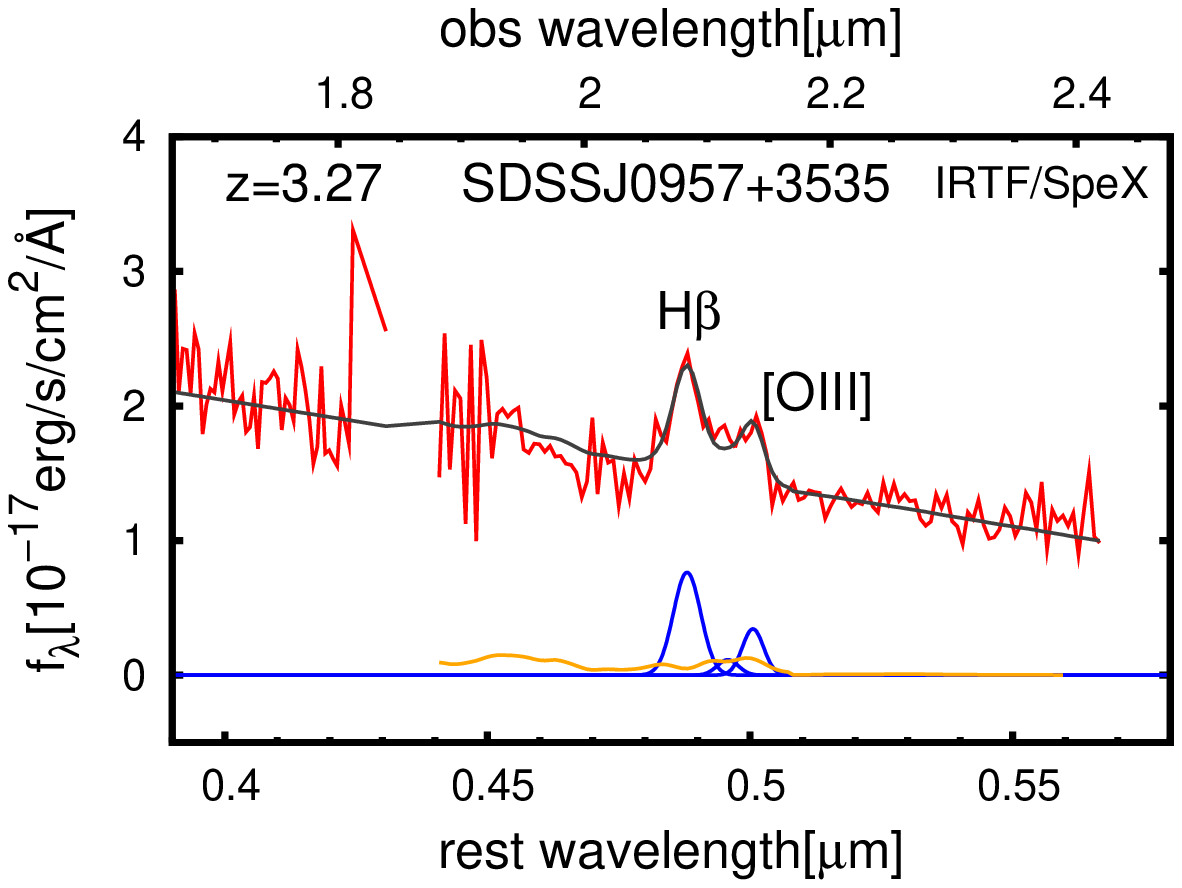} 
  \end{minipage} 
  \end{center}

\hspace{20mm}
   
  \begin{center}
  \begin{minipage}{.45\linewidth} 
  \FigureFile(75mm, 75mm){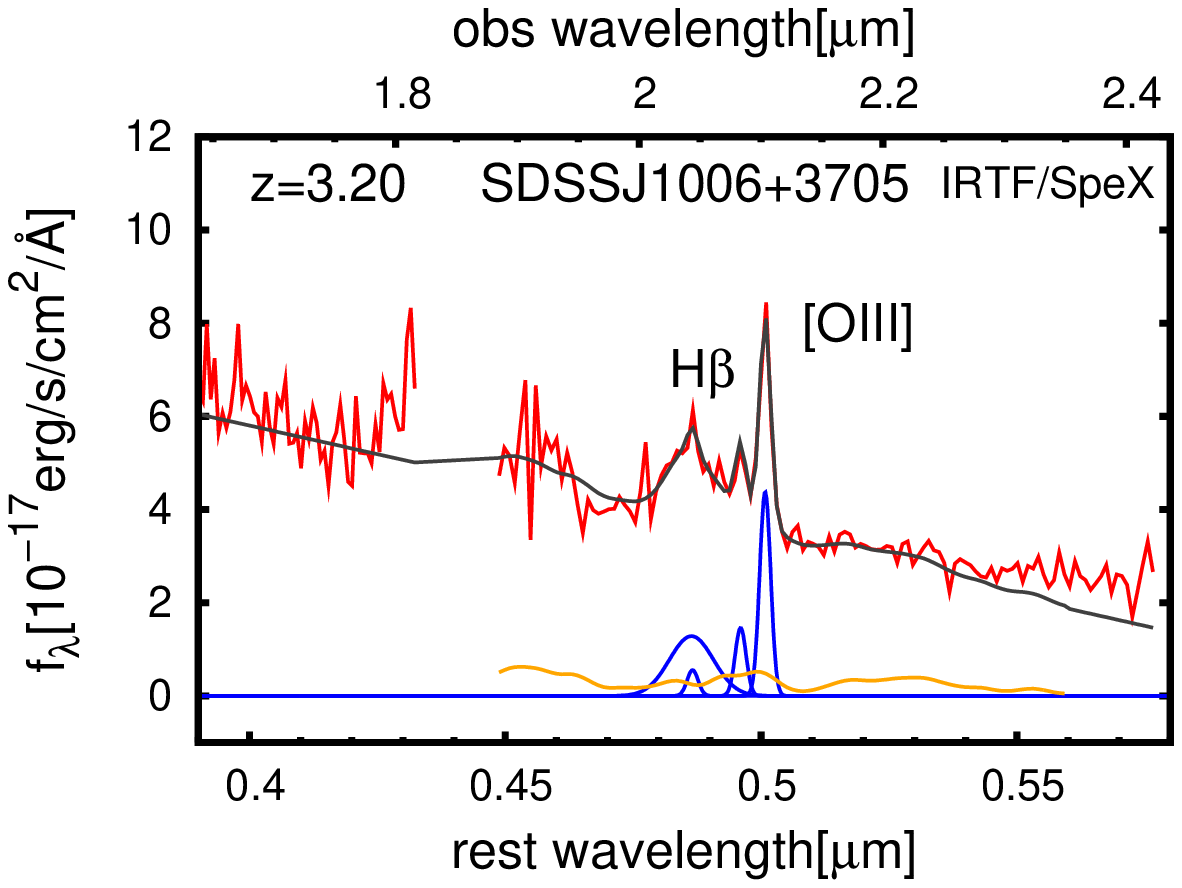} 
  \end{minipage}
    \hspace{2.0pc} 
   \begin{minipage}{.45\linewidth} 
   \FigureFile(75mm, 75mm){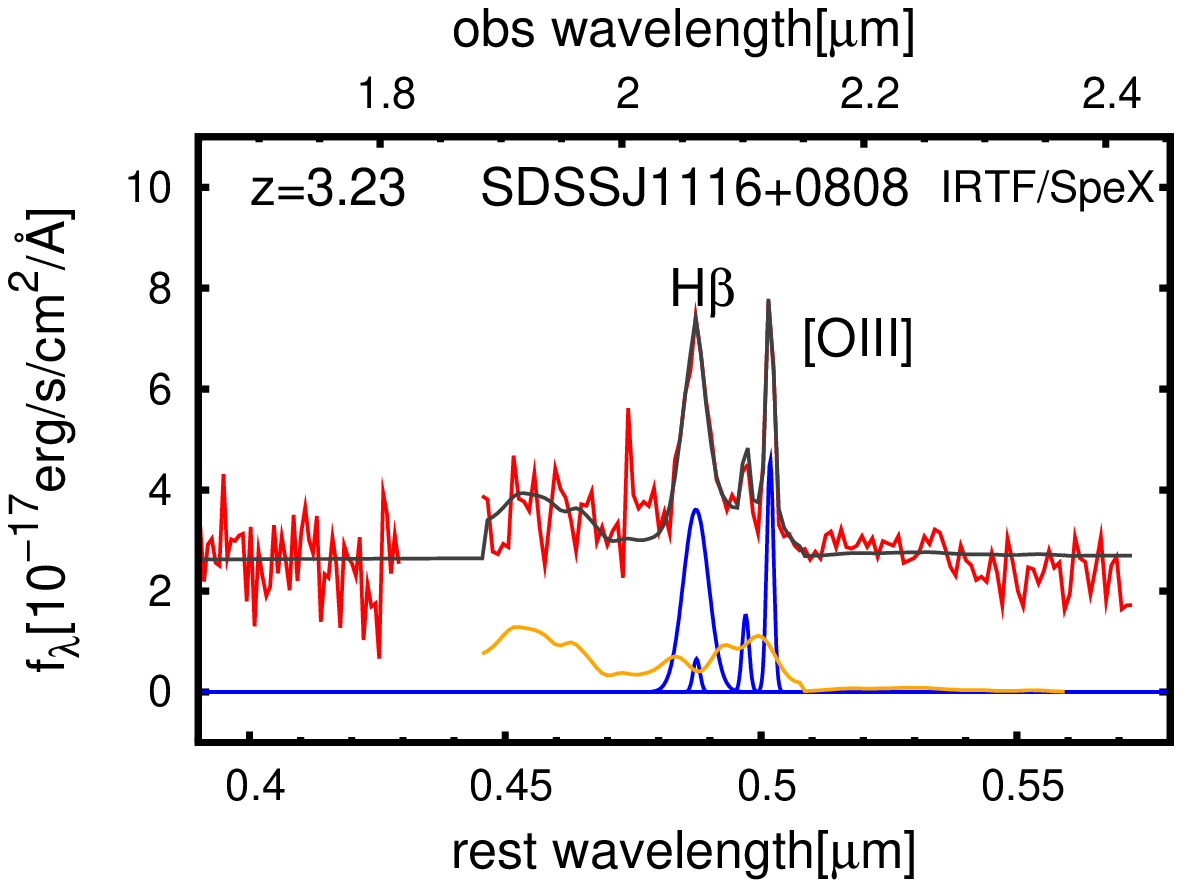} 
  \end{minipage} 
  \end{center}
  \caption{continued}
  \end{figure}

\newpage
\setcounter{figure}{0}  
  \begin{figure}[htbp] 
  \begin{center}
  \begin{minipage}{.45\linewidth} 
  \FigureFile(75mm, 75mm){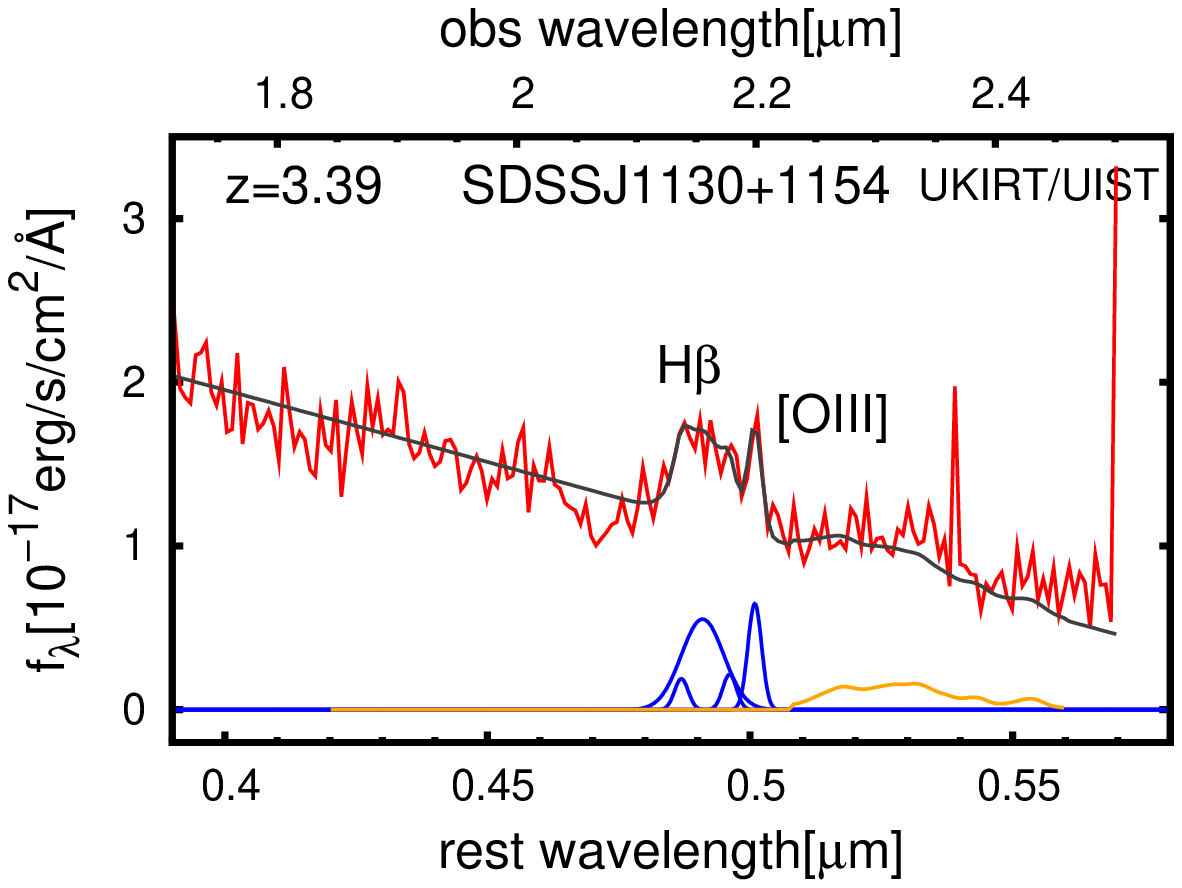} 
  \end{minipage}
    \hspace{2.0pc} 
   \begin{minipage}{.45\linewidth} 
   \FigureFile(75mm, 75mm){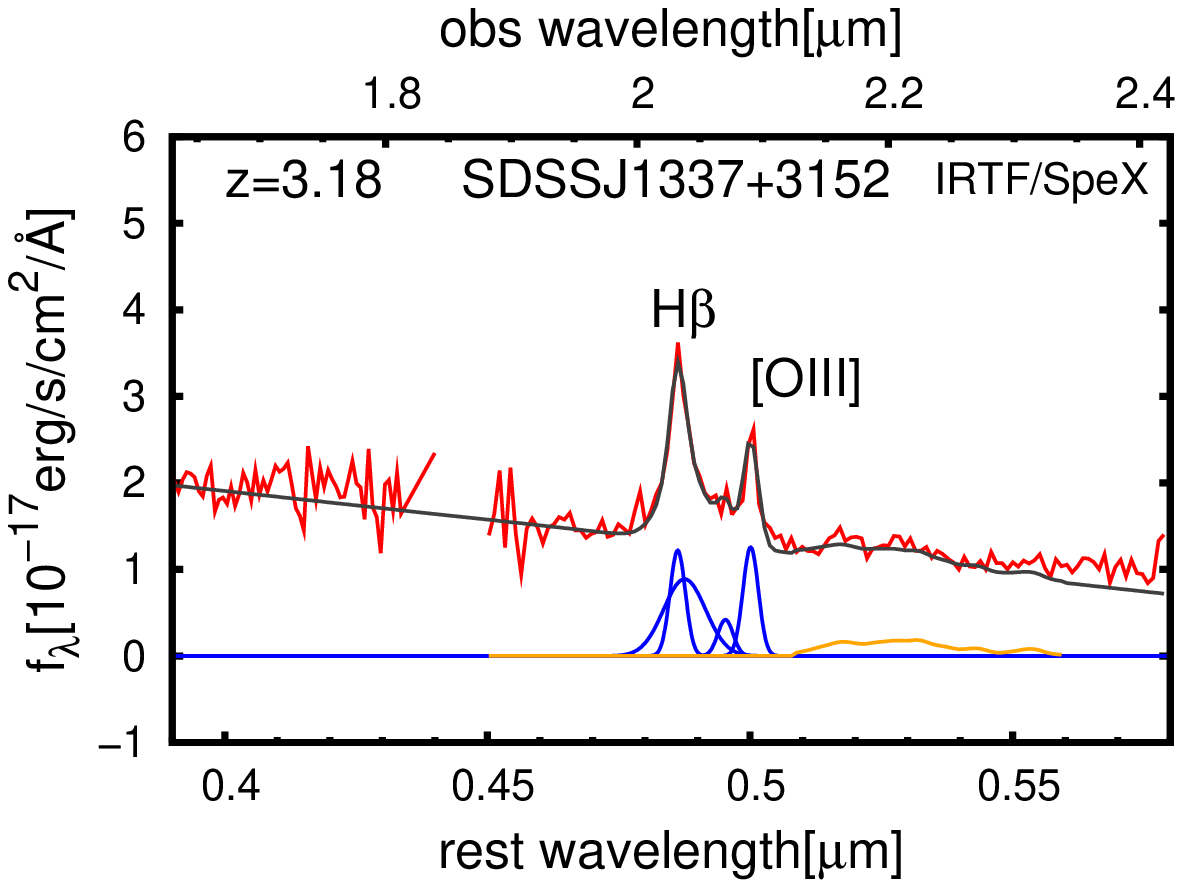} 
  \end{minipage} 
  \end{center}

\hspace{20mm}

\begin{center}
  \begin{minipage}{.45\linewidth} 
  \FigureFile(75mm, 75mm){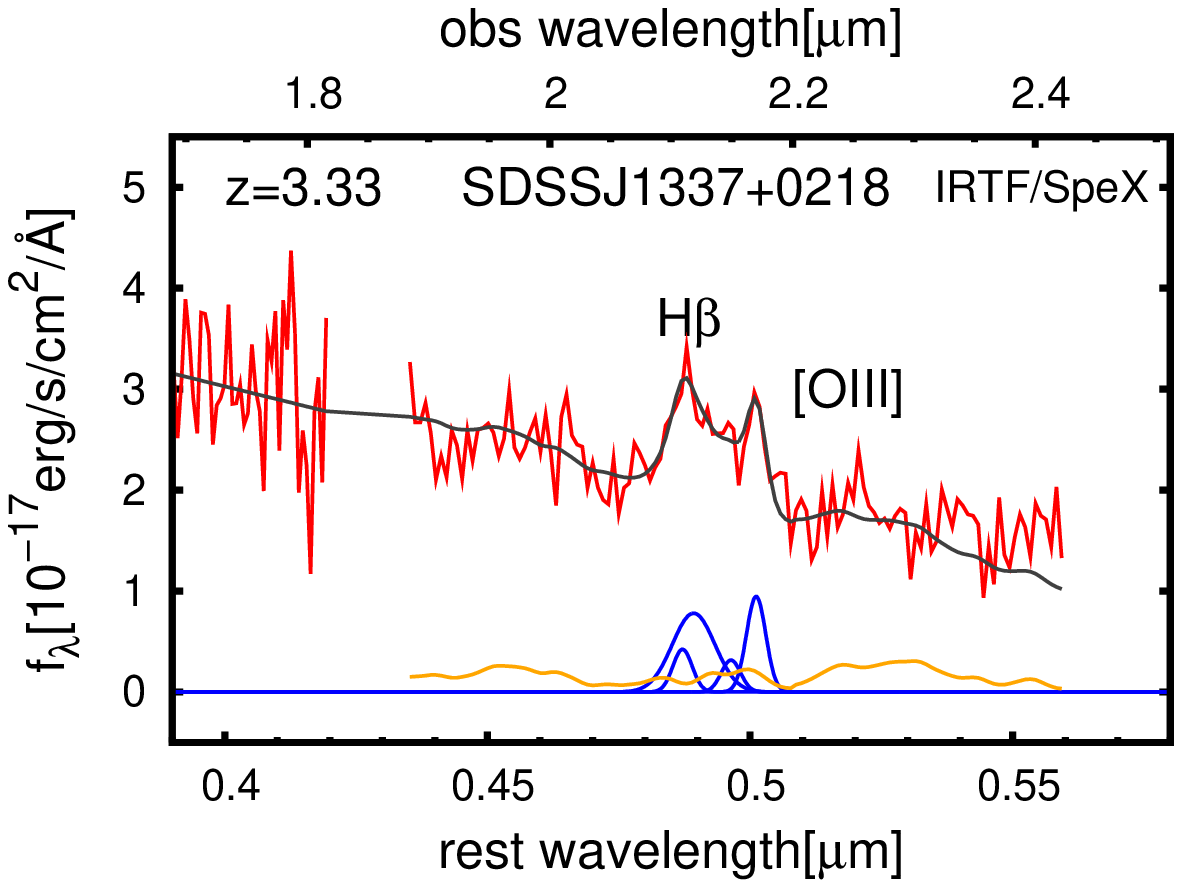} 
  \end{minipage}
   \hspace{2.0pc} 
   \begin{minipage}{.45\linewidth} 
   \FigureFile(75mm, 75mm){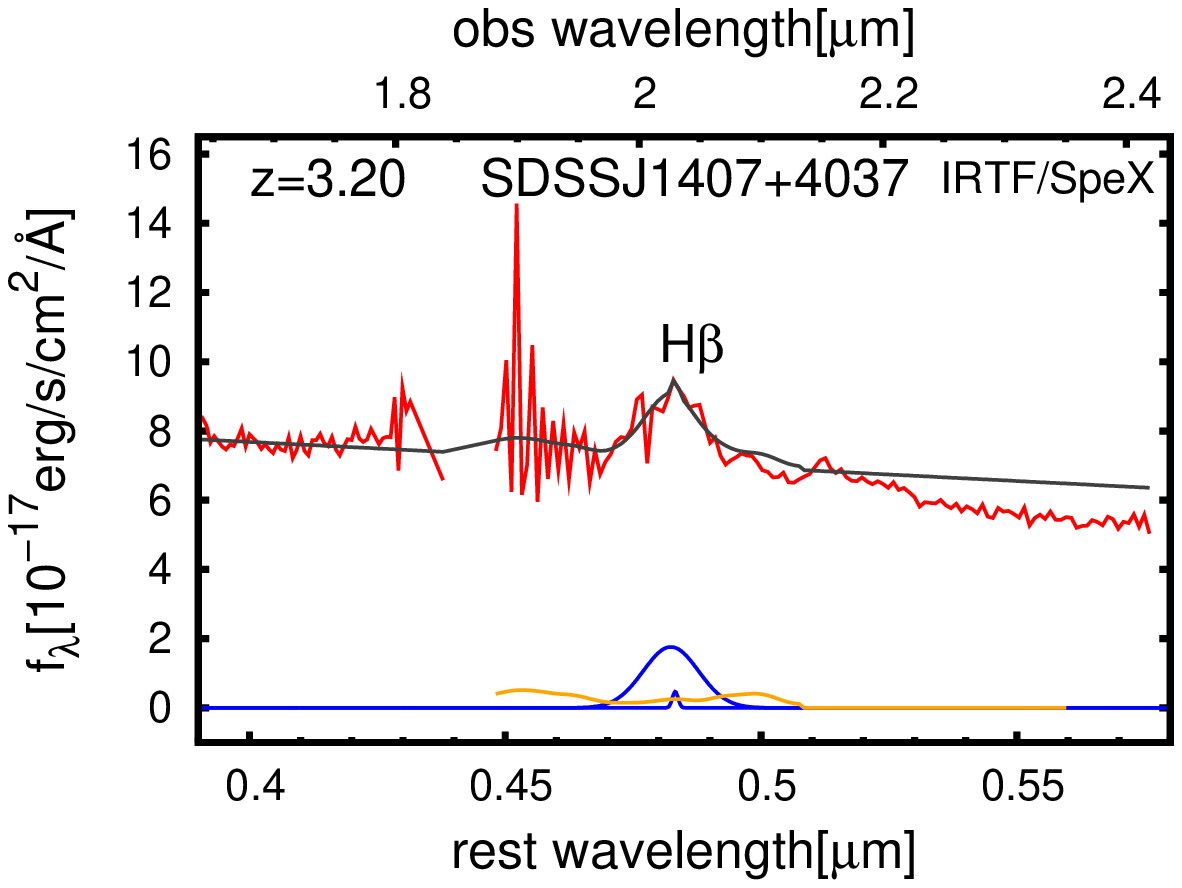} 
  \end{minipage} 
  \end{center}

\hspace{20mm}

  \begin{center}
  \begin{minipage}{.45\linewidth} 
  \FigureFile(75mm, 75mm){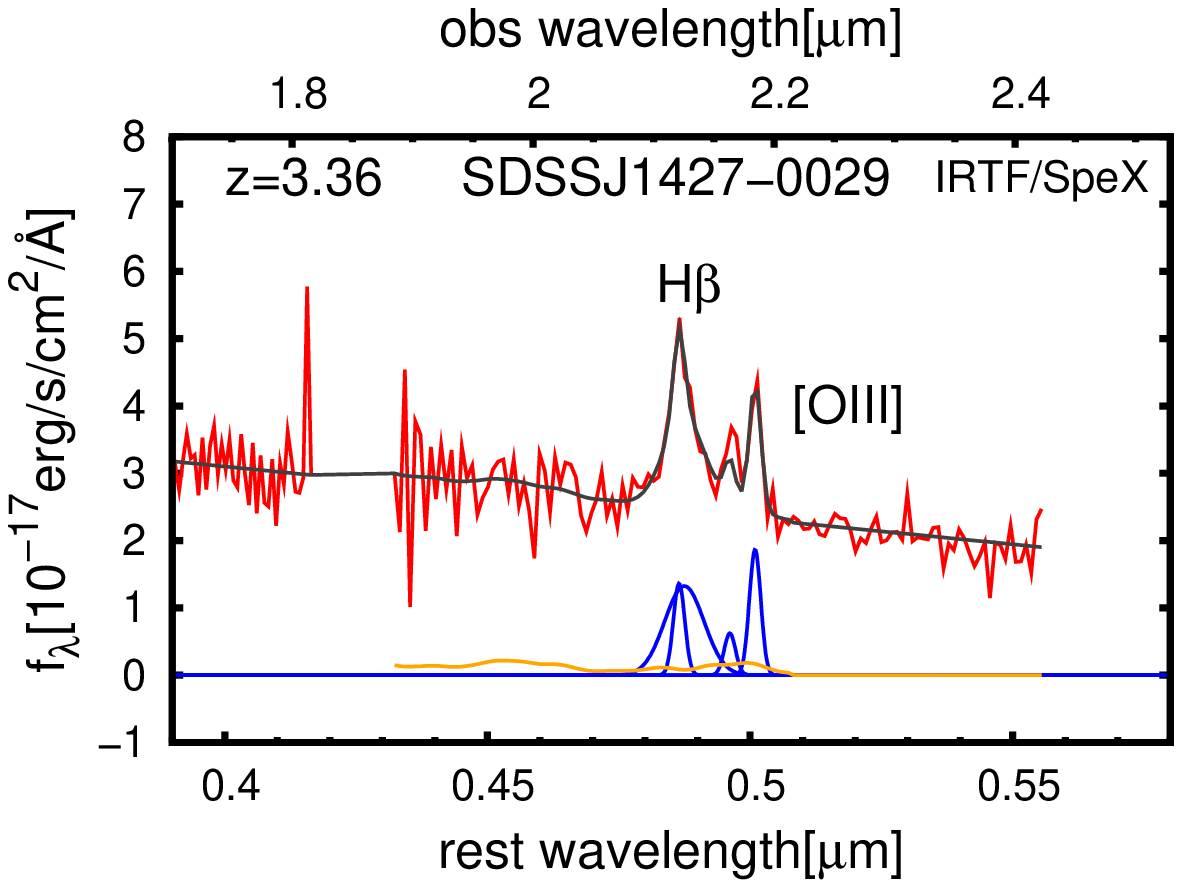} 
  \end{minipage}
   \hspace{2.0pc} 
  \begin{minipage}{.45\linewidth} 
   \FigureFile(75mm, 75mm){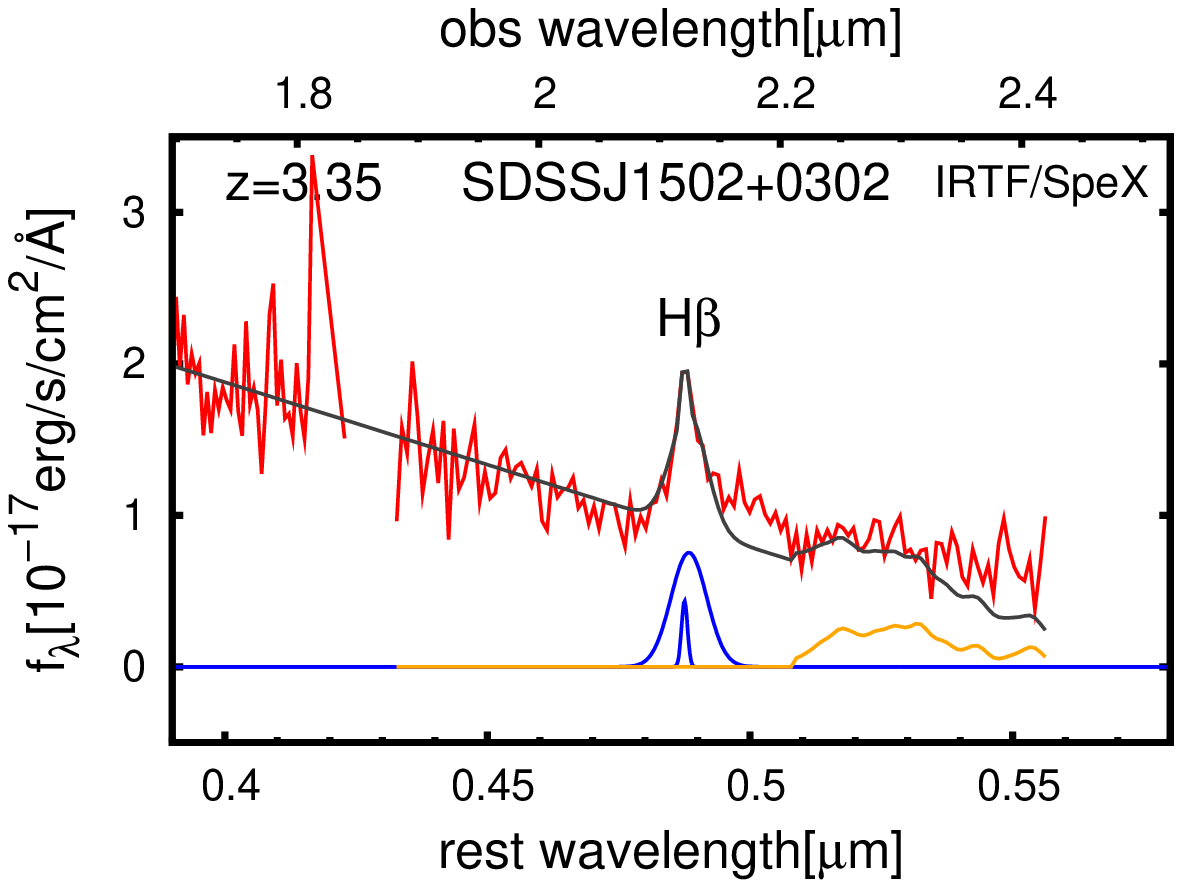} 
  \end{minipage} 
  \end{center}
  \caption{continued}
  \end{figure}

\newpage
\setcounter{figure}{0}  
\begin{figure}[htbp] 
  \begin{center}
   \begin{minipage}{.45\linewidth} 
   \FigureFile(75mm, 75mm){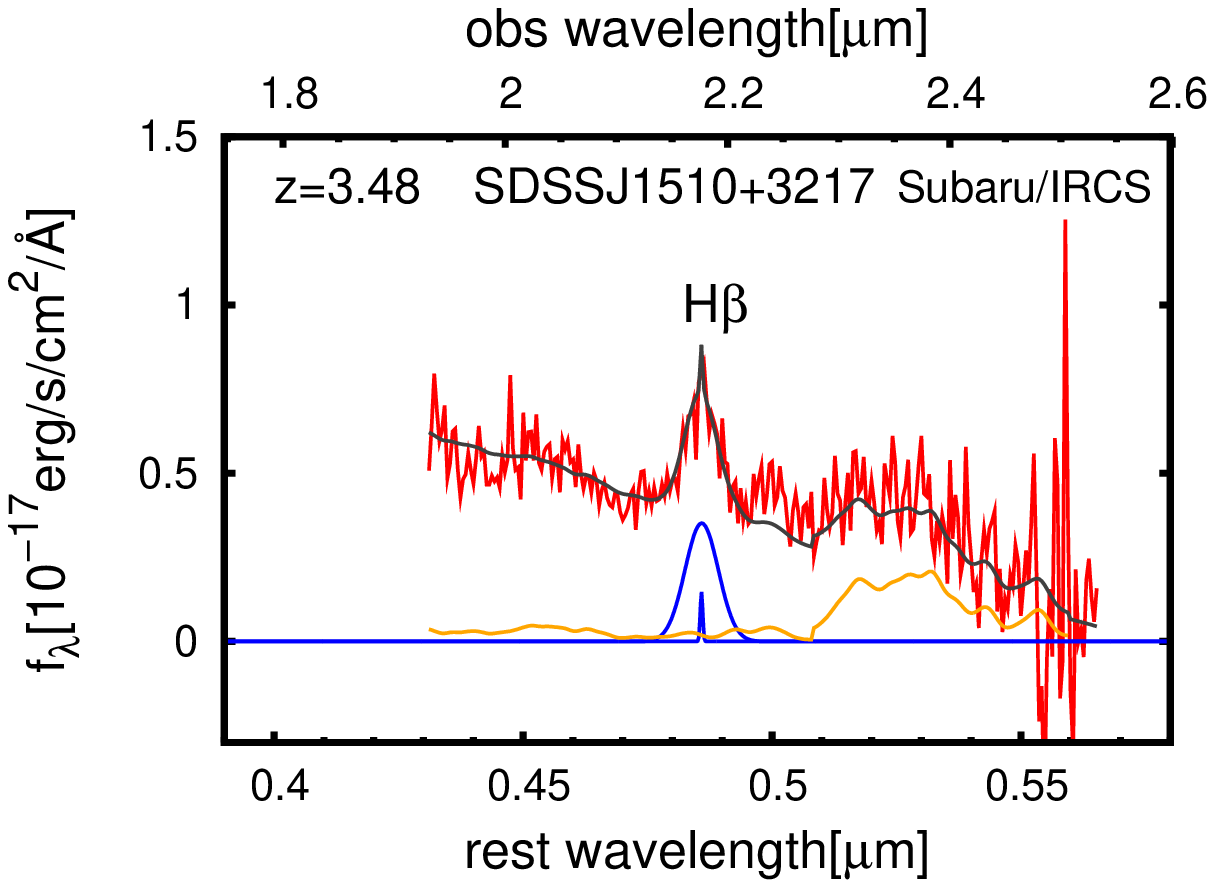} 
  \end{minipage} 
  \hspace{2.0pc} 
   \begin{minipage}{.45\linewidth} 
   \FigureFile(75mm, 75mm){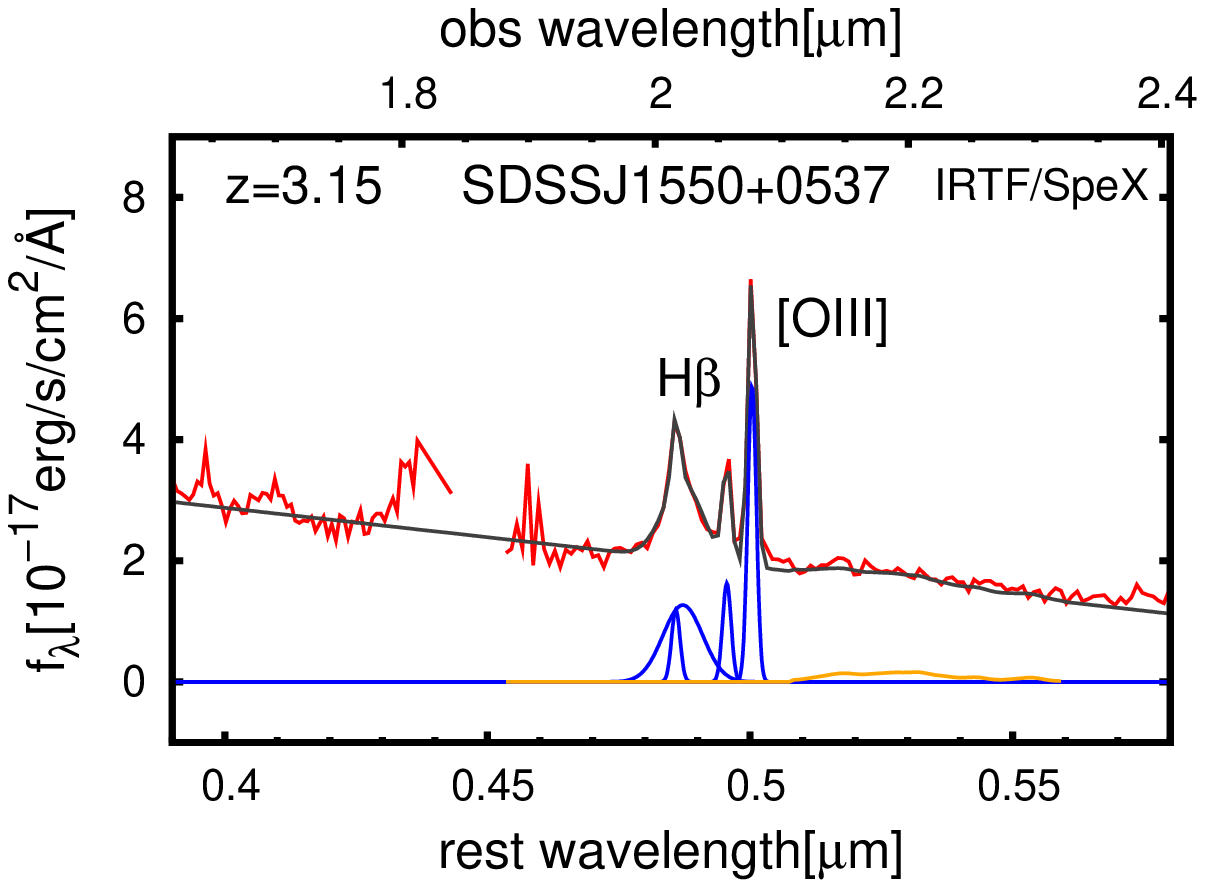} 
  \end{minipage} 
  \end{center}
  
  \hspace{20mm}
  
   \begin{center}
   \begin{minipage}{.45\linewidth} 
   \FigureFile(75mm, 75mm){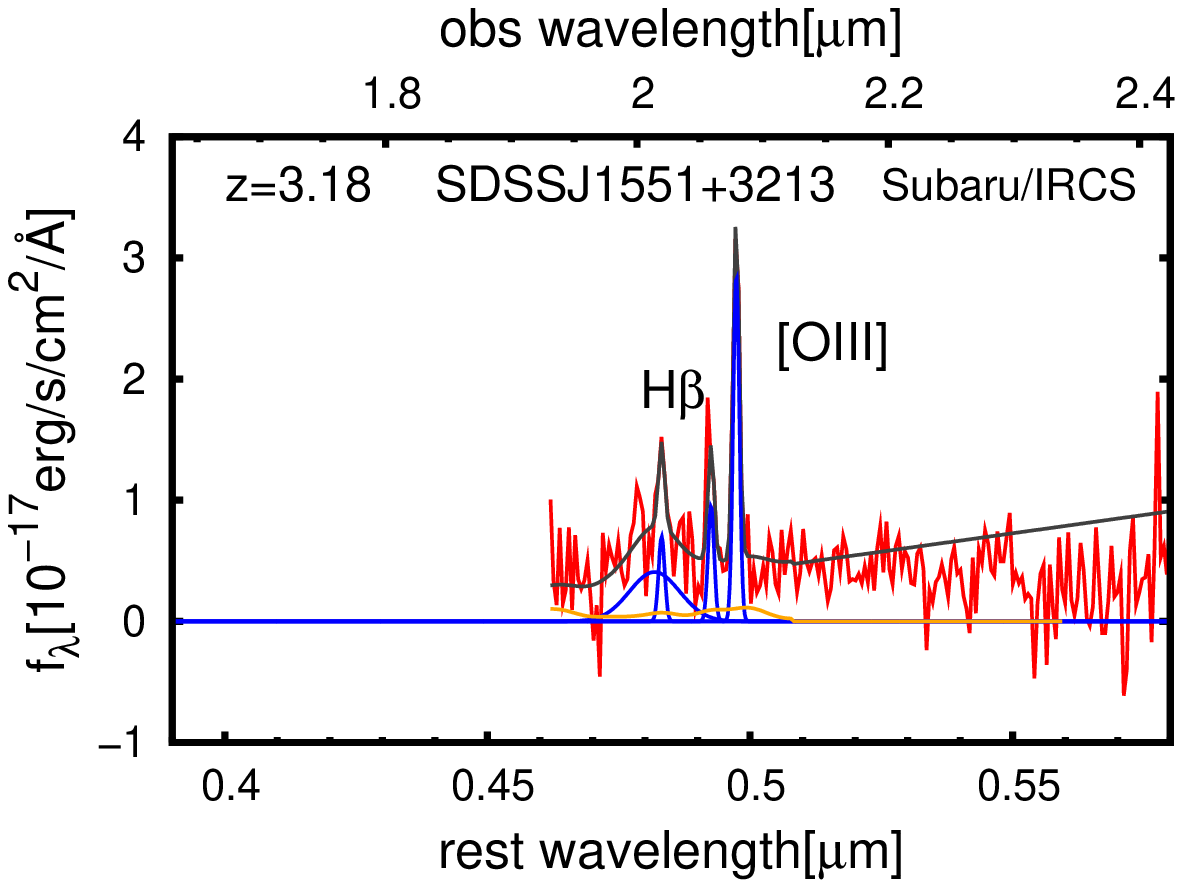} 
  \end{minipage} 
  \hspace{2.0pc} 
   \begin{minipage}{.45\linewidth} 
   \FigureFile(75mm, 75mm){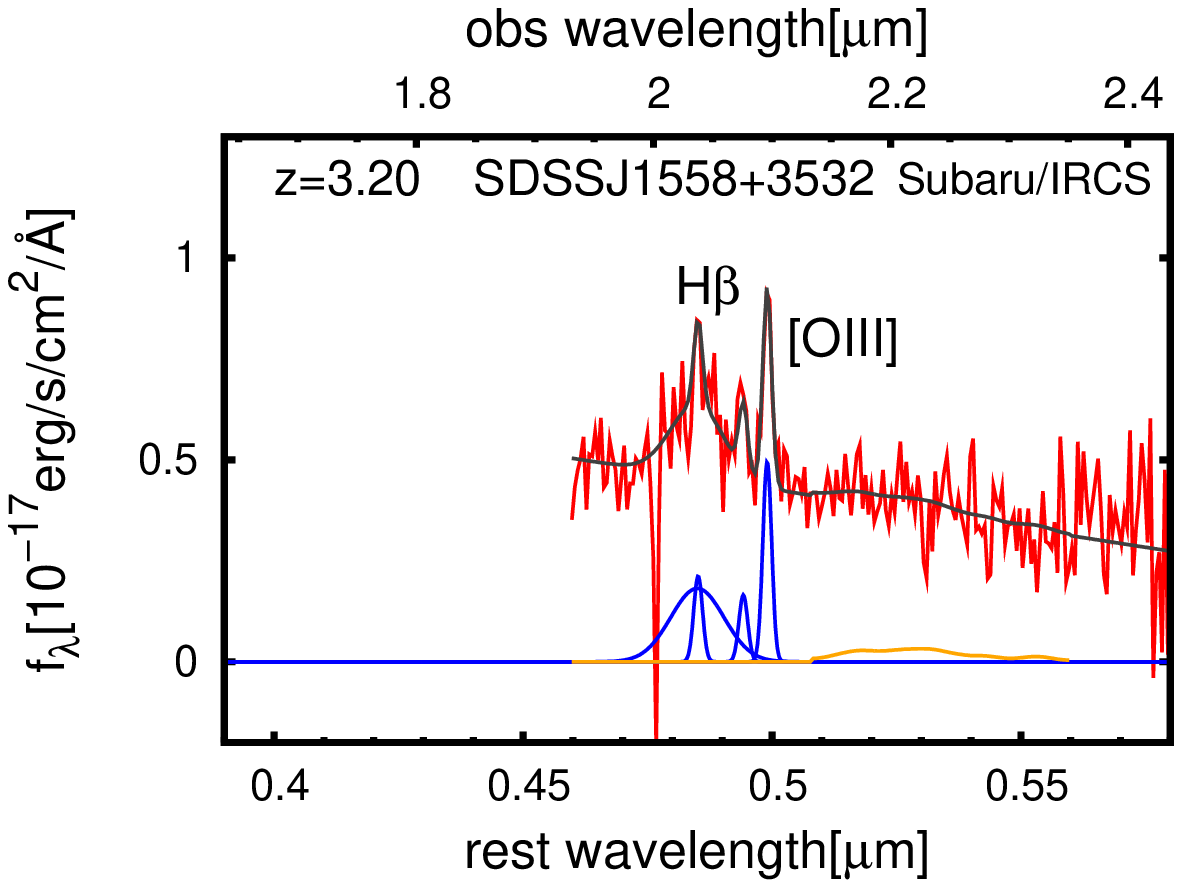} 
  \end{minipage} 
  \end{center}
  
  \hspace{20mm}
  
  \begin{center}
  \begin{minipage}{.45\linewidth} 
  \FigureFile(75mm, 75mm){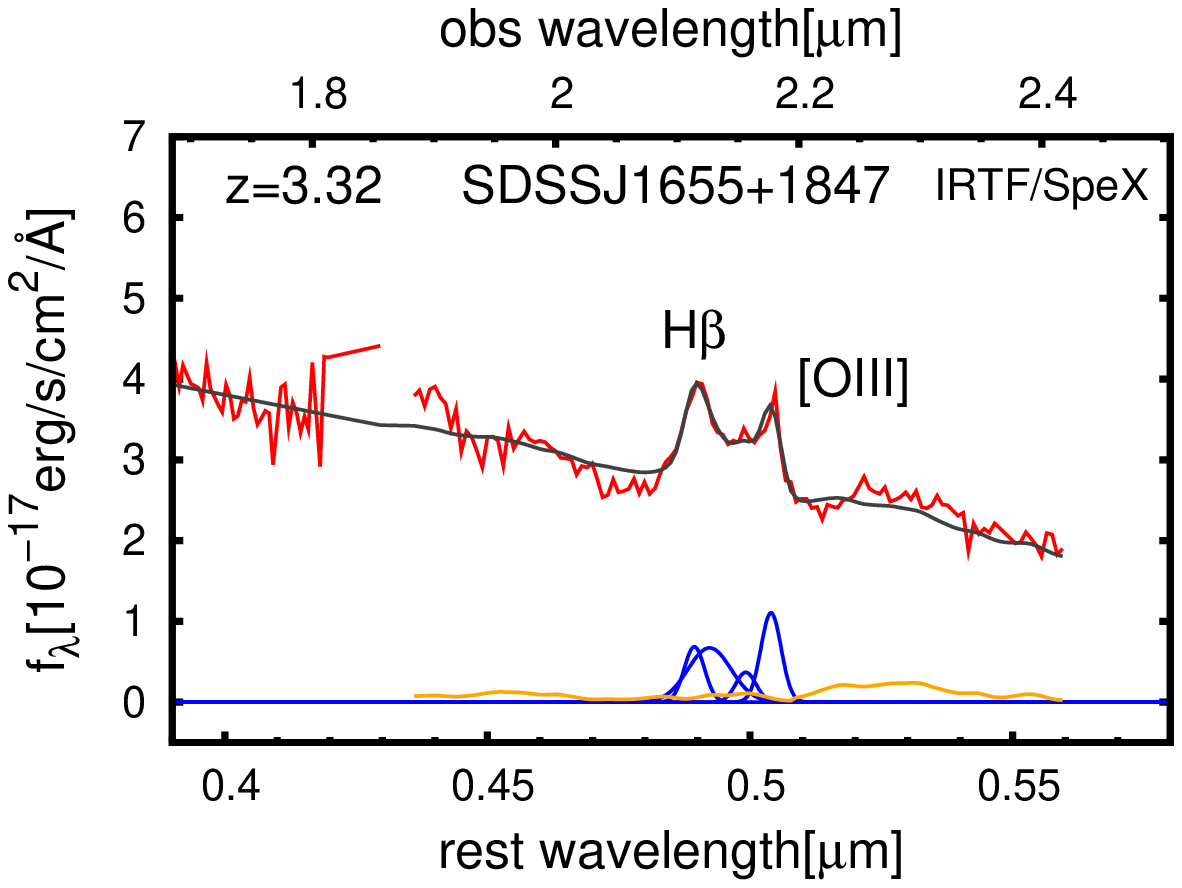} 
  \end{minipage}
  \hspace{2.0pc} 
  \begin{minipage}{.45\linewidth} 
   \FigureFile(75mm, 75mm){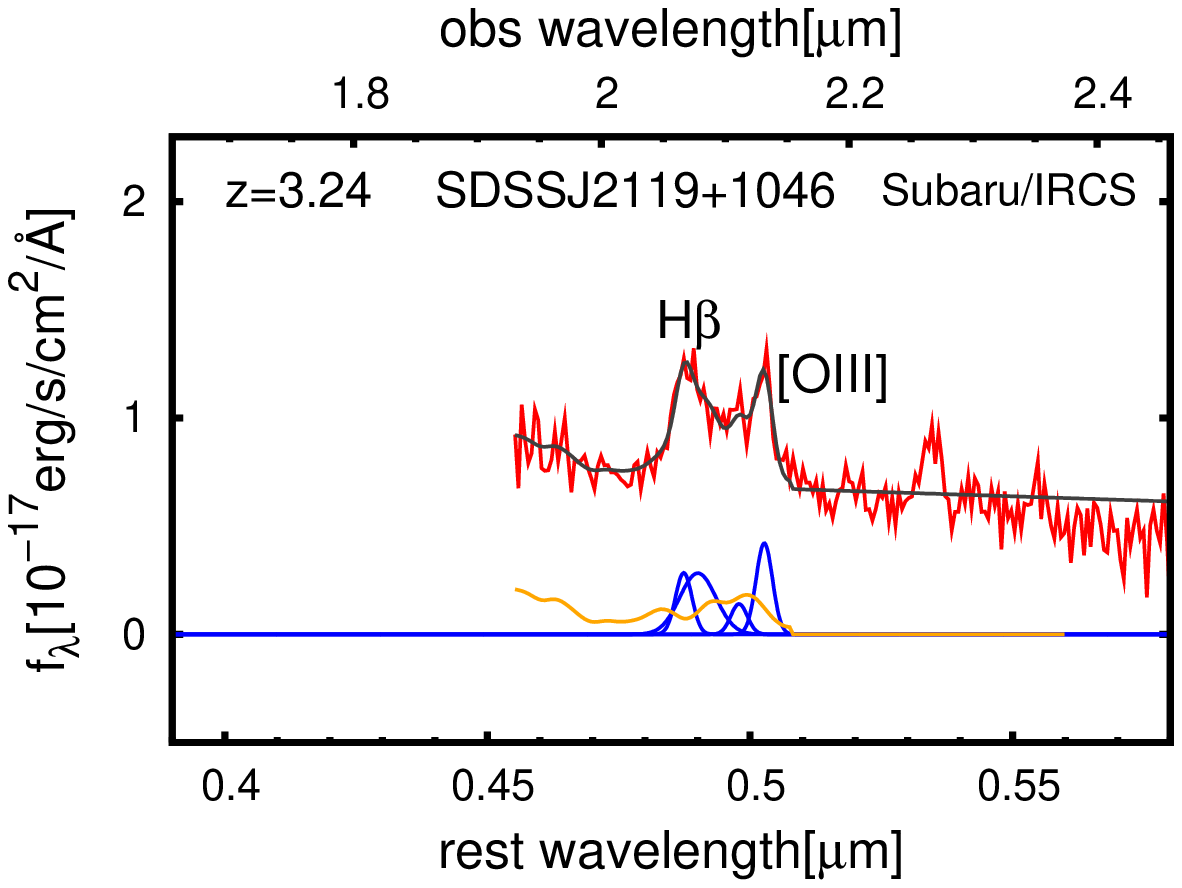} 
  \end{minipage} 
  \end{center}
    \caption{continued}
  \end{figure}

\newpage
\setcounter{figure}{0}  
\begin{figure}[htbp] 
  \begin{center}
  \begin{minipage}{.45\linewidth} 
   \FigureFile(75mm, 75mm){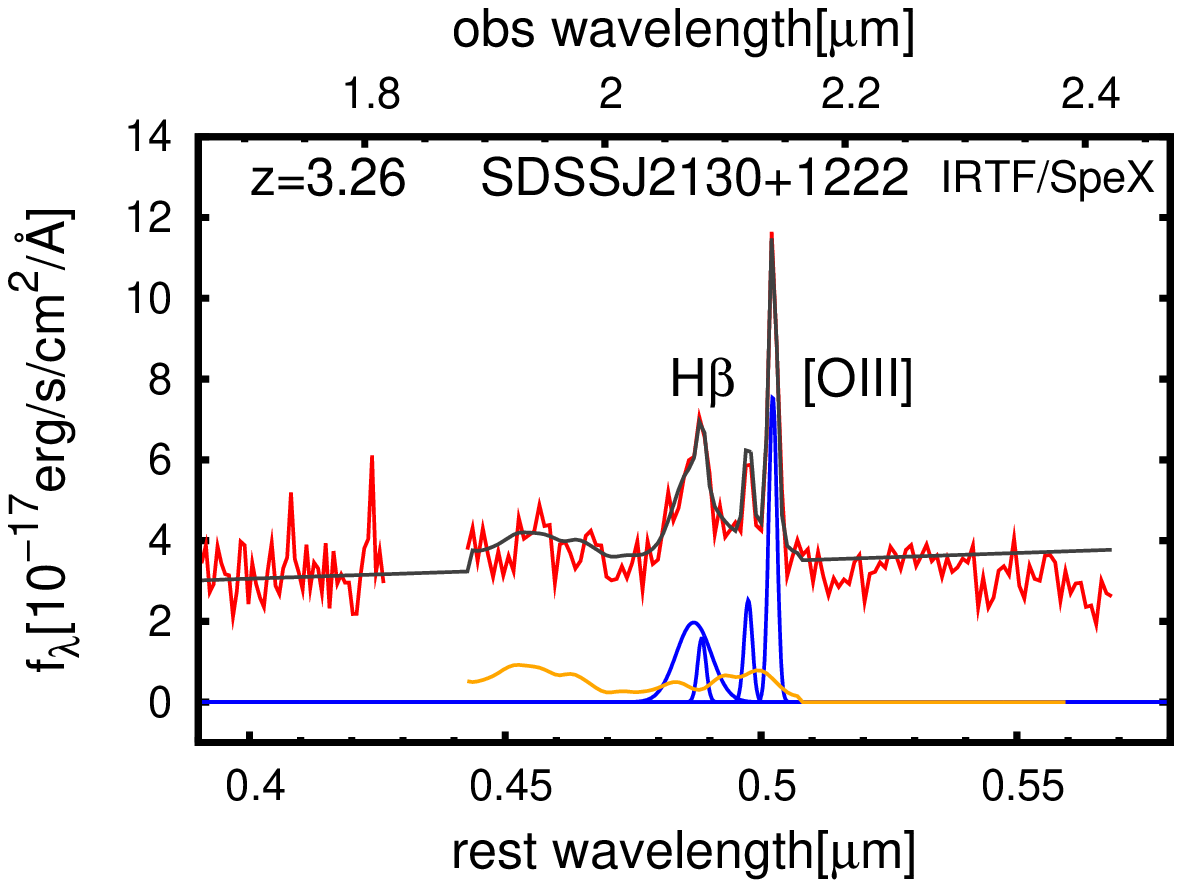} 
  \end{minipage}
  \hspace{2.0pc} 
  \begin{minipage}{.45\linewidth} 
   \FigureFile(75mm, 75mm){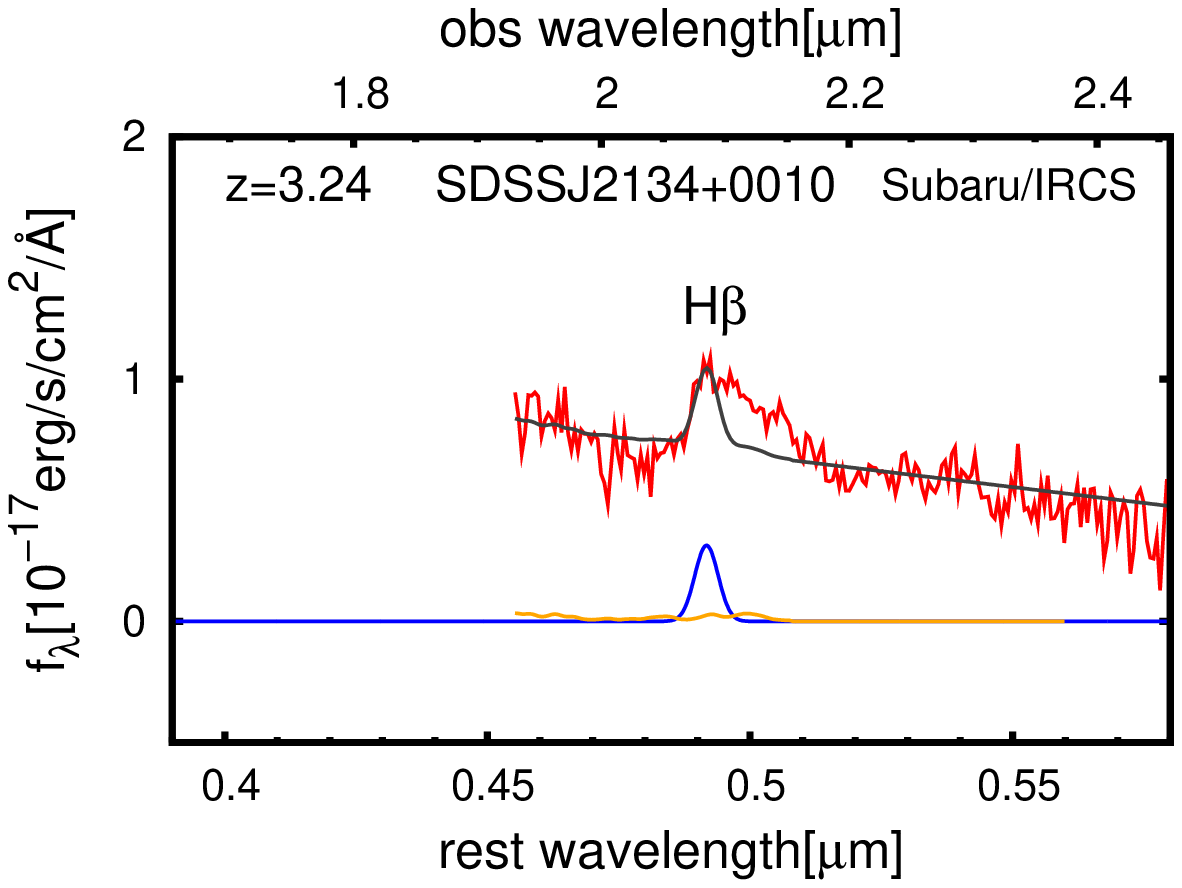}
  \end{minipage}
  \end{center}

\hspace{20mm}
   
  \begin{center}
   \begin{minipage}{.45\linewidth} 
   \FigureFile(75mm, 75mm){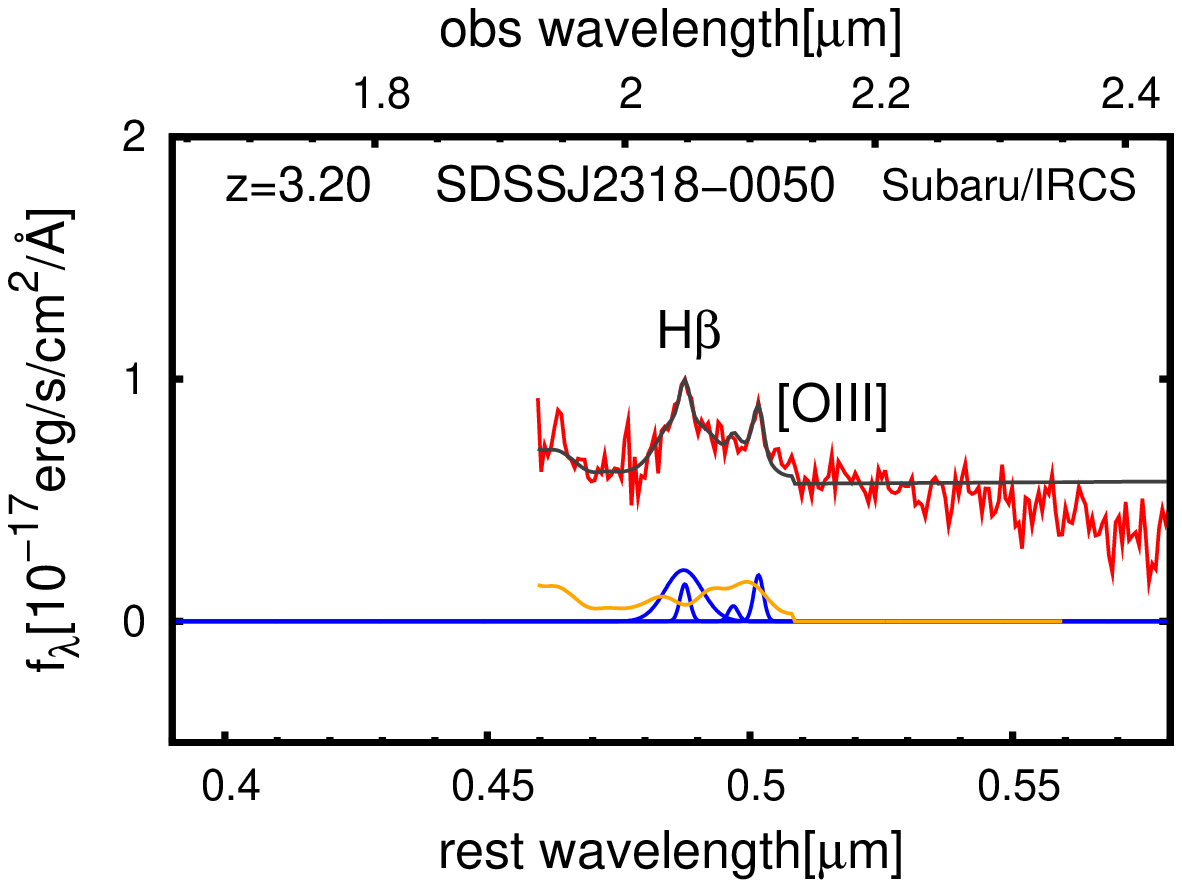} 
  \end{minipage} 
  \hspace{2.0pc} 
  \begin{minipage}{.45\linewidth} 
   \FigureFile(75mm, 75mm){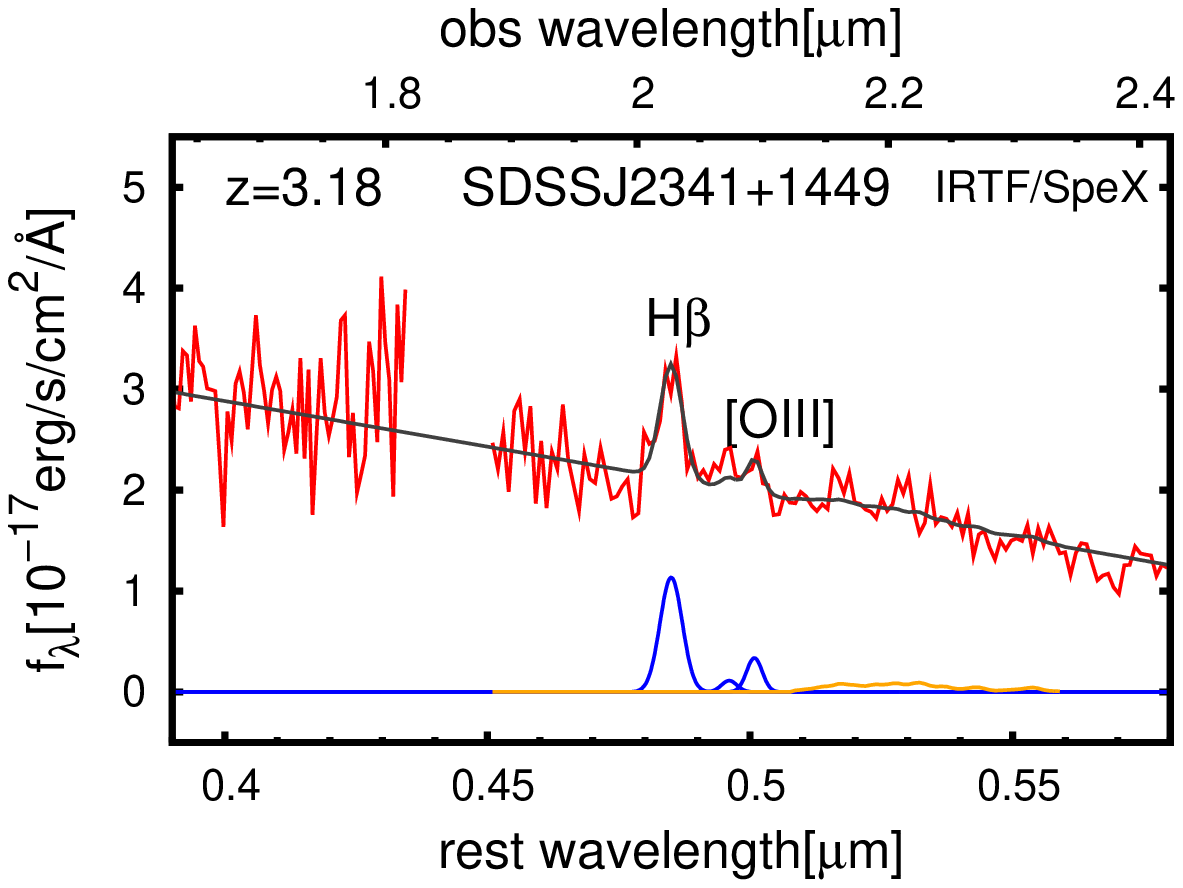} 
  \end{minipage}
  \end{center}
  \caption{continued}
  \end{figure}

  \begin{figure}[htbp]
  \begin{center} 
  \begin{minipage}{.45\linewidth} 
   \FigureFile(75mm, 75mm){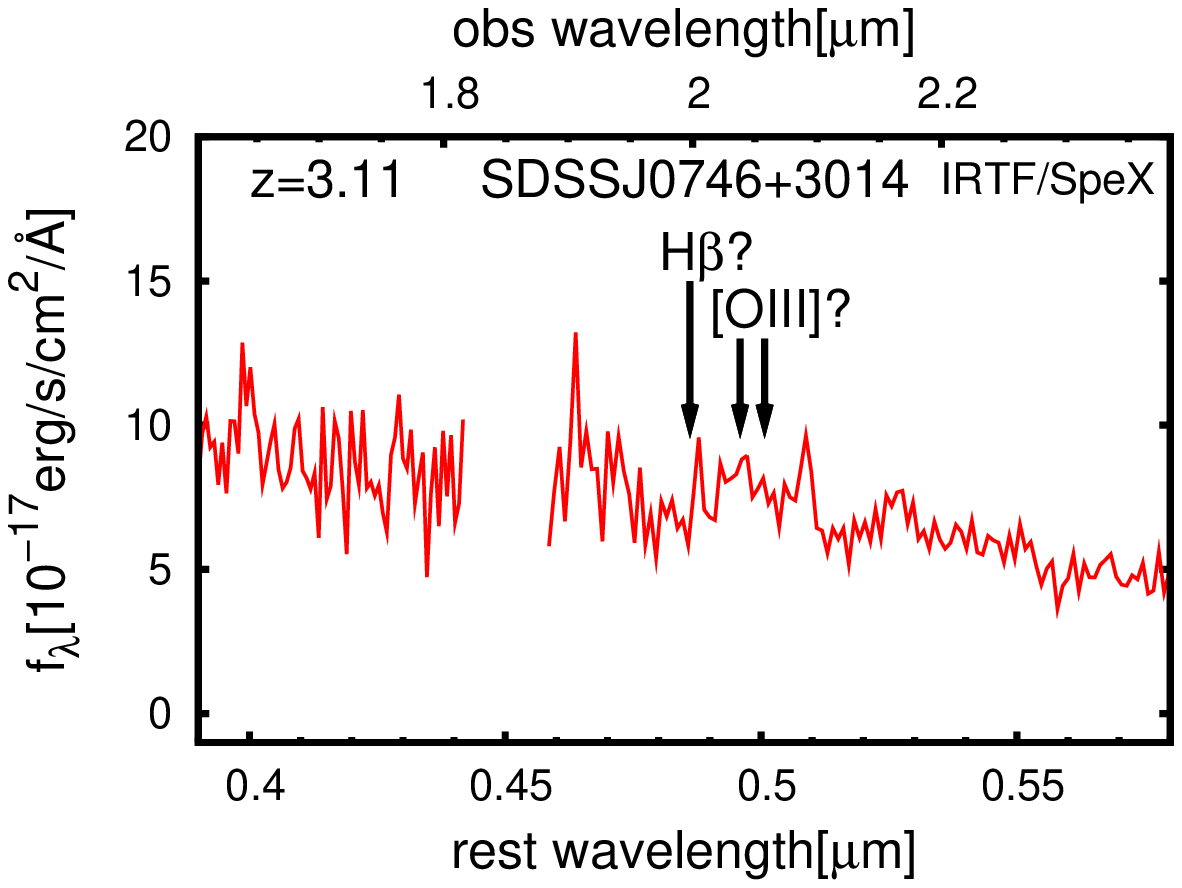} 
\end{minipage}
 \hspace{2.0pc} 
  \begin{minipage}{.45\linewidth} 
   \FigureFile(75mm, 75mm){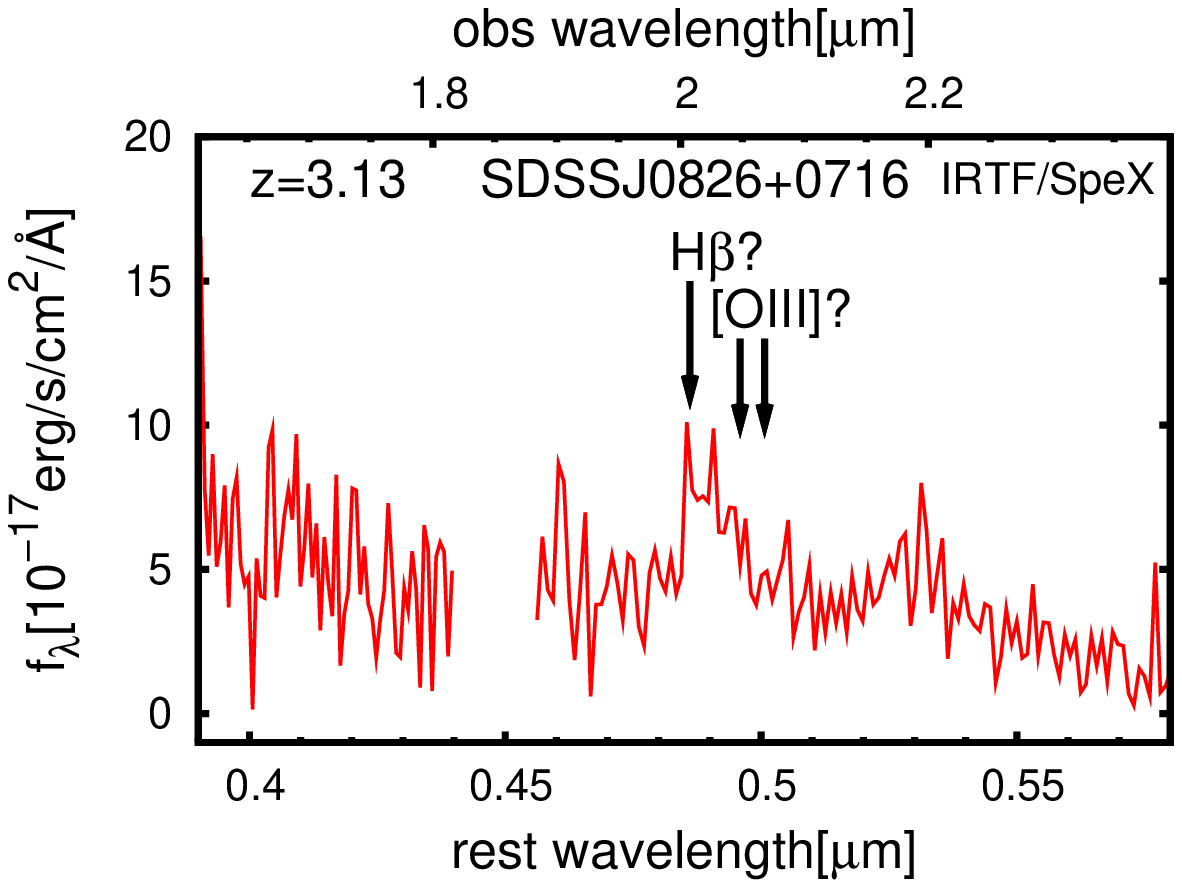} 
  \end{minipage} 
 \end{center}
  
  \hspace{20mm}
  
  \begin{center} 
  \begin{minipage}{.45\linewidth} 
   \FigureFile(75mm, 75mm){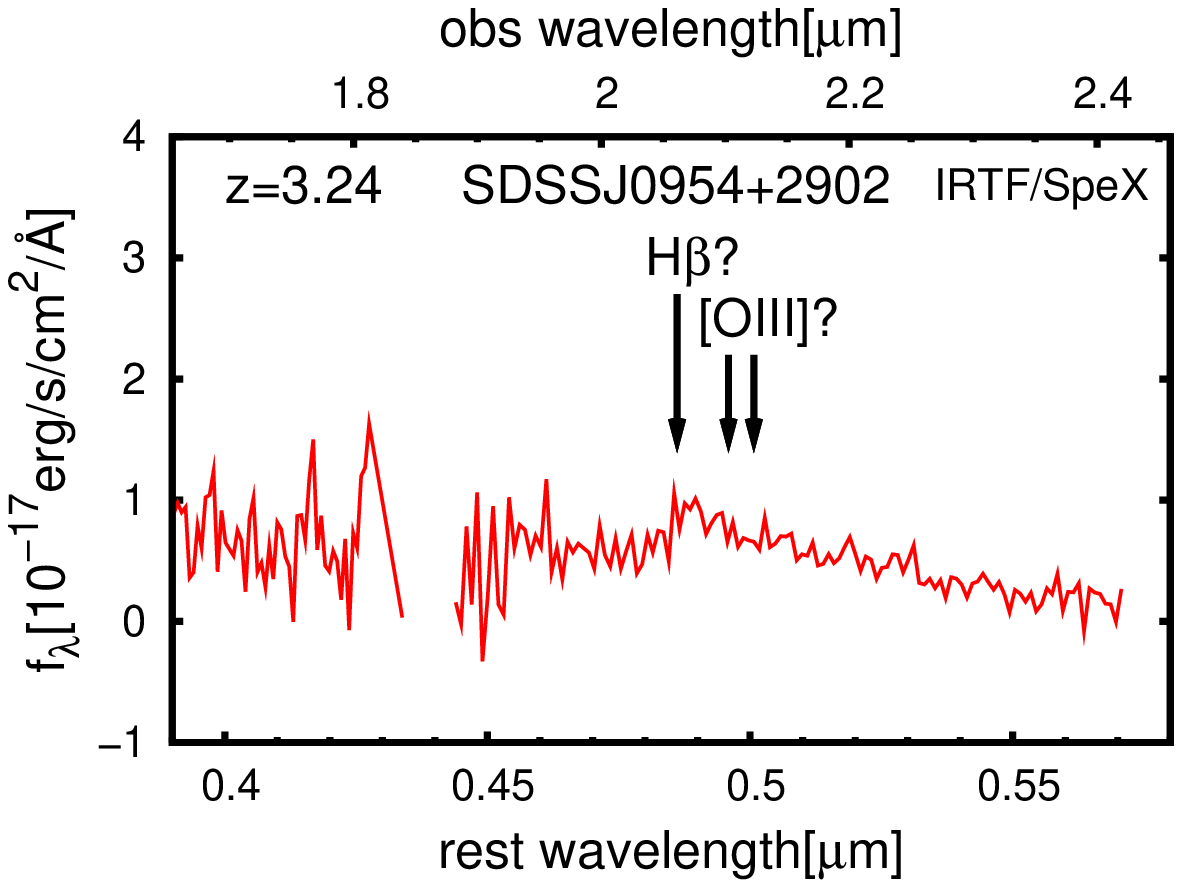} 
\end{minipage}
 \hspace{2.0pc} 
  \begin{minipage}{.45\linewidth} 
   \FigureFile(75mm, 75mm){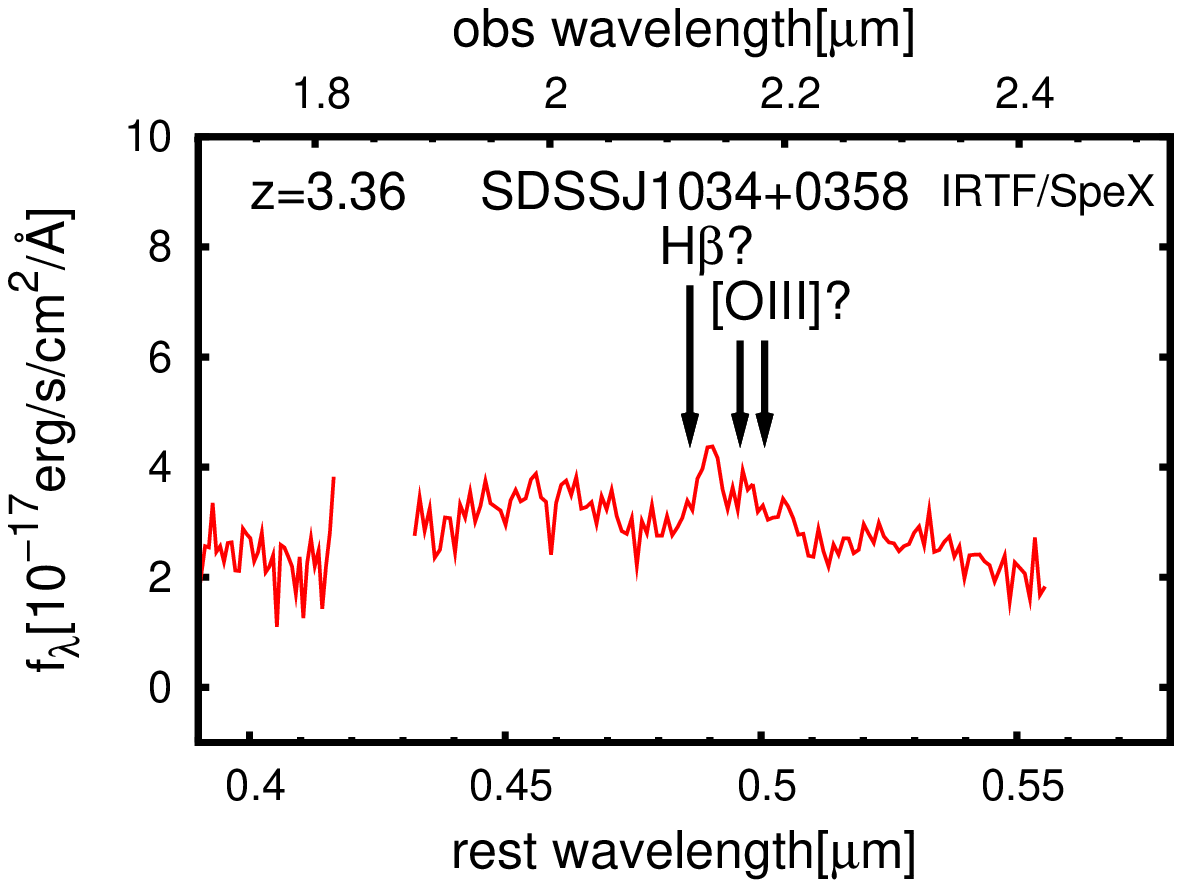} 
  \end{minipage} 
 \end{center}
  
  \hspace{20mm}
  
  \begin{center} 
  \begin{minipage}{.45\linewidth} 
   \FigureFile(75mm, 75mm){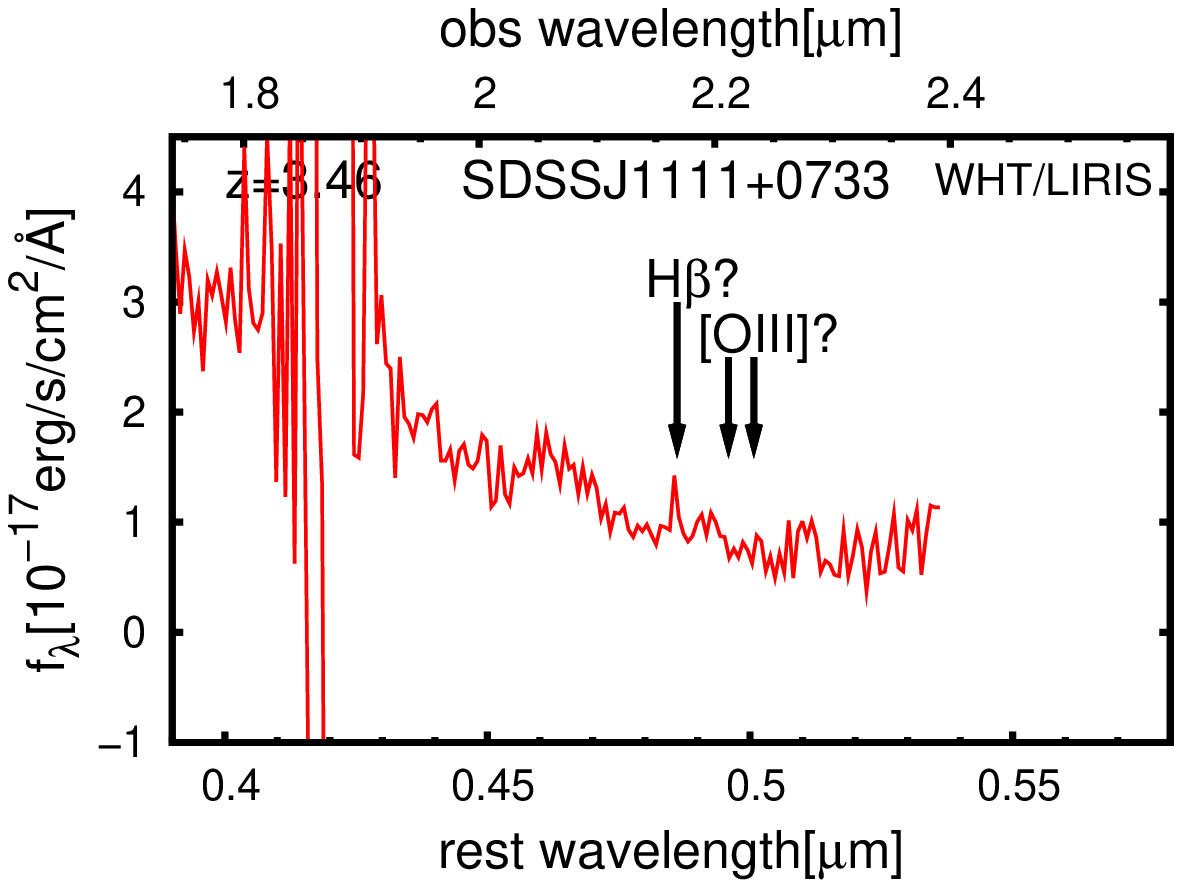} 
\end{minipage}
\hspace{2.0pc} 
 \begin{minipage}{.45\linewidth} 
   \FigureFile(75mm, 75mm){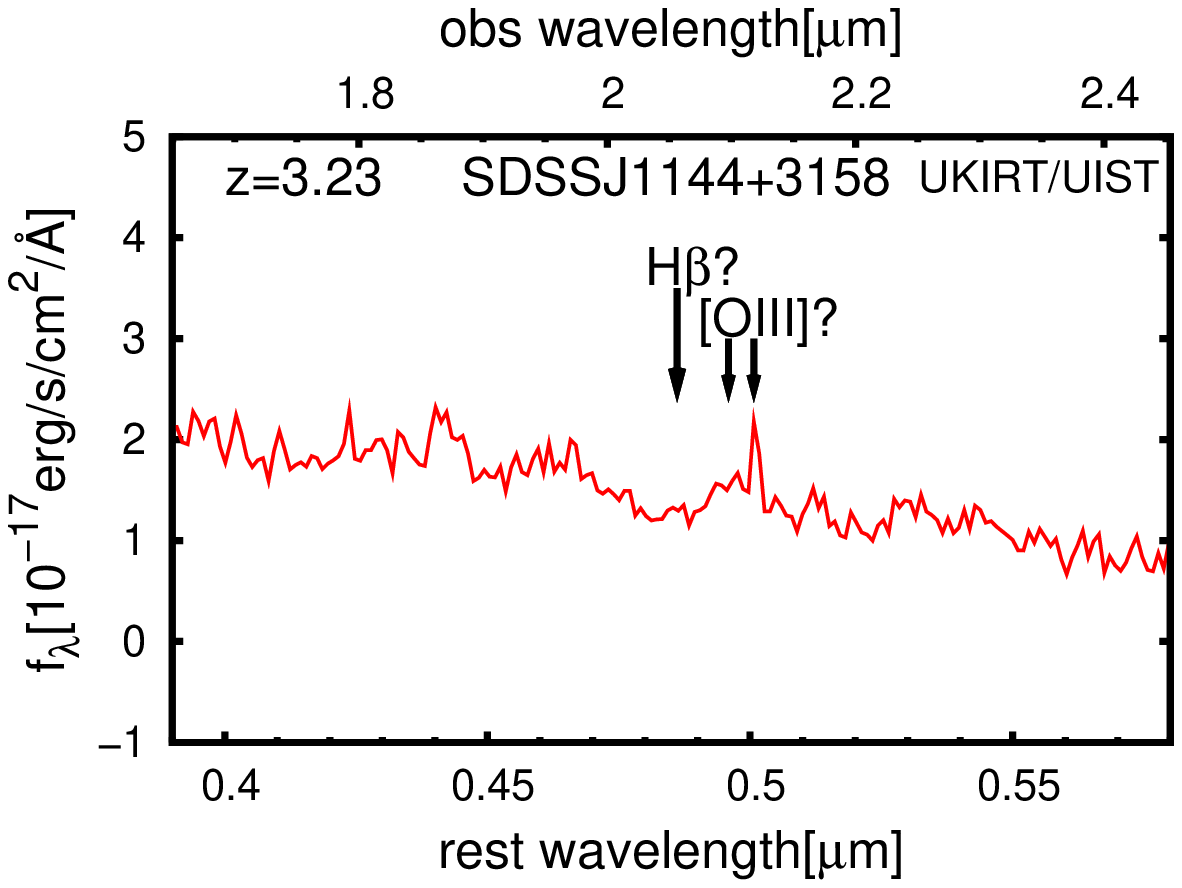} 
  \end{minipage} 
  \end{center}
 \caption{Spectra without clearly detected emission lines. The abscissa is the rest frame (bottom) and the observed (top) wavelength in [$\mu$m]. The ordinate is the flux in [$\mathrm{10^{-17} erg/s/cm^{2}/ \mathrm{\AA}}$]. H$\beta$ and [O\emissiontype{III}] emission lines should be positioned as indicated by the downward arrows in each panel.}\label{fig:2}
  \end{figure}

  \begin{figure}[htbp]
  \setcounter{figure}{1}
  \begin{center}
  \begin{minipage}{.45\linewidth} 
   \FigureFile(75mm, 75mm){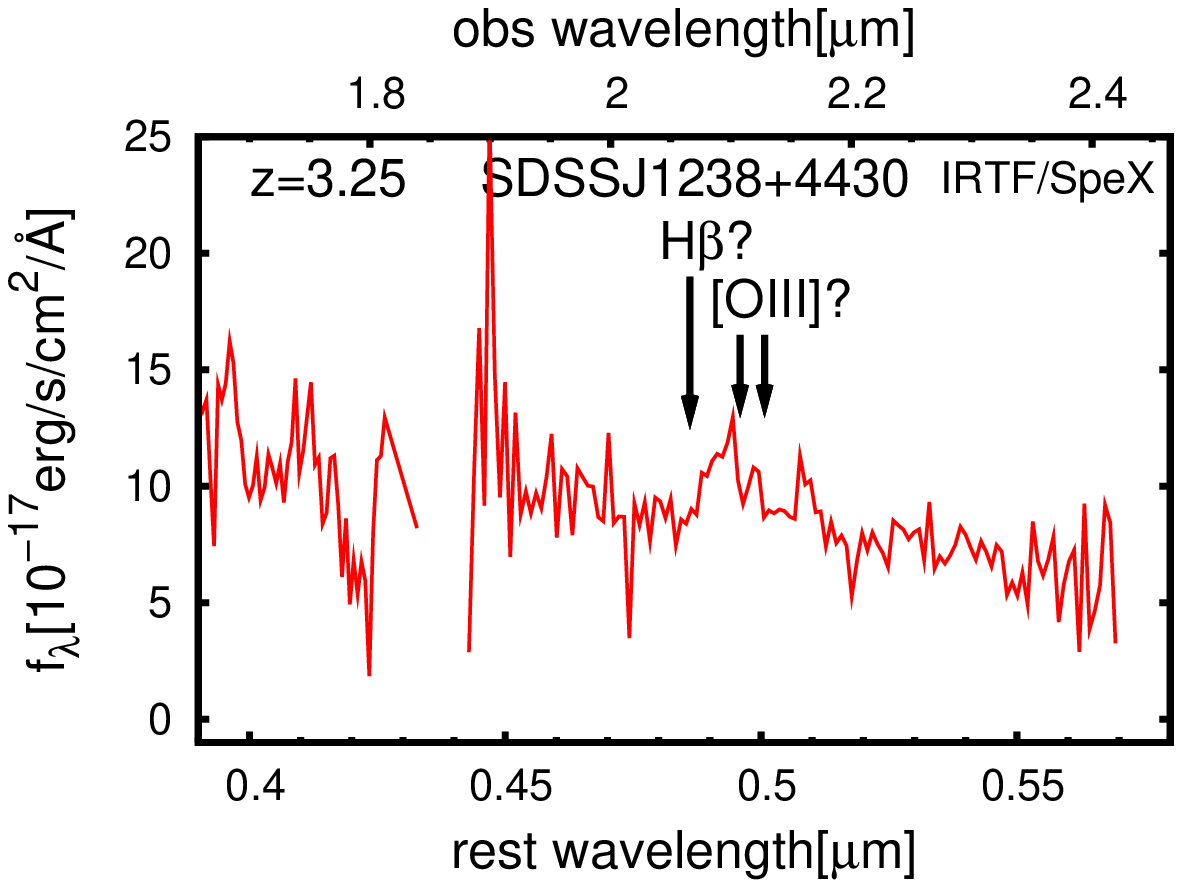} 
  \end{minipage}
   \hspace{2.0pc} 
    \begin{minipage}{.45\linewidth} 
   \FigureFile(75mm, 75mm){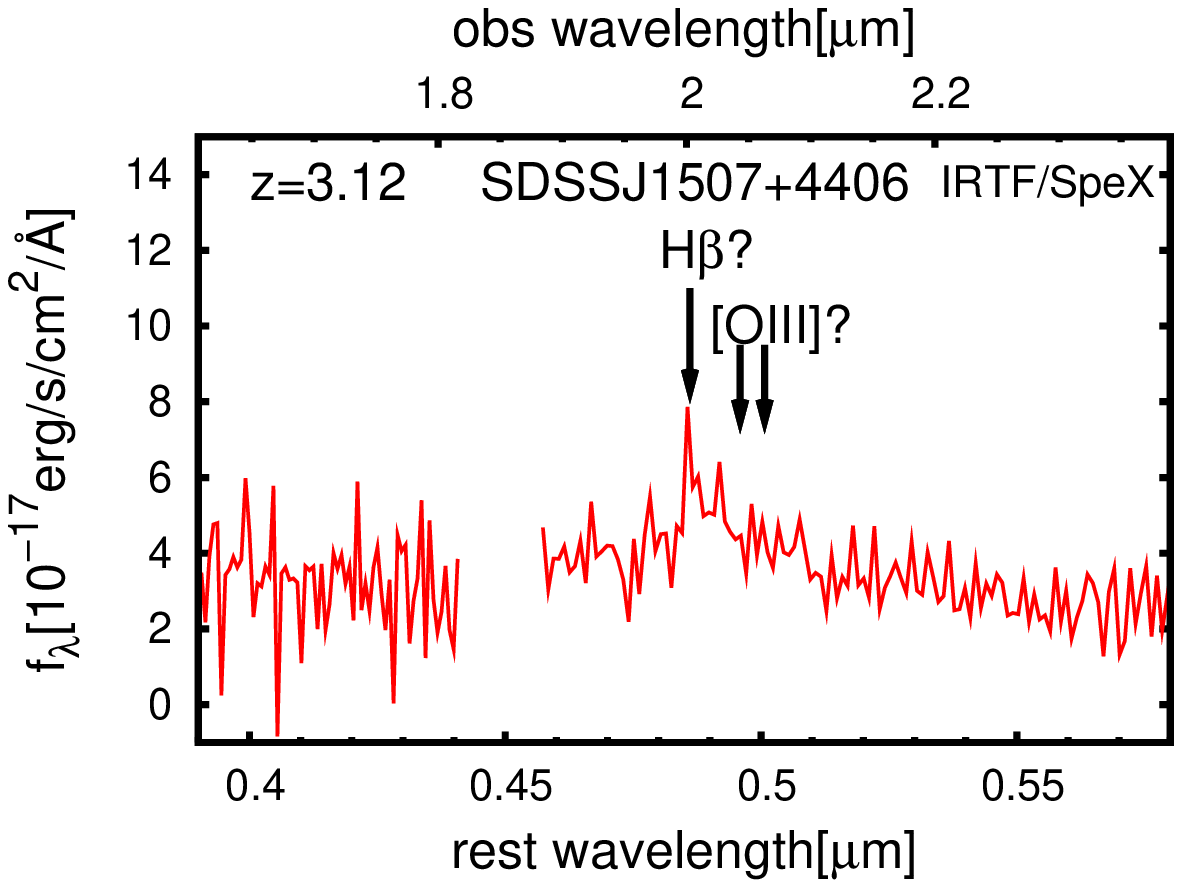} 
  \end{minipage}
  \end{center}

\hspace{20mm}  

  \begin{center}
  \begin{minipage}{.45\linewidth} 
   \FigureFile(75mm, 75mm){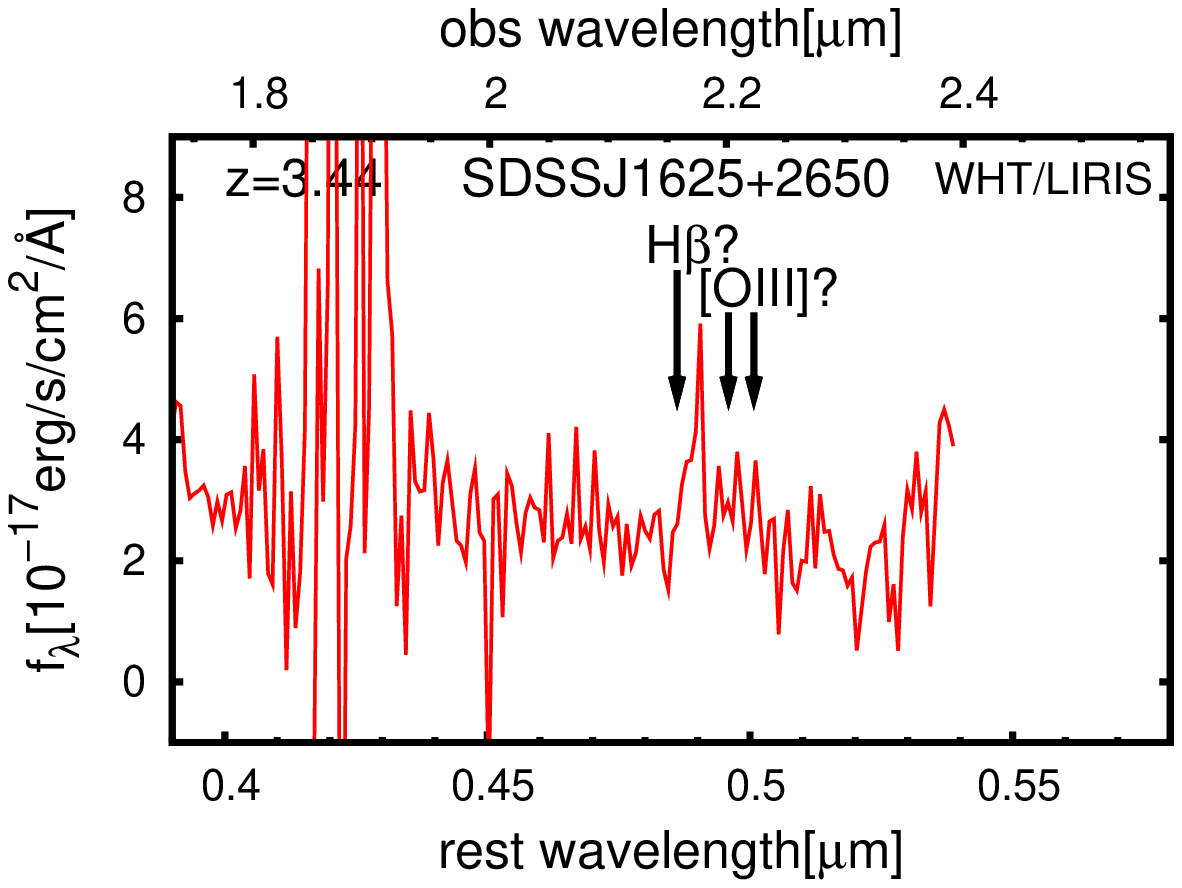} 
  \end{minipage} 
  \end{center}
    \caption{continued}
  \end{figure}

 \begin{figure}
  \begin{center}
    \FigureFile(75mm,140mm){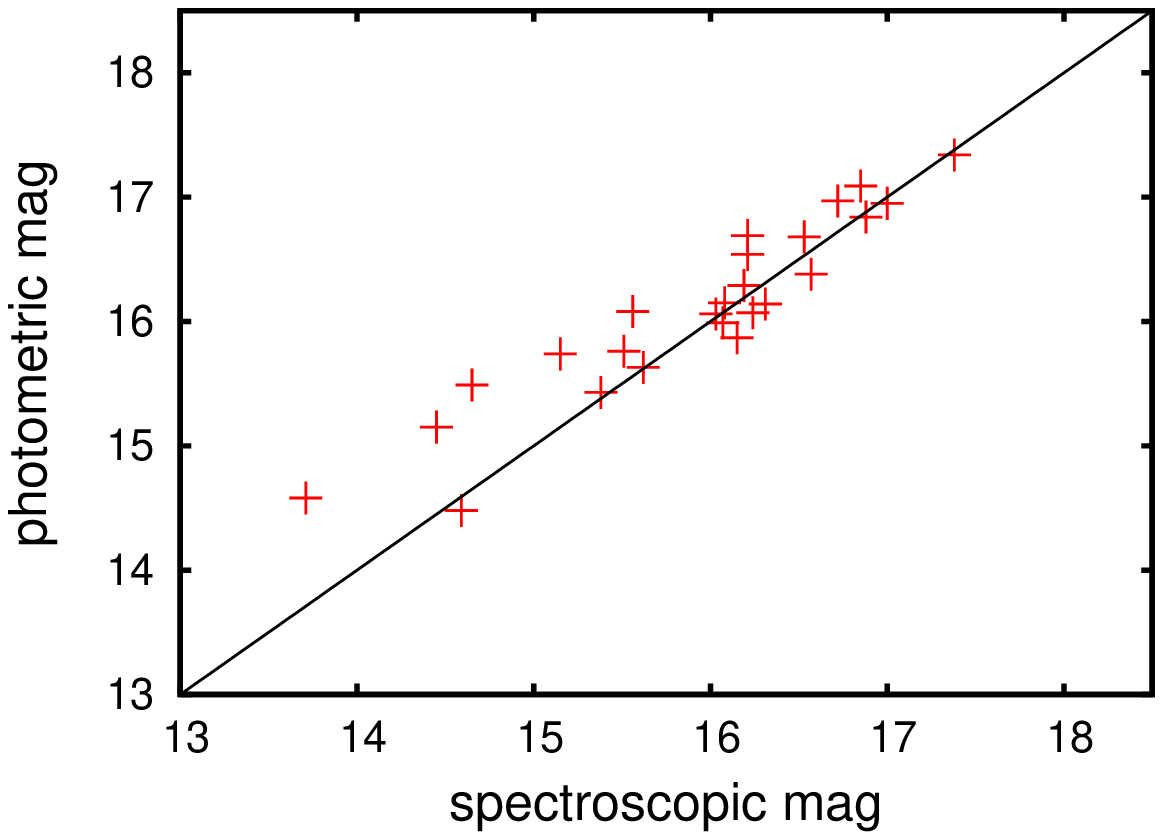}
  \end{center}
  \caption{Comparison of $K$-band magnitudes estimated using spectroscopic data (abscissa) and imaging data (ordinate). The dashed line represents a 1:1 correspondence.}\label{fig:3}
\end{figure}

 \begin{figure}
  \begin{center}
    \FigureFile(75mm,140mm){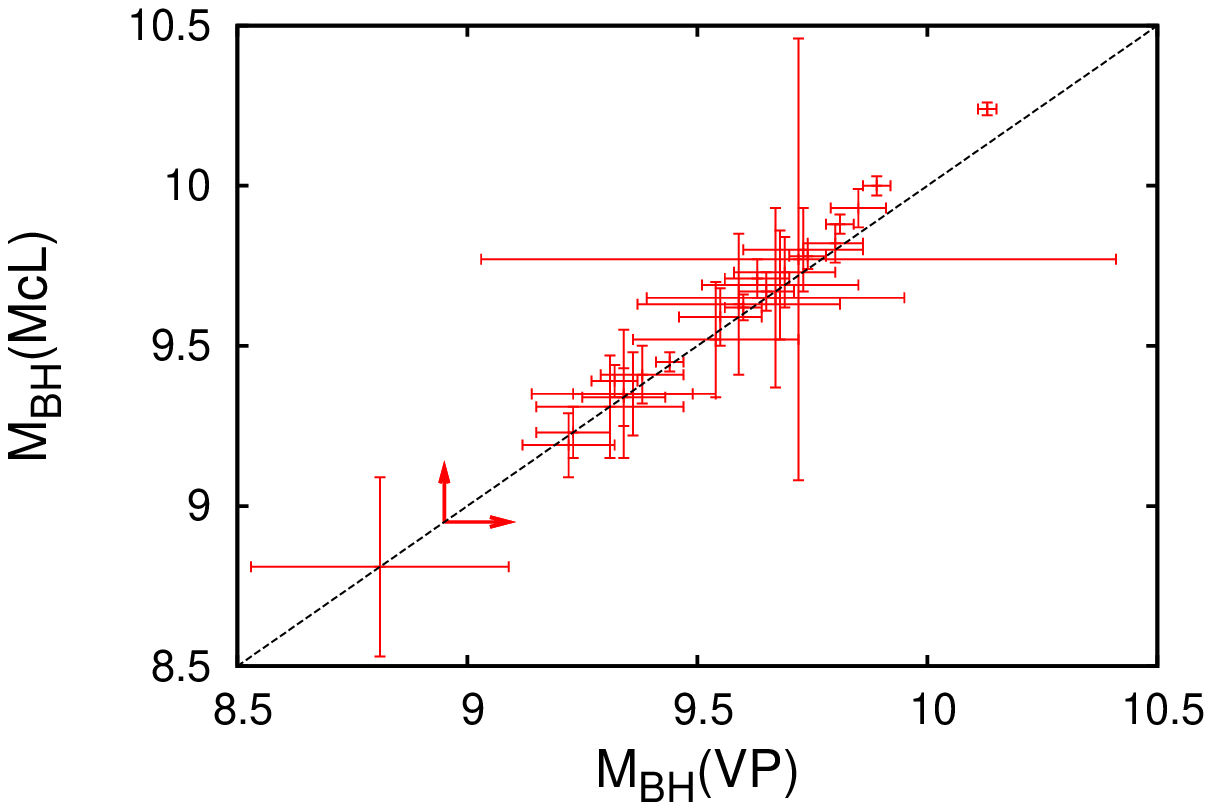}
  \end{center}
  \caption{Comparison of $M_\mathrm{BH}$ derived from a different formula (Table~\ref{table:6}). The abscissa and ordinate are the values derived based on \citet{Vestergaard2006}, and \citet{McLure2002}, respectively. Arrows mean lower limits. The dashed line represents a 1:1 correspondence.}\label{fig:4}
\end{figure}

\begin{figure}[htbp] 
  \begin{center}
  \begin{minipage}{.45\linewidth} 
  \FigureFile(75mm, 75mm){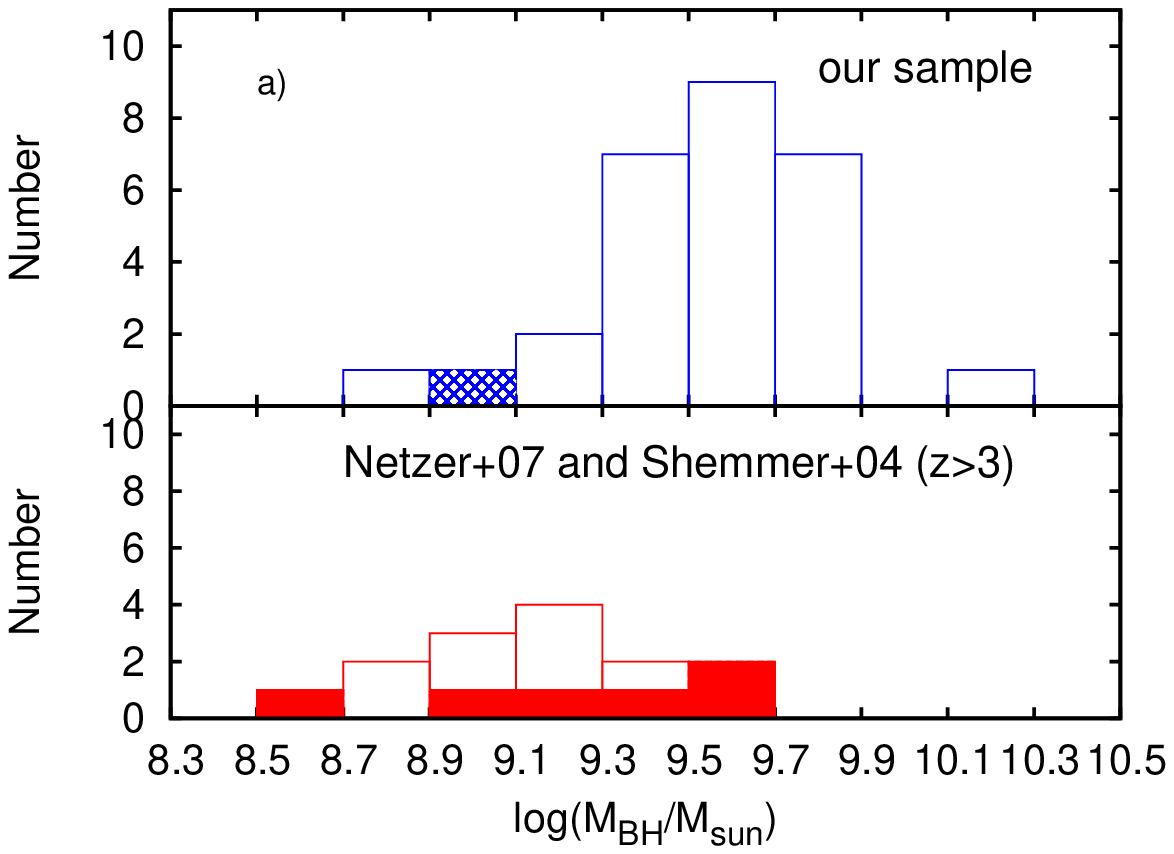} 
  \end{minipage}
  \hspace{2.0pc} 
  \begin{minipage}{.45\linewidth} 
   \FigureFile(75mm, 75mm){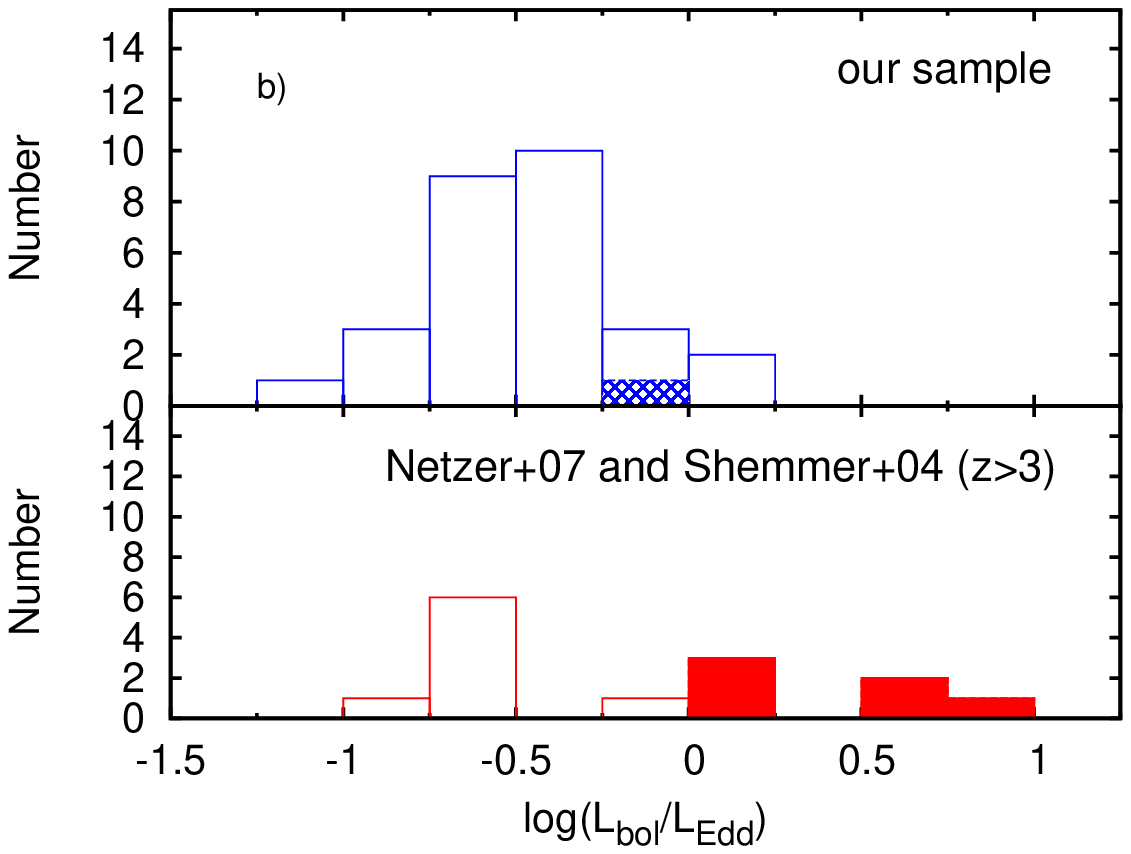} 
  \end{minipage} 
  \end{center}
   \caption{(a) Histogram of $M_\mathrm{BH}$ estimated in this study (upper panel) and in the literature (\cite{Netzer2007}; \cite{Shemmer2004}) for only $3<z<4$ sources (lower panel). A lower limit object is shown as a shaded area. (b) Comparison of the $L_\mathrm{bol}/L_\mathrm{Edd}$ distribution for our sample (upper panel) and Netzer's sample at $z>3$ (lower panel). An upper limit object is displayed by a shaded area. In the lower panels of both a) and b), open and filled histograms correspond to Netzer's sample and Shemmer's sample, respectively.}\label{fig:5}
  \end{figure}

  \begin{figure}
  \begin{center}
    \begin{minipage}{.45\linewidth} 
    \FigureFile(75mm,140mm){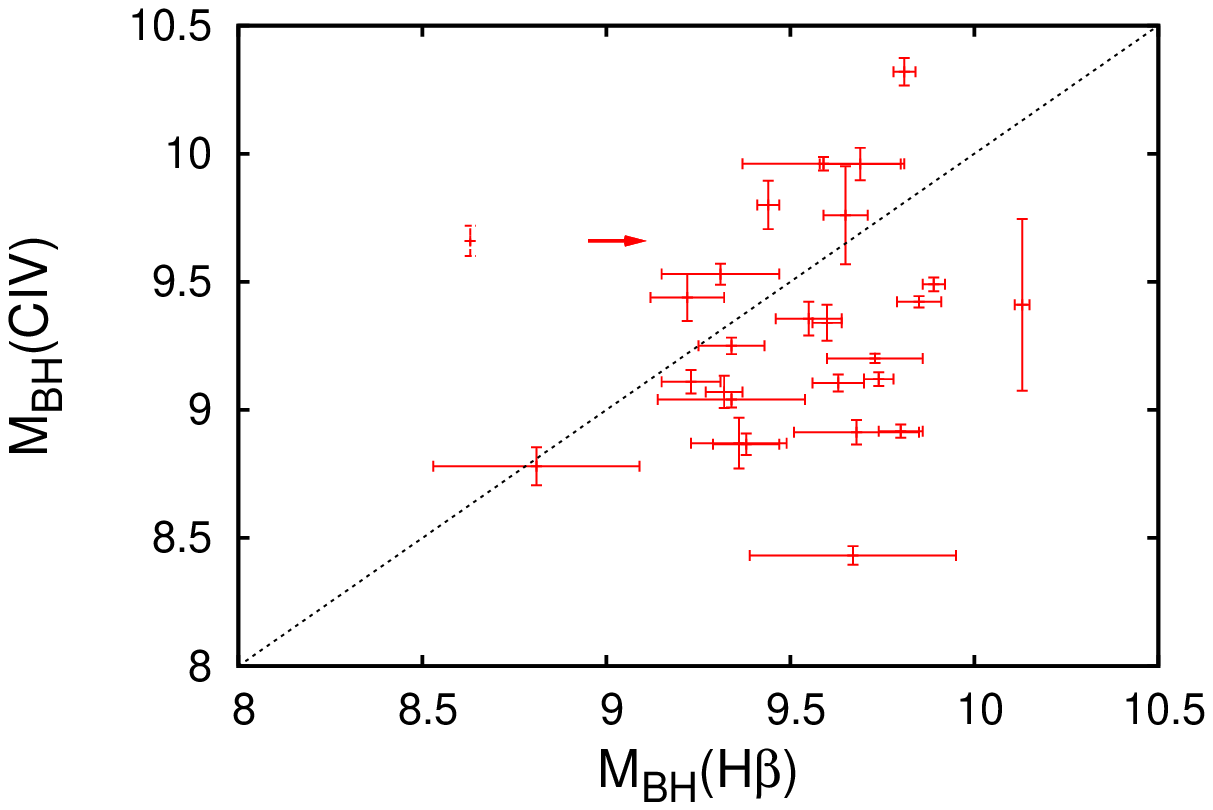}
    \end{minipage}
  \hspace{2.0pc} 
  \begin{minipage}{.45\linewidth} 
  \FigureFile(75mm,140mm){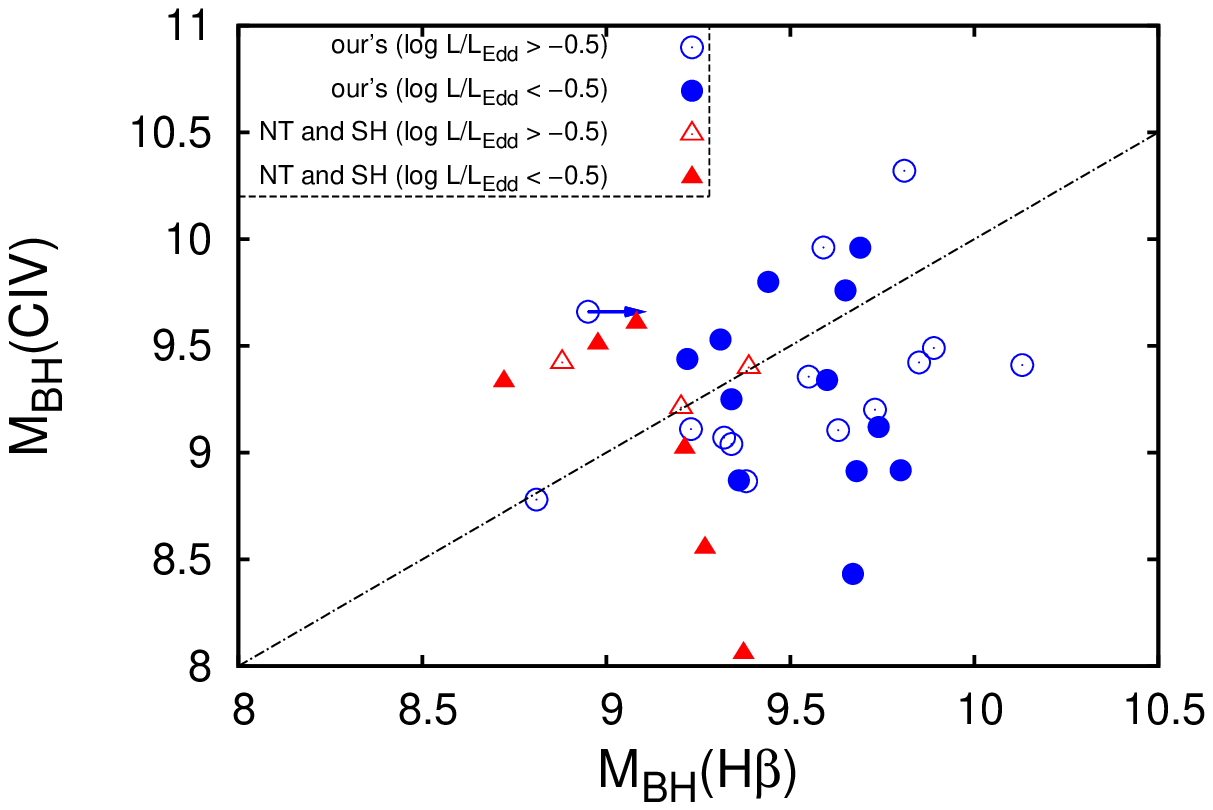}
   \end{minipage}
  \end{center}
   \caption{Left: Comparison of the SMBH masses ($M_\mathrm{BH}$) estimated from the H$\beta$ method (abscissa) and the C\emissiontype{IV} method (ordinate). C\emissiontype{IV} data are from \citet{Shen2011}. Two QSOs are not plotted here due to lack of C\emissiontype{IV} data. The right arrow means a lower limit. The dashed line represents a 1:1 correspondence. Right: The same plot as the left figure but with the Eddington ratio information added. The open blue circles are targets with high Eddington ratios with log($L_\mathrm{bol}/L_\mathrm{Edd}) > -0.5$. The filled blue circles are those with low Eddington ratios with log($L_\mathrm{bol}/L_\mathrm{Edd}) < -0.5$. The open red triangles correspond to Netzer and Shemmer's sample which has a high Eddington ratio of log($L_\mathrm{bol}/L_\mathrm{Edd}) > -0.5$. The filled red triangles are those with low Eddington ratios of log($L_\mathrm{bol}/L_\mathrm{Edd}) < -0.5$. For two QSOs in our sample and five QSOs in Netzer and Shemmer's sample, C\emissiontype{IV} data are not available in \citet{Shen2011}. The right arrow indicates a lower limit. We find that samples with high Eddington ratios appear to be distributed in the upper-left side compared to those with low Eddington ratios which distribute in both the upper-left and lower-right sides.}\label{fig:6}
\end{figure}

\begin{figure}
  \begin{center}
    \FigureFile(75mm,140mm){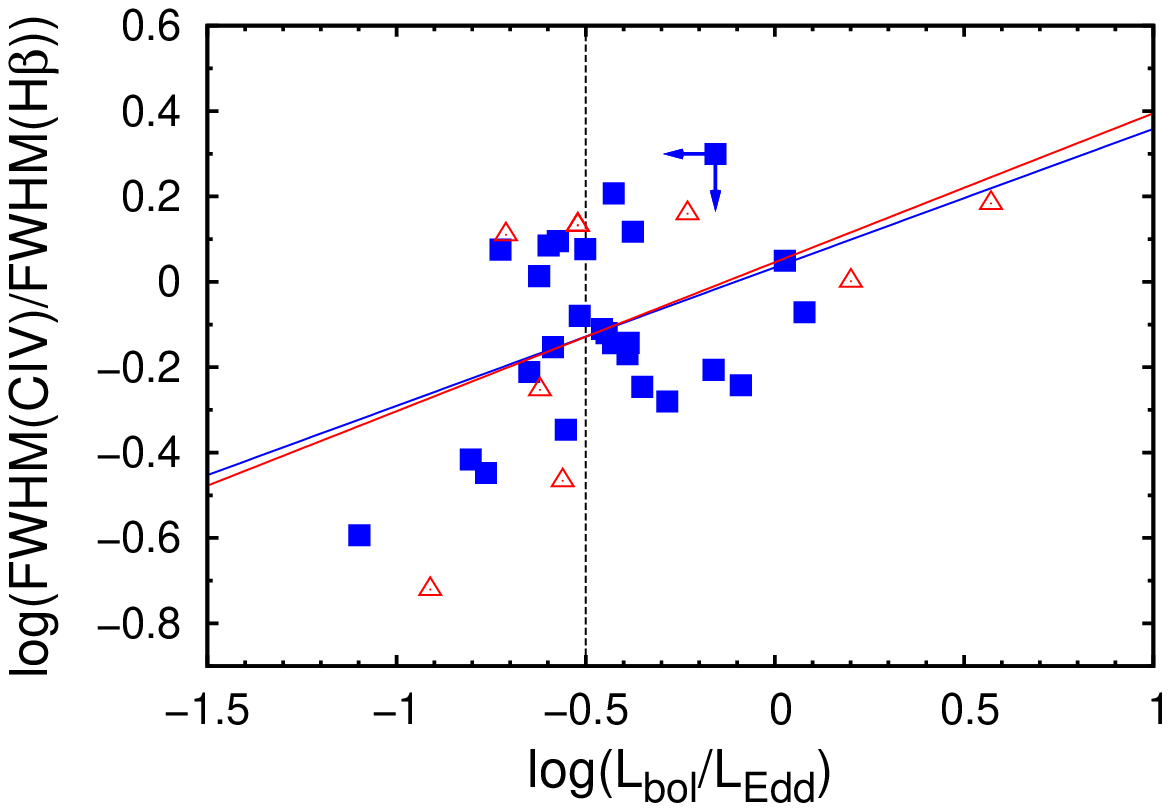}
  \end{center}
   \caption{Relationship between Eddington ratio (abscissa) and the FWHM(C\emissiontype{IV}) to FWHM(H$\beta$) ratio (ordinate) . The filled blue boxes show our sample, and the open red triangles show the sample of \citet{Netzer2007} and \citet{Shemmer2004}. Two QSOs in our sample and five QSOs in Netzer and Shemmer's sample are not plotted here because C\emissiontype{IV} data are not available in \citet{Shen2011}. The blue line represents the best fit for our sample only. The red line shows the best fit for combined samples (ours, Netzer's, and Shemmer's). Arrows mean upper limits. The dashed line represents the border between the high and low Eddington ratios.}\label{fig:7}
\end{figure}

\end{document}